\documentclass{JFM-FLM_Au}
\usepackage{tikz}
\usepackage{xcolor}
\usepackage[x11names]{xcolor}
\usepackage{booktabs}

\newcommand{\bu}{\mathbf{u}}

\newcommand{\Ra}{\mathrm{Ra}}
\newcommand{\Nu}{\mathrm{Nu}}

\definecolor{deepskyblue}{RGB}{0,191,255}
\definecolor{gold}{RGB}{255,215,0}
\definecolor{pink}{RGB}{255,192,203}
\definecolor{green}{RGB}{0,128,0}
\definecolor{darkorange}{RGB}{255,140,0}
\definecolor{purple}{RGB}{128,0,128}

\newcommand{\EmptyTriangle}[1]{%
\tikz{\draw[#1, thick] (0,0)--(0.18,0.3)--(0.36,0)--cycle;}
}

\newcommand{\FilledTriangle}[1]{%
\tikz{\fill[#1] (0,0)--(0.18,0.3)--(0.36,0)--cycle;}
}

\newcommand{\HalfTriangle}[1]{%
\tikz{
\fill[#1] (0,0)--(0.18,0.3)--(0.18,0)--cycle;
\draw[#1, thick] (0,0)--(0.18,0.3)--(0.36,0)--cycle;}
}
\newcommand{\EmptySquare}[1]{%
\tikz{\draw[#1, thick] (0,0) rectangle (0.25,0.25);}
}

\newcommand{\FilledSquare}[1]{%
\tikz{\fill[#1] (0,0) rectangle (0.25,0.25);}
}

\newcommand{\HalfSquare}[1]{%
\tikz{
\fill[#1] (0,0) rectangle (0.125,0.25);
\draw[#1, thick] (0,0) rectangle (0.25,0.25);
}
}

\lefttitle{A. Tulekeyev and P. Garaud.}
\righttitle{Journal of Fluid Mechanics}

\title{Fluxes through a uniform double-diffusive staircase at low Prandtl number.}

\author{A. Tulekeyev\aff{1} and P. Garaud\aff{1}}

\affiliation{\aff{1}Department of Applied Mathematics, Baskin School of Engineering, University of California Santa Cruz, 1156 High Street, Santa Cruz, CA 95064, USA}

\corresau{A. Tulekeyev, atulekey@ucsc.edu}

\begin{document}
\maketitle

\begin{abstract}

 Double-diffusive convection in the diffusive regime is often observed to be in a long-lived  layered state, well-known  
in the Arctic ocean and in volcanic lakes, and thought to exist in the interiors of giant planets. 
 The density profile resembles a ‘staircase', 
with stacks of convective layers of uniform density separated by strongly stratified 
interfaces. The governing parameters are the Prandtl number $\Pr$, 
diffusivity ratio $\tau$,  
density ratio $R_\rho$, and Rayleigh number $\Ra$. In this work, we perform direct numerical simulations in triply periodic domains to systematically investigate the properties of statistically stationary double-diffusive staircases with $\tau \le \Pr \ll 1$. 
We demonstrate that they can exist for significantly larger density ratios  than previously reported, 
and propose a new criterion for the existence of long-lived layered states. 
Notably, this criterion depends weakly on $\Ra$. In all simulations, we measure the total fluxes of density $F^{\rm tot}_\rho$, temperature $F^{\rm tot}_T$ and composition $F^{\rm tot}_C$, and the flux ratio $\gamma =F^{\rm tot}_C/F^{\rm tot}_T$. We confirm that $\gamma$ is approximately constant for moderate to large $R_\rho$ but also depends weakly on $\Ra$.
We show that the Nusselt numbers for composition and temperature scale as $\Ra^{1/3}$ in this limit, which notably differs from
the $(\Ra \Pr)^{1/3}$ scaling previously found 
at lower $R_\rho$. 
 We propose a new model for these findings that is based on the rise and turbulent mixing of density anomalies that form at the edges of the interfaces. The model explains the existing low $\Pr$  data with reasonable accuracy. 
\end{abstract}


\section{Introduction}

\subsection{Oscillatory double-diffusive convection} \label{sec:intro-motiv}

Double-diffusive convection refers to the nonlinear outcome of a class of instabilities that occur in stably-stratified fluids specifically because their density depends on two components that diffuse at different rates \citep{stern1960}. In many cases, temperature is the rapidly diffusing component, and the concentration of a particular chemical is the slowly diffusing one. In oceanography, the chemical considered is salt \citep{Schmitt1994}; in the interiors of giant planets and stars, it can be any element heavier than the dominant constituent of the fluid by mass, a common example being Helium in a Hydrogen-rich fluid, and another one being a so-called `heavy element' in a Hydrogen-Helium mixture \citep{Garaud2020}.   
Several types of double-diffusive instabilities exist depending on the signs and orientations of the background temperature and composition gradients \citep[see the review by][]{radko2013double}. 

In this paper, we focus on a type of double-diffusive convection that takes place in the presence of a stabilizing vertical composition gradient and a destabilizing vertical temperature gradient \citep{Walin1964}. 
This situation is common in the Arctic \citep{TimmermansEtal2008} and Antarctic regions of the ocean \citep{Muench1990}, and in volcanic lakes \citep{Wüestetal2012}. 
In the astrophysical context, it can occur in the deep interiors of stars \citep{Kato1966, Spiegel1969,Spruit1992} and giant planets \citep{Stevenson1982,MankovichFuller2021}, which often contain stabilizing composition gradients as a result of nuclear fusion (in the case of stars) or as a legacy of their formation (in the case of giant planets).

In what follows, we refer to the linear instability of this type of stratification, and its nonlinear evolution into small-scale, weakly nonlinear wave-like turbulence, as Oscillatory Double-Diffusive Convection (ODDC). We shall ignore the effects of rotation and magnetic fields for simplicity. It is customary to study ODDC by considering perturbations on a linearly stratified background with a constant temperature gradient $dT^*_b/dz<0$ and a constant composition gradient $dC^*_b/dz<0$. 
The instability takes the form of gravity waves supported by the stable compositional stratification, that are destabilized by thermal diffusion. 
When the size of the domain is much larger than the intrinsic length scale of the instability (so it can be considered infinite for practical purposes), ODDC is mainly governed by three non-dimensional parameters: the Prandtl number $\Pr$, the diffusivity ratio $\tau$, and the density ratio $R_{\rho}$, defined as:
\begin{equation}\label{eq:parameters}
\Pr = \frac{\nu^*}{\kappa_T^*}, \quad \tau = \frac{\kappa_C^*}{\kappa_T^*}, \quad R_{\rho} = \frac{\alpha_{C}^*  | dC^*_b/dz |}{\alpha_{T} | dT^*_b/dz|},
\end{equation}
where $\nu^*$ is the kinematic viscosity of the fluid, $\kappa_{C}^*$ and $\kappa_{T}^*$ are compositional and thermal diffusivities, respectively, $\alpha_{T}^*$ is the coefficient of thermal expansion, and $\alpha_{C}^*$ is the coefficient of compositional contraction. Here, and henceforth, dimensional quantities are denoted with star superscripts, while dimensionless quantities are not. In a finite domain of height $L_z^*$, a fourth dimensionless parameter appears, namely the thermal Rayleigh number defined as 
\begin{equation}
    \Ra = \frac{\alpha^*_T g^* |dT^*_b/dz| L_z^{*4}}{\kappa_T^* \nu^*},
    \label{eq:Radef}
\end{equation}
where $g^* $ is the local gravity. Two more parameters can be added to describe the domain aspect ratio, although these will not be varied in this work. 

The stability properties of ODDC to infinitesimal perturbations in this model setup were established by \citet{baines1969}. They showed that for large $\Ra$ ODDC is linearly unstable provided
\begin{equation}\label{eq:oddc_crit}
    1\le R_{\rho}\le \lim_{\Ra \rightarrow \infty} R_{c}(\Pr,\tau,\Ra) = \frac{\Pr +1}{\Pr+\tau} ,
\end{equation}
where the lower limit $R_{\rho} = 1$ is the threshold for instability to regular convection, and the upper limit $R_\rho= R_{c}(\Pr,\tau,\Ra)$ is the marginal stability limit for ODDC.

In the oceanographic context, and in any fluid that has a relatively high Prandtl number, $R_c$ is close to 1 so the region of parameter space that is linearly unstable is quite small (for example, $R_{c}(\Pr = 7, \tau = 0.01, \Ra\rightarrow \infty) \approx 1.1$ in salt water). 
In the low $\Pr$ and $\tau$ regime characteristic of astrophysical fluids, by contrast, the linear instability range is much broader.  
Typically, both $\Pr$ and $\tau$ are $O(10^{-3} - 10^{-1})$ in planetary interiors, and $\le O(10^{-5})$ in stellar interiors, so $R_{c}$ ranges from $O(1 - 10^{2})$ for giant planets and exceeds $O(10^{5})$ for stellar interiors, suggesting that ODDC is much more prevalent in these contexts \citep{Garaud2018,Garaud2020}.

\subsection{ODDC vs. layered diffusive convection}

As mentioned above, ODDC evolves into a weakly nonlinear state of wave-like turbulence when arising from the linear instability of a fluid with constant temperature and composition gradients  \citep[see][for a study of nonlinear ODDC in the astrophysical regime]{Mirouh2012,Moll2016}. However, another fully nonlinear state also exists, that will be referred to as Layered Diffusive Convection (LDC) hereafter, in which fully convective layers (with almost constant horizontally-averaged density, temperature, and composition) are separated by stably-stratified horizontal interfaces. This results in staircase-like density, temperature and composition profiles, with one or more steps. These have been observed in the polar oceans and in volcanic lakes \citep{Wüestetal2012, TimmermansEtal2008,Muench1990}, and have long been hypothesized to exist in the 
low $\Pr$ interiors of giant planets and stars \citep{Stevenson1982,GoughToomre1982}.
These staircases are sometimes called diffusive-convective staircases \citep[see the review by][]{TimmermansCarpenter2026}, a terminology we shall adopt here too. In this  paper, our goal is to model the properties of diffusive-convective staircases at low $\Pr$. 

\cite{Veronis1965} demonstrated that LDC can arise from a subcritical branch of instability that persists as long as $R_\rho \le R_L$, where the subcritical threshold $R_L$ satisfies
\begin{equation}
    R_c(\Pr,\tau,\Ra) \le R_L(\Pr,\tau,\Ra).
\end{equation}
He further argued that 
\begin{equation}
    \lim_{\Ra\rightarrow \infty} R_L(\Pr,\tau,\Ra)= \tau^{-1}.
\end{equation}
This subcritical branch explains why diffusive-convective staircases in the polar oceans and in volcanic lakes are allowed to exist even though they almost always have $R_\rho > R_c$. Formation mechanisms for these staircases at $\Pr \ge O(1)$ are reviewed by \citet{TimmermansCarpenter2026}.

At low $\Pr$, interestingly, LDC can be achieved through two completely different routes. On the one hand, the same subcritical branch of instability also exists, and directly leads to LDC with a suitable choice of initial conditions \citep[see][]{Molletal2017}. On the other hand, \citet{rosenblumal2011} and \citet{Mirouh2012} demonstrated that ODDC  can spontaneously evolve into LDC through a secondary instability called the $\gamma$-instability \citep{radko2003mechanism}, whenever 
\begin{equation}
    R_\rho < R_\gamma.
\end{equation}
The marginal stability threshold for the $\gamma-$instability, $R_\gamma$, is a function of $\Pr$ and $\tau$ (and possibly of $\Ra$ as well at low $\Ra$), and is always smaller than $R_c$ because the $\gamma-$instability first requires ODDC to develop, so
\begin{equation}
R_\gamma (\Pr,\tau,\Ra) \le R_c(\Pr,\tau,\Ra).
\end{equation}
Specific values of $R_\gamma$ for selected parameter pairs $(\Pr,\tau)$ at high $\Ra$ were computed by \citet{Mirouh2012}, and are presented in table \ref{tab:simulations}. Note that \citet{Mirouh2012} (and others since then) have used the notation $R_L$ for what we call $R_\gamma$ here, but we now reserve $R_L$ more generally for the overall LDC threshold. 

Parameter space at low $\Pr$ is thus partitioned into the following domains (in the absence of rotation and magnetic fields):
\begin{itemize}
\item For $R_\rho \le 1$, the stratification is statically unstable, and always fully convective.
\item For $1 < R_\rho < R_\gamma$, the stratification is statically stable but ODDC unstable. ODDC itself is $\gamma-$unstable, which means that the system always eventually reaches a LDC state \citep{rosenblumal2011,Mirouh2012,Woodal13}.
\item For $R_\gamma \le R_\rho <  R_c$, the stratification is ODDC unstable and ODDC is $\gamma-$stable. The ultimate nonlinear state is either ODDC starting from infinitesimal perturbations \citep{Moll2016} or LDC starting from a suitable choice of initial conditions through a subcritical instability
\citep{Molletal2017}. 
\item For $ R_c \le R_\rho < R_L$, the stratification is ODDC-stable, but LDC can be achieved with a suitable choice of initial conditions through a subcritical instability.
\item for $R_\rho \ge R_L$, the only possible state is the conductive state. 
\end{itemize} 

\subsection{Vertical fluxes and long-term evolution of diffusive-convective staircases}

Geophysical diffusive-convective staircases (for $\Pr \ge O(1)$) are observed to be long-lived and coherent over large horizontal scales. 
A long-standing problem in the field is to quantify the rate at which temperature, composition, and density are transported through the layers and interfaces within the staircase.
A common approach used by many previous experimental studies considers a simpler model with a single interface between two convective layers.  
In that case the density ratio is defined from the temperature jump $\Delta T^*$ and composition jump $\Delta C^*$ across the interface, as 
\begin{equation}
    R_{\rho}  = \frac{\alpha^*_C \Delta C^* }{ \alpha_T^* \Delta T^*}.
\end{equation} 

One of the first experimental studies of this type of two-layer system was carried out by \citet{TURNER1965}, who measured fluxes through the double-diffusive interface in a salt solution tank heated from below. He argued on dimensional grounds that the heat and salt fluxes through the interface ought to be proportional to $\Delta T^{*4/3}$. He also found that the ratio of the density flux due to salt transport $\alpha_C^* F^*_{C}$, to (minus) the density flux due to heat transport $\alpha_T^* F^*_{T}$, a quantity often called the flux ratio $\gamma$, is approximately constant with 
\begin{equation}
    \gamma = \frac{\alpha_C^* F^*_{C}}{\alpha_T^* F^*_{T}} \simeq 0.15 \quad {\rm  for} \quad R_{\rho} \gtrsim 2,
\end{equation}
and increases as $R_{\rho} \rightarrow 1$. Not long thereafter, \citet{Shirtcliffe_1973} experimentally formed a double-diffusive interface by using a salt solution layered over a sugar solution ($\tau \simeq 1/3$, $\Pr \simeq 1000$), and found that $\gamma \simeq 0.6$ for $R_{\rho}\ge 1.7$. Taken together, the measured values of the flux ratio in both experiments appear to satisfy:
\begin{equation}
    \gamma  = \tau^{1/2}, \quad \mbox{for}\quad R_{\rho} \gtrsim 2.
\end{equation}

\citet{Linden_Shirtcliffe_1978} later provided some theoretical rationale for this finding by creating a model of a double-diffusive interface, and further argued that their interface can only be in a statistically stationary state provided $R_\rho < \tau^{-1/2}$. Subsequent laboratory and numerical experiments aimed at measuring $\gamma$ in LDC and testing the \citet{Linden_Shirtcliffe_1978} theory were inconclusive.
While some of these studies agreed that $\gamma \simeq \tau^{1/2}$ \citep{Crapper1975,Stampetal1998}, others found significantly different results \citep{MARMORINO1976,Takao1980,Newell1984,Fernando1989,Carpenter_Sommer_Wüest_2012}, and the source of the discrepancy remains to be fully understood \citep[although see][for a possible resolution]{WORSTER2004}.


At low $\Pr$, by contrast, diffusive-convective staircases appear to rapidly coarsen with time in all non-rotating and non-magnetic simulations performed to date \citep{rosenblumal2011,Mirouh2012,Woodal13,Fuentes2022,Tulekeyev2024}, regardless of the domain shape (Cartesian or spherical) or boundary conditions applied (periodic, fixed temperature and composition, or fixed flux of temperature and composition): layers rapidly merge until a single convective zone remains. We note that rotation and/or magnetic fields can both greatly slow down the merger rate but do not seem to halt it altogether \citep{MollGaraud2017,Sanghi_2022,Fuentes_2023,Pruzina2025,Pruzina2026}. 

Quantifying the fluxes of temperature and composition through LDC is also a key question in stellar and planetary astrophysics. 
Using Direct Numerical Simulations (DNS) in a triply-periodic domain, with a fixed background density ratio $R_\rho$ (see \S\ref{sec:method} for more on this model setup), \citet{rosenblumal2011} showed that the vertical transport of heat, composition and density is strongly enhanced in LDC compared with ODDC, and increases with the layer height as the staircase coarsens. Subsequent work by \citet{Woodal13} measured the turbulent fluxes through the staircase as functions of the average layer height. Focusing on the $\gamma$-unstable regime ($R_\rho<R_\gamma$), their results suggest that the thermal Nusselt number is proportional to $(\Ra\Pr)^{1/3}$, where $\Ra$ in their work is based on the average layer height rather than the computational domain size. 
 
 \cite{Molletal2017} expanded upon the work of \citet{Woodal13} by investigating LDC in the $\gamma$-stable regime ($R_\rho>R_\gamma$), while restricting their analysis to the case where $\Pr = \tau$. To trigger the subcritical branch of instability discussed above, they 
 initialized the simulations with a pre-existing staircase structure, and let it relax to a statistically stationary LDC state.   
Notably, \citet{Molletal2017} did not find statistically stationary layered states for $R_\rho > R_c$ when $\Pr = \tau$. They did not report on the dependence of the temperature flux through the staircase, and focused instead on  the flux ratio. They
showed that the previously suggested model 
by \citet{Linden_Shirtcliffe_1978}, which predicts $\gamma \simeq \tau^{1/2}$, does not hold for low $\Pr$ LDC. 

While the result of \citet{Woodal13} and \citet{Molletal2017}  provided important information about low $\Pr$ diffusive-convective staircases, it is crucial to continue exploring parameter space more comprehensively. Indeed, \citet{Woodal13} focused on the small range of $R_\rho$ that is $\gamma$-unstable, and the DNS of \citet{Molletal2017} were limited to  cases where $\Pr = \tau$ while in reality, astrophysical fluids generally have $\tau < \Pr$. 
In this work, we therefore continue to explore  properties of LDC at low $\Pr$ by adding new simulations to the existing dataset, and exploring the data more systematically.

Furthermore, because the temperature and composition fluxes are strongly dependent on the average layer height, and the latter evolves via mergers, it is important to develop a theory for the merger process itself and to quantify the typical merger timescale. \citet{radko2007} developed a layer merger theory for a single-component `infinite uniform density staircase', i.e., an infinite staircase with equal layer heights $L_z^*$, and equal density jumps across each interface $\Delta \rho^*$. This theory predicts the merger timescale as long as the functional dependence of the density flux through the staircase on both $L_z^*$ and $\Delta \rho^*$ is known. Extending the theory to the doubly-stratified case thus requires a model for the density flux in an infinite uniform staircase as a function of input parameters. 

In what follows, we therefore focus on studying the properties of LDC in an infinite uniform staircase at low $\Pr$. This has the dual advantage of having statistically stationary solutions, and will yield information on the density flux that can then be used in conjunction with the theory of \citet{radko2007} to predict the layer merger timescale.

 In \S\ref{sec:method}, we present the model setup and define the concept of a uniform density staircase. In \S\ref{sec:results}, we present results on the properties of interfaces and convective layers in these staircases. In \S\ref{sec:model}, we propose an almost ad-hoc theoretical model of LDC to predict the fluxes through the density staircase as a function of input parameters. 
Finally in \S\ref{sec:discussion}, we summarize and discuss our results.

\section{Methodology}\label{sec:method}

In what follows, we use the Boussinesq approximation, assuming that the typical double-diffusive layer height is much smaller than a pressure or density scale height, and that the flow velocities are much smaller than the sound speed. Within this approximation, temperature fluctuations $T^*$ away from the local mean $T_m^*$, and composition fluctuations $C^*$ away from the local mean $C_m^*$, are both small enough that the equation of state can be linearized as 
\begin{equation}\label{eq:setup_eos}
    \frac{\rho^*}{\rho_m^*} = -\alpha^*_T T^* + \alpha^*_C C^*,
\end{equation}
where $\rho^*$ is the density fluctuation away from the reference density. The governing equations of motion are: 
\begin{eqnarray}
&& \nabla^* \cdot \bu^*  = 0 , \label{eq:dim_divu} \\
&& \frac{\partial \bu^* }{\partial t^* } + \bu^*  \cdot \nabla^*  \bu^*   =  - \frac{1}{\rho_m^* } \nabla^*  p^*  + (\alpha^* _T T^*  - \alpha^* _C C^* ) g^*  {\bf e}_z  +  \nu^*  \nabla^{*2} \bu^* , \label{eq:dim_mom} \\
&&  \frac{\partial T^* }{\partial t^* } + \bu^*  \cdot \nabla^*  T^*   =  \kappa_T^*  \nabla^{*2} T^* , \label{eq:dim_T} \\
&& \frac{\partial C^* }{\partial t^* } + \bu^*  \cdot \nabla^*  C^*  =  \kappa_C^*  \nabla^{*2} C^* , \label{eq:dim_C}
\end{eqnarray}
where $\bu^*  = (u^* ,v^* ,w^* )$ is the velocity field, and $p^* $ is the pressure perturbation away from hydrostatic equilibrium. 

Infinite uniform staircases can ideally be modeled using DNS in domains with triply-periodic boundary conditions, using the same setup as in \citet{Molletal2017}. We assume that there is a stratified background state (denoted as $T^*_b(z)$ and $C^*_b(z)$) with a constant temperature gradient $dT^*_b/dz^*$ and a constant composition gradient $dC^*_b/dz^*$ (both gradients are negative). We then decompose the total temperature and composition fields as the sum of these background contributions plus triply-periodic perturbations $\tilde T^*$ and $\tilde C^*$: 
\begin{equation} \label{eq:T-C_fullState}
         T^*  =  z^*\frac{d T^*_{b}}{dz^*} + \tilde{T}^*,  C^* =  z^*\frac{d C^*_{b}}{d z^*} + \tilde{C}^*,
   \end{equation}
   where $z^*$ is the vertical coordinate (increasing upwards). The governing equations for $\tilde T^*$ and $\tilde C^*$, in addition to (\ref{eq:dim_divu}), are:
   \begin{eqnarray}
&& \frac{\partial \bu^*}{\partial t^*} + \bu^* \cdot \nabla^* \bu^*  =  - \frac{1}{\rho_m^*} \nabla^* \tilde{p}^* + (\alpha^*_T \tilde{T}^* - \alpha_C^* \tilde{C}^*) g^* {\bf e}_z  +  \nu^* \nabla^{*2} \bu^*, \label{eq:dim_momtilde} \\
&&  \frac{\partial \tilde{T}^*}{\partial t^*} + \bu^* \cdot \nabla^* \tilde{T}^*  + w^* \frac{dT^*_b}{dz^*} =  \kappa_T^* \nabla^{*2} \tilde{T}^*, \label{eq:dim_Ttilde} \\
&& \frac{\partial \tilde{C}^*}{\partial t^*} + \bu^* \cdot \nabla^* \tilde{C}^* + w^* \frac{dC^*_b}{dz^*} =  \kappa^*_C \nabla^{*2} \tilde{C}^*, \label{eq:dim_Ctilde}
\end{eqnarray}
where $\tilde p^* = p^* - p^*_b(z)$ and $p^*_b(z)$ satisfies 
\begin{equation}\label{eq:density_tilde}
    \frac{1}{\rho^*_m} \frac{dp^*_b}{dz^*} = - g^*z^* \frac{d\rho^*_b}{dz^*} =  g^* z^* \left(\alpha^*_T\frac{dT^*_b}{dz^*} - \alpha^*_C  \frac{dC^*_b}{dz^*} \right).   
\end{equation}

While this formulation of the equations is entirely equivalent to (\ref{eq:dim_divu})-(\ref{eq:dim_C}), we can now apply triply-periodic boundary conditions to $\bu^*$, $\tilde{p}^*$, $\tilde{C}^*$ and $\tilde{T}^*$, to model either ODDC or LDC in the absence of solid boundaries. We use a standard non-dimensionalization \citep{radko2013double} based on the thermal diffusive length scale 
\begin{equation}
d^*  = \left( \frac{\kappa^*_T \nu^*}{\alpha_T^* g^* | dT^*_b/dz^*| } \right)^{1/4}. \label{eq:nondim-d} 
\end{equation}
 The units of length, time, velocity, temperature, and composition are:
\begin{eqnarray}
&& [l] = d^*,\quad  [t] = \frac{d^{*2}}{\kappa_T^*},\quad [u] = \frac{\kappa_T^*}{d^*},\label{eq:nondim_scales} \\
&&[\tilde T] = d^* | dT_b^*/dz^*|,\quad [\tilde C] = \frac{\alpha^*_T}{\alpha^*_C} [\tilde T].\label{eq:nondim-Cscale}
\end{eqnarray}
In these units,
\begin{eqnarray}
&& \nabla \cdot \bu = 0, \label{eq:nondim_divu}\\
&& \frac{\partial \bu}{\partial t} +  \bu \cdot \nabla  \bu  =  - \nabla p + \Pr ( \tilde T - \tilde C) {\bf e}_z  +  \Pr \nabla^2  \bu, \label{eq:nondim_mom}\\
&&  \frac{\partial \tilde T}{\partial t} +  \bu \cdot \nabla \tilde T -  w    =  \nabla^2 \tilde T,\label{eq:nondim_T} \\
&& \frac{\partial \tilde{C}}{\partial t} +  \bu \cdot \nabla \tilde{C} - R_\rho  w   = \tau \nabla^2 \tilde{C}, \label{eq:nondim_C} 
\end{eqnarray}
where $\Pr$, $\tau$, $R_\rho$ have been defined in \S\ref{sec:intro-motiv}. The dimensionless total temperature, composition and density fields away from the constant mean are
\begin{eqnarray}\label{eq:nondim-eos}
    \rho = -  T + C , \mbox{ with } \\
    T = - z + \tilde{T}, \mbox{ and } C = - R_\rho z + \tilde{C}.
 \end{eqnarray}

All of the DNS discussed in this paper have been run using the pseudo-spectral PADDI code \citep{TRAXLER_STELLMACH_GARAUD_RADKO_BRUMMELL_2011}, which solves equations (\ref{eq:nondim_divu})-(\ref{eq:nondim_C}) in a triply-periodic domain. The computational domain is always cubic, with size $L_x = L_y = L_z$ (expressed in units of $d^*$). In what follows, we primarily consider two cases: $L_z = 100$ and $L_z = 200$. We note that the thermal Rayleigh number defined in equation \eqref{eq:Radef} is related to the dimensionless domain height as 
\begin{equation}
    \Ra = L_z^4,
\end{equation}
so the $L_z = 100$ domain has $\Ra = 10^8$, while the $L_z = 200$ domain has $\Ra = 1.6 \times 10^9$. The resolution used depends on the input parameters, and is highest at low $\Pr$, $\tau$ and $R_\rho$ (see table \ref{tab:simulations} below for details). 

Some of data presented and studied in this work was previously published in \citet{Mirouh2012}, \citet{Woodal13}, and \citet{Molletal2017}, while the rest of the data is new (see table \ref{tab:simulations}). 
In the simulations of \citet{Mirouh2012} and \citet{Woodal13}, low-amplitude white noise on the grid scale was used to seed the perturbations $\tilde{T}$ and $\tilde{C}$, which then  evolved into ODDC. At low $\Pr$ and low $R_\rho$, ODDC spontaneously transitions into LDC through the $\gamma-$instability. The layers always merge until a single interface remains, at which point the flow reaches the statistically stationary layered state we are interested in. Given the periodic boundary conditions, that state models an exactly uniform staircase (i.e  a staircase with equal size steps).  

At higher $R_\rho$, the $\gamma-$instability is not active, and the layers always have to be ‘seeded' instead. This can be done in two ways, which were both used in \citet{Molletal2017}, and in the new simulations run for this work. The first option is to initialize $\tilde{T}$ and $\tilde{C}$ with vertical profiles specifically chosen to contain one convectively-unstable region (i.e. regions where the total density increases with height), and one convectively stable region, and let the system evolve into a statistically stationary layered state with a single interface. The second option is to start from the end of a simulation that had achieved such a layered state, and change the input parameters $R_\rho$, $\Pr$ or $\tau$ by a small amount, then wait until the new statistically stationary single-interface state is reached. 

\begin{figure}
    \centering
    \includegraphics[width=0.9\textwidth]{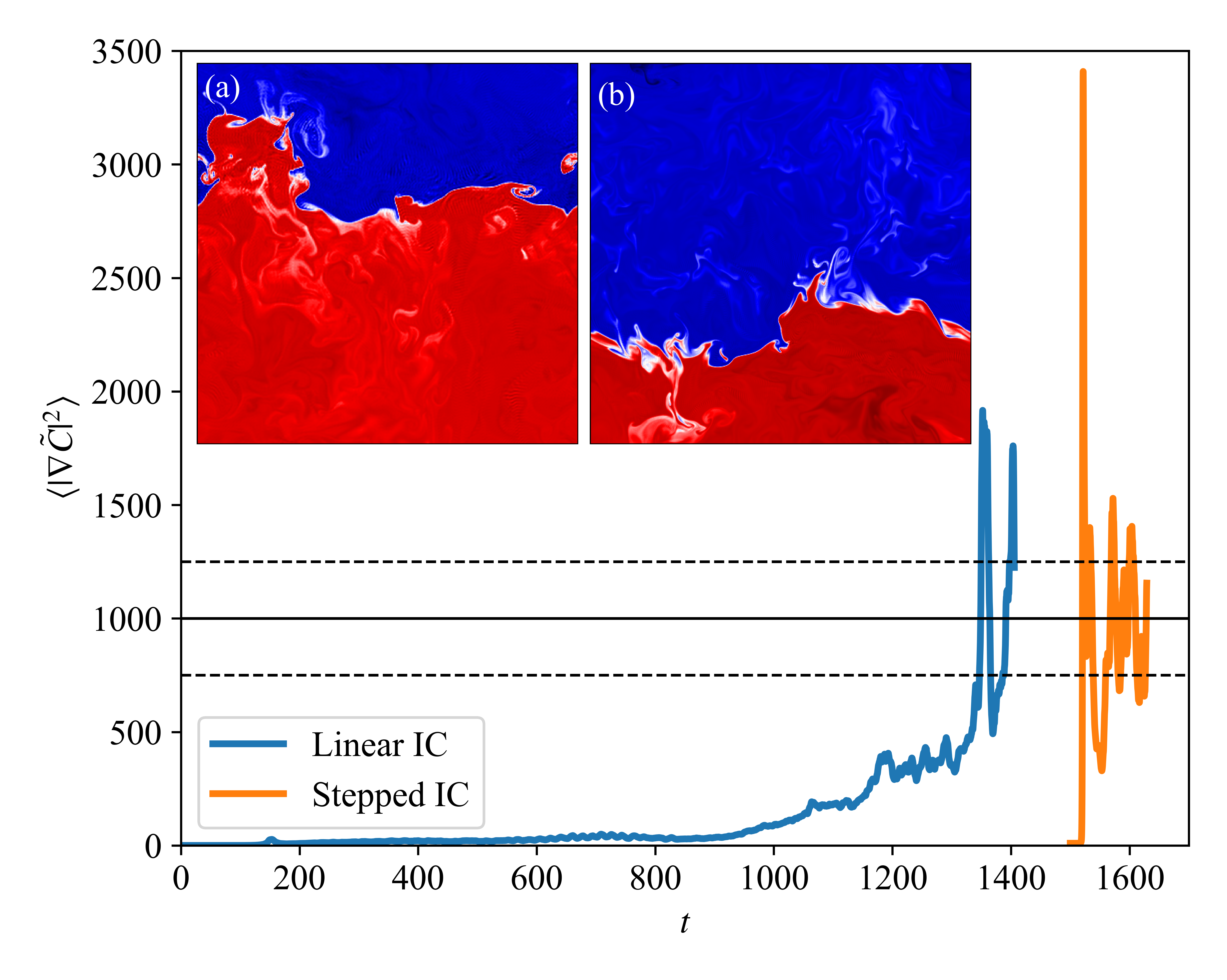}
    \caption{Comparison of two simulations at the same parameters ($\Pr = 0.3, \tau = 0.03, R_\rho = 1.5, L_z = 100$) but with different initial conditions. Time series of $\langle | \nabla \tilde C |^2 \rangle$ are shown for an initially linear density profile (blue line), that first goes through a long ODDC phase, then spontaneously develops layers through the $\gamma$-instability, and an initially stepped profile (orange line), that immediately develops into LDC. The solid horizontal line represents the temporal mean $\langle | \nabla \tilde C |^2 \rangle$ in the statistically stationary single-interface state, and the dashed lines correspond to this temporal mean plus or minus the temporal rms variability of $\langle | \nabla \tilde C |^2 \rangle$ in the statistically stationary state. The two simulations have approximately the same mean in the single-interface layered state, regardless of the initial conditions. The insets show the total composition field for ({\it a}) the linear initial conditions (with slight under-resolution leading to mild Gibbs oscillations visible near the interface) and ({\it b}) the stepped initial conditions (fully resolved). }
\label{fig:sampletimeseries}
\end{figure}

We have verified that the statistically stationary single-interface LDC state reached is independent of the way the simulation was initialized -- as long as all input parameters $\Pr,\tau,R_\rho$ and $L_z$ are the same. An example is provided in figure \ref{fig:sampletimeseries}, which shows the temporal evolution of the quantity $\langle |\nabla \tilde C|^2 \rangle$ (where $\langle \cdot \rangle$ denotes a volume average), for 
two simulations at the same parameters, namely $\Pr=  0.3$, $\tau = 0.03$, $R_\rho=  1.5$ and $L_z = 100$. At these parameters, ODDC is $\gamma-$unstable \citep{Mirouh2012}. In the first case (figure \ref{fig:sampletimeseries}({\it a}), blue line), the simulation was initialized with small amplitude white noise on top of a linearly stratified background, and layers naturally appear through the $\gamma-$instability, then merge until a single layer is left. In the second case (figure \ref{fig:sampletimeseries}({\it b}), orange line), the simulation was initialized with a convectively-unstable region, that rapidly becomes violently convective, then settles down into the same statistically stationary LDC state. We note that the first simulation had a lower resolution and was slightly under-resolved in the single-layer phase (which is the reason why its integration was halted shortly after reaching that state), while the second is fully resolved (see insets). This does not affect the quantity $\langle |\nabla \tilde C|^2 \rangle$ measured, however, in as much as the means are consistent with one another in the two simulations. For the purpose of this study, and whenever two simulations are available at the same input parameters, we only use the higher-resolution simulation.

\begin{table}
    \centering
    \begin{tabular}{|c|c|c|c|c|c|c|}
    \toprule
    $\Pr$, $\tau$  & $R_{c}$ & $ R_{\gamma} (\simeq) $ & $L_z$ & $R_{\rho}$ & $N_{x,y}$,$N_{z}$ & Marker \\
    \midrule
           0.03, 0.03 & 17.1667 &  3.25  &100  & $1.5^{*}$  & (576, 768) & $\FilledTriangle{pink}$ \\
        0.03, 0.03 & 17.1667& 3.25  & 100  &  $3.0^{\dagger}$, $5.25^{\dagger}$ & (576,576), (384,384) &$\HalfTriangle{pink}$\\
         0.03, 0.03 & 17.1667&  3.25 & 100  &  $7.87^{\dagger}$, $10.0^{\dagger}$, 13.0 & (384,384) &$\EmptyTriangle{pink}$\\
          0.1, 0.1 & 5.5& 2  & 100  &  $1.25^{*}$, $1.4^{\dagger}$, $1.8^{\dagger}$ & (384,384) & $\FilledTriangle{gold}$\\
        0.1, 0.1 &  5.5& 2  &100 &  $2.2^{\dagger}$, $2.6^{\dagger}$, $3.0^{\dagger}$ & (384,384) &$\HalfTriangle{gold}$\\
         0.1, 0.1 & 5.5&  2  & 100 &  $3.5^{\dagger}$, $4.25^{\dagger}$, $4.5^{\dagger}$ & (384,384) &$\EmptyTriangle{gold}$\\
         0.1, 0.1 & 5.5& 2  &200 &  $1.5^{\dagger}$, $2.2^{\dagger}$, $3.0^{\dagger}$, $3.5^{\dagger}$ & (384,384)  & $\HalfSquare{gold}$\\
         0.1, 0.1 & 5.5& 2 &200 &  $4.0^{\dagger}$, $4.5^{\dagger}$, $5.0^{\dagger}$ & (384,384)   & $\EmptySquare{gold}$\\
          0.3, 0.3 &  2.1667&  1.35   &100  &  $1.2^{*}$, $1.25^{*}$  & (192,192)  & $\FilledTriangle{deepskyblue}$\\
        0.3, 0.3 &    2.1667& 1.35 & 100 &  $1.3^{\dagger}$, $1.5^{\dagger}$, $1.7^{\dagger}$ &  (144,144) &$\HalfTriangle{deepskyblue}$\\
         0.3, 0.3 &    2.1667 & 1.35  & 100 & $1.75^{\dagger}$, $1.8^{\dagger}$, $1.9^{\dagger}$, $2.0^{\dagger}$, $2.05^{\dagger}$ &   (144,144) &   $\EmptyTriangle{deepskyblue}$\\
         0.3, 0.3 &  2.1667& 1.35 & 200 &  $1.2^{*}$ &  (384,384)&$\FilledSquare{deepskyblue}$\\
         0.3, 0.3 &  2.1667& 1.35 & 200 &  $1.5^{\dagger}$, $1.75^{\dagger}$& (288,288) & $\HalfSquare{deepskyblue}$\\
         0.3, 0.3 &  2.1667 &1.35 & 200 &  $2.0^{\dagger}$, $2.05^{\dagger}$ & (288,288) & $\EmptySquare{deepskyblue}$\\
          0.3, 0.1 &3.25& 2 (?)   & 100  &  $1.1^{*}$, $1.2^{*}$, $1.4^{*}$  & (384, 384), (240, 240), (192,192) &$\FilledTriangle{green}$ \\
        0.3, 0.1 &3.25 &  2 (?)  &100 & 2.0, 2.5, 3.0  & (384, 384)& $\HalfTriangle{green}$\\
         0.3, 0.1 &3.25&  2 (?)  &100  &  3.25, 3.5 & (384, 384) & $\EmptyTriangle{green}$   \\
        0.1, 0.03 & 8.4615& ? &100  & 3.0,  5.5 & (576, 576) & $\HalfTriangle{darkorange}$ \\
         0.1, 0.03 & 8.4615& ? &100 &  6.0, 8.0, 10.0 & (384, 384) &  $\EmptyTriangle{darkorange}$\\
          0.3, 0.03 & 3.9394& 2 (?)  &100  &  1.25, 1.5  & (576, 576)  &  $\FilledTriangle{purple}$\\
        0.3, 0.03 & 3.9394& 2 (?) & 100  &  3.0 &  (384, 384) & $\HalfTriangle{purple}$\\
         0.3, 0.03 & 3.9394& 2 (?) & 100  &  5.5, 6.0, 8.0 & (384, 384) & $\EmptyTriangle{purple}$\\
    \bottomrule
    \end{tabular}
    \caption{List of simulations at different $\Pr$, $\tau$, $R_{\rho}$ and $L_{z}$, with corresponding $R_{c}$ (column 2) and $R_{\gamma}$ (column 3) and colour/marker-coding (column 7) used in all the figures. In all cases, the computational domain is cubic. In column 5, the $*$ superscript indicates $\gamma-$unstable simulations from \citet{Mirouh2012}  and \citet{Woodal13}, and the $\dagger$ superscript marks the $\gamma$-stable simulations from \citet{Molletal2017}. The simulations with no superscript are new. In column 3, question marks indicate cases where the measurement is uncertain, or is not available.}
    \label{tab:simulations}
\end{table}

Table \ref{tab:simulations} contains a summary of all the simulations available, including their input parameters. Those previously published by \citet{Mirouh2012} and \citet{Woodal13} are restricted to $\gamma-$unstable cases, for which $R_\rho$ is relatively small. The simulations of 
\citet{Molletal2017} include much larger values of $R_\rho$ but are restricted to cases where $\Pr = \tau$. We have run additional simulations with both $\Pr = \tau$ and $\Pr > \tau$ (which is the common situation in most astrophysical systems) to improve parameter space coverage. We have limited the exploration to $0.3 \ge \Pr \ge  \tau \ge 0.03$, which is a range of parameter appropriate for liquid metals, some white dwarf stars, and for the interiors of giant planets. 

\section{Results of the numerical experiments}\label{sec:results}

In this section, we present detailed aspects of the data obtained from these numerical experiments. In \S\ref{sec:gamma+fluxes}-- \S\ref{sec:interface-thickness}, we analyse the interface properties. \S\ref{sec:convlayers} focuses on the  convective layers and their properties, and \S\ref{subsec:densityfluxlaw} measures the density flux through the staircase. These results  will later be modeled in \S\ref{sec:model}.  

To simplify notation, we define the following averages for a quantity $Q$:
\begin{eqnarray}
    \overline{Q}(z) = \frac{1}{A(t_{f}-t_{s})}\int_{t_{s}}^{t_f}\int_{A}Q(x,y,z,t)dAdt,\\ \langle Q\rangle(t) = \frac{1}{V}\int_{V}Q(x,y,z,t)dV,\\
    \langle Q\rangle_{t} = \frac{1}{(t_{f}-t_{s})}\int_{t_{s}}^{t_f}\langle Q\rangle(t) dt,
\end{eqnarray}
where $A = L_xL_y$, $dA = dxdy$, $V = A L_{z}$ and $dV = dAdz$. The time average is taken over a time-interval $[t_s,t_f]$ where the simulation is deemed to be in a statistically stationary state.

\subsection{Fluxes through the uniform staircase}\label{sec:gamma+fluxes}

We begin by measuring the fluxes of temperature and composition through the statistically stationary uniform staircase for all available simulations. We leverage the well-known fact \citep[see, e.g.][]{Woodal13} that in a statistically stationary state, and in a triply-periodic domain, the average convective fluxes of temperature and composition are related to the corresponding temperature and composition dissipations as:
\begin{eqnarray}
 F^{\rm turb}_T =  \langle w  T \rangle_t =  \langle w  \tilde T  \rangle_t \simeq  \langle |\nabla \tilde T|^2 \rangle_t, \label{eq:fluxT_SSState}\\
  F^{\rm turb}_C =   \langle w C \rangle_t = \langle w \tilde C \rangle_t \simeq \tau R_\rho^{-1}   \langle |\nabla \tilde C|^2 \rangle_t.
  \label{eq:fluxC_SSState}
\end{eqnarray}
As discussed by \citet{Woodal13}, the instantaneous dissipations are usually much less variable than the instantaneous fluxes in LDC, because they filter out reversible transport due to waves riding on the stable interface. In what follows, we therefore always measure the time-averaged total fluxes of temperature and composition through the domain, using 
\begin{eqnarray}\label{eq:fluxC_timeAVG}
F_T^{\rm tot} = F^{\rm diff}_T + F_T^{\rm turb} = 1 +   \langle |\nabla \tilde T|^2 \rangle_t,  \label{eq:fluxT_timeAVG}\\
F_C^{\rm tot} = F^{\rm diff}_C+ F^{\rm turb}_C = \tau R_\rho + \tau R_\rho^{-1} \langle |\nabla \tilde C|^2 \rangle_t,  
\end{eqnarray}
where $F^{\rm diff}_T = 1$ and $F^{\rm diff}_C =  \tau R_\rho$ are the diffusive fluxes associated with the background temperature and composition stratification, respectively.

\begin{figure}
    \centering
    \includegraphics[width=0.9\textwidth]{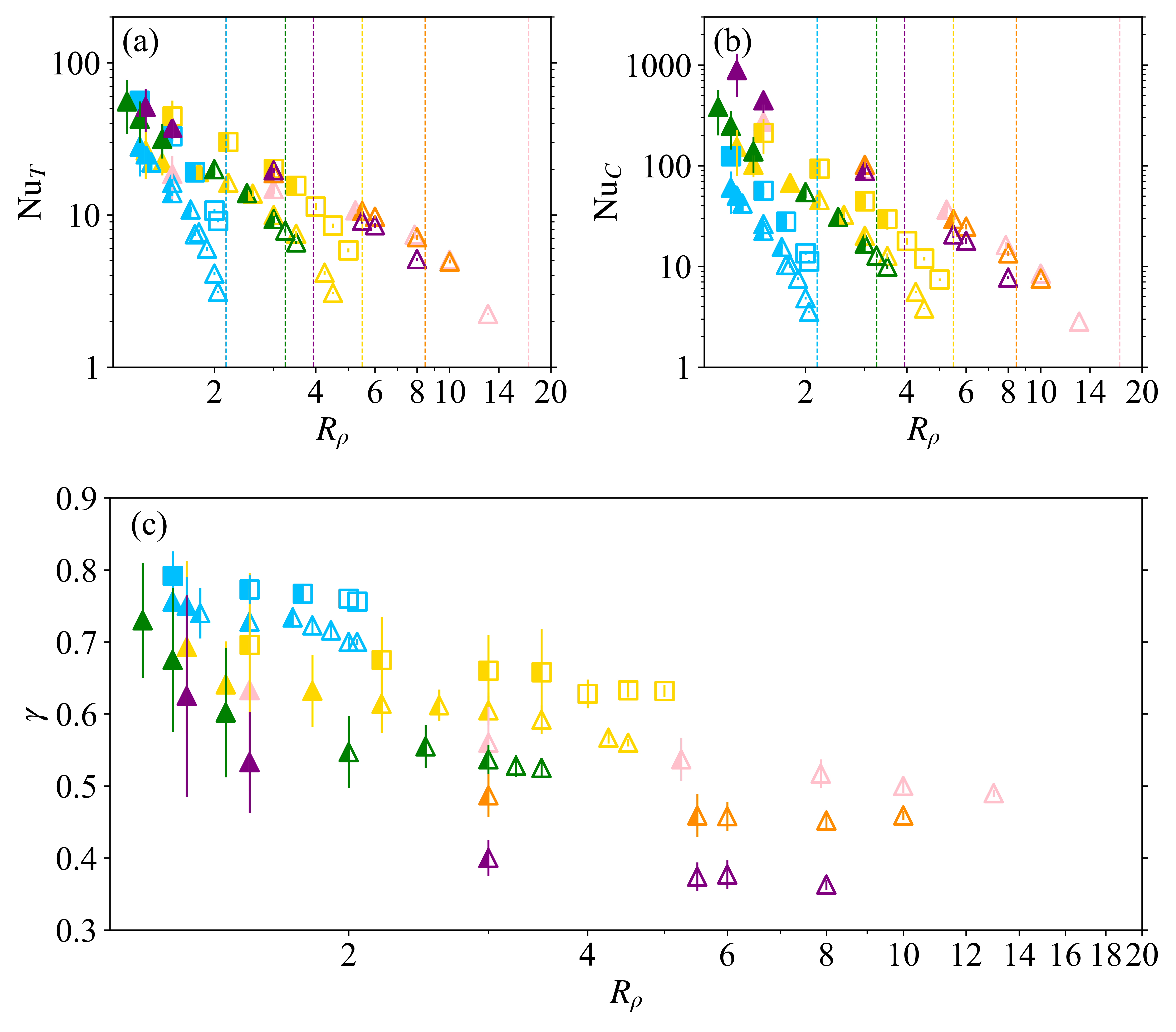}
    \caption{Variation of ({\it a}) $\Nu_{T}$  and ({\it b}) $\Nu_C$  with $R_{\rho}$ for all available simulations. Symbol colour varies according to the selected $(\Pr,\tau)$ values, symbol shape according to $L_z$, and symbol shading according to the inferred interface type (see table \ref{tab:simulations} and \S\ref{sec:interfacetype} for detail).  
    The temporal rms variability of the Nusselt numbers is shown as errorbars, but these are often smaller than the symbol size when $R_\rho >2$. The coloured dashed lines show $R_{c}$ for the corresponding $\Pr$ and $\tau$, as in table \ref{tab:simulations}.
   ({\it c}) Flux ratio $\gamma$ (given in equation \ref{eq:gammadef}) as a function of $R_{\rho}$ with the same  colour and marker coding. We see that $\gamma$ decreases monotonically with $R_{\rho}$, reaches a plateau, then drops again slightly as $R_\rho$ approaches the critical density ratio above which layers cease to be stable. }
\label{fig:TCfluxes}
\end{figure}

Figure \ref{fig:TCfluxes}({\it a,b}) shows the temperature and composition Nusselt numbers, defined as
\begin{equation}
\Nu_T = \frac{F^{\rm tot}_T}{F^{\rm diff}_T}, \mbox{  and } \Nu_C = \frac{F^{\rm tot}_C}{F^{\rm diff}_C}, 
\end{equation}
respectively, for all available simulations as  functions of $R_\rho$. Errorbars capture the rms temporal variability of the instantaneous dissipations. Different colours denote different values of the input parameters $(\Pr,\tau)$, as shown in table \ref{tab:simulations}. 
Triangles are used for domains of size $L_z = 100$, while  squares are used for domains of size $L_z = 200$. Each symbol is either fully filled, partially filled or open, depending on the interface type (see more on this in \S\ref{sec:interfacetype}).

We see several clear trends in the data. First, we note that, all other parameters being held constant, $\Nu_T$ and $\Nu_C$ decrease with increasing $R_\rho$, and increase with increasing layer height $L_z$ (i.e. square symbols are higher than triangular ones for a given colour). Both $\Nu_T$ and $\Nu_C$ also seem to depend more obviously on $\tau$ than on $\Pr$: for instance, all $\tau = 0.03$ datasets with varying $\Pr$ (pink, orange and purple symbols) lie very close to one another, while those with $\Pr = 0.3$ but varying $\tau$ (blue, green and purple datasets) are clearly distinct. Some of these trends will be explained by the model proposed in \S\ref{sec:model}.

For given $\Pr$, $\tau$ and $L_z$, we find that layered solutions cease to exist above a certain threshold in $R_\rho$.  \citet{Molletal2017} had already shown that this threshold is strictly greater than $\tau^{-1/2}$ at low Prandtl number, contrary to the $\Pr = O(1)$ model predictions of \citet{Linden_Shirtcliffe_1978}. They also suggested that this threshold is consistent with the critical density ratio $R_c$ given in equation (\ref{eq:oddc_crit}), at least for their own simulations. 
We have marked $R_c$ in dashed vertical lines of the same colour as the corresponding symbols in figure \ref{fig:TCfluxes}({\it a,b}). We see that when $\Pr = \tau$ (blue, yellow, pink symbols and lines), which was the dataset created by \citet{Molletal2017}, it indeed seems that layered states only exist for $R_\rho<R_c$, and disappear when $R_\rho>R_c$. For  the new data with $\Pr > \tau$, however, we find that stable layered states exist for $R_\rho > R_c$, demonstrating the existence of nonlinear states beyond the threshold for linear instability, as in the $\Pr = O(1)$ oceanographic case, and as predicted by \citet{Veronis1965}. 

Another quantity of interest is the total flux ratio $\gamma$, which we define here as the time-average of the instantaneous total flux ratio:
\begin{equation} 
\gamma = \frac{1}{t_f - t_s} \int_{t_s}^{t_f} \frac{\tau R_\rho + \tau R_\rho^{-1} \langle |\nabla \tilde C|^2 \rangle }{1 +\langle |\nabla \tilde T|^2 \rangle } dt . \label{eq:gammadef}
\end{equation}
It is shown in figure \ref{fig:TCfluxes}({\it c}), with errorbars denoting the rms time-variability of the integrand in (\ref{eq:gammadef}). 
We see that for a given choice of $\Pr$, $\tau$ and $L_z$, $\gamma$ first rapidly decreases with increasing $R_\rho$, then settles down to a (roughly) constant value as $R_\rho$ continues to increase, before dropping a little bit again when $R_\rho$ reaches the maximum possible value for layering  \citep{Molletal2017}. The fact that $\gamma$ is more--or-less constant over a wide range of density ratios is consistent with previous experimental findings at $\Pr = O(1)$ \citep{Shirtcliffe_1973, Linden_Shirtcliffe_1978, MARMORINO1976, Takao1980, Stampetal1998}, and with the results of \citet{GoughToomre1982}. The value of $\gamma$ in this plateau is henceforth called $\gamma_p$. We see that $\gamma_p$ increases slightly with increasing $L_z$ and increasing $\tau$, but decreases with increasing $\Pr$. These findings will be revisited in \S\ref{subsec:Ridata} and \S\ref{sec:convlayers}. 

Pertinent to the study of giant planets, which originally motivated the work of \citet{Molletal2017}, we thus confirm that $\gamma$ generally lies between $0.35$ and $1$ for all parameter values explored, and is significantly larger than the $\tau^{1/2}$ prediction originally put forward by \citet{Linden_Shirtcliffe_1978} for $\Pr = O(1)$ fluids. It is also different from other model predictions and experimental results that suggest $\gamma \propto R_\rho$ \citep{Newell1984,Fernando1989,Carpenter_Sommer_Wüest_2012}.

We now systematically explore various aspects of each simulation to try and understand the origin of these trends in the Nusselt numbers and flux ratio data. We begin by looking at the interfaces, and later look at the convective layers.

\subsection{Types of interfaces}
\label{sec:interfacetype}

Most prior models of layered convection have focused on the structure of the double-diffusive interface, so we begin our investigation by taking a qualitative look at its properties.
\citet{Molletal2017} distinguished between two different types of interfaces 
depending on the dominant contribution to the interfacial flux: diffusive interfaces, at high $R_\rho$, and turbulent interfaces, at lower $R_\rho$. As we argue below, this classification needs to be revised to include two categories of ‘turbulent' interfaces.  

We begin with the diffusive interfaces (henceforth denoted as D-type). They are marked with open symbols on all of the figures. Transport through D-type interfaces is dominated by diffusion, and in \S\ref{subsec:diffturbinterfaces} we shall formally define them as interfaces for which the interfacial Nusselt number (i.e. the
ratio of the total  to diffusive interfacial fluxes) for composition does not exceed 1.15.

\begin{figure}
    \centering
    \includegraphics[width=\textwidth]{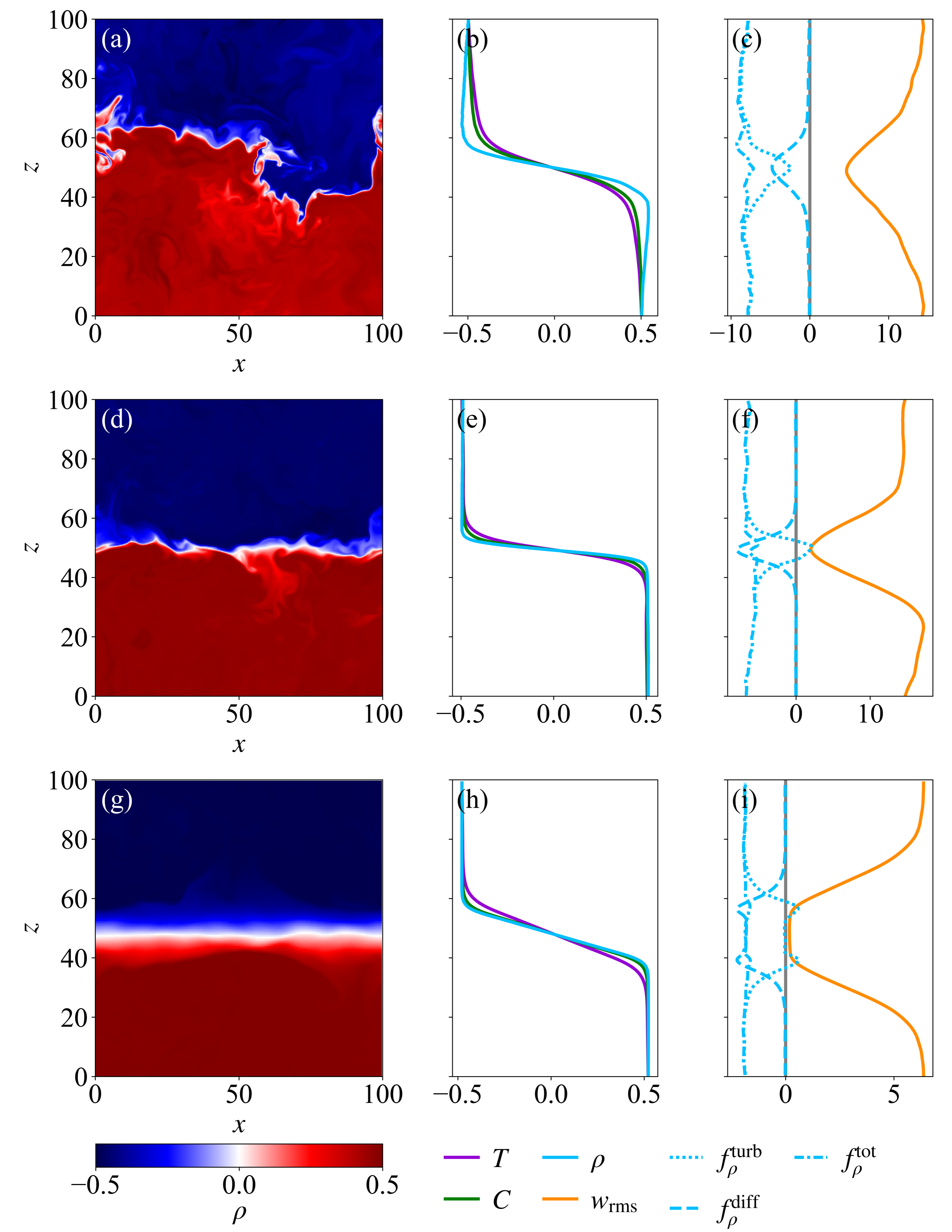}
    \caption{ Sample simulations at $\Pr = \tau = 0.1$, with $R_{\rho}=1.4$ (top row), $R_{\rho}=2.2$ (middle row), and $R_{\rho}=4.25$ (bottom row). In each row, the left panels show sample snapshots of the $C$ field with corresponding colour-bars. The center panels show profiles of $\overline{T} / \Delta T$(purple), $\overline{C} / \Delta C$(green), and $\overline{\rho} / \Delta \rho$ (blue) as a function of $z$ on the vertical axis. The right panels show profiles of $f_{\rho}^{\rm tot}$ (dot-dashed), $f_{\rho}^{\rm diff}$ (dashed), $f_{\rho}^{\rm turb}$ (dotted), and $w_{\rm rms}$ (orange).}.
    \label{fig:interface-structure}
\end{figure}

The typical structure of a D-type interface is illustrated in figure \ref{fig:interface-structure}({\it g-i})  \citep[see also figure 2 of][]{Molletal2017}, which shows results from a simulation at $\Pr = \tau = 0.1$, $R_\rho = 4.25$, $L_z = 100$. Figure \ref{fig:interface-structure}({\it g}) shows a snapshot of $C$ in the $(x,z)$ plane of the simulation, at some arbitrary position $y$ and time $t$ during the statistically stationary state. We see a thick interface, from which domain-size convective plumes emerge on either side, which is typical of this regime. Profiles of the time-averaged and horizontally-averaged total temperature, composition and density fields, $\overline{T}(z)$, $\overline{C}(z)$, and 
\begin{equation}
    \bar \rho(z) = - \bar T(z) + \bar C(z),\label{eq:DensityProfile}
\end{equation}
are shown in figure \ref{fig:interface-structure}({\it h}) for the same simulation. They are normalized by their respective jumps across the interface (or equivalently, jump from one interface to the next in this uniform staircase model), which are 
\begin{equation}
    \Delta T = L_z, \Delta C = R_\rho L_z, \mbox{ and } \Delta \rho = (R_\rho-1) L_z.
\label{eq:iterface-jumps}
\end{equation}

We see that the interface has a linearly stratified core \citep{Molletal2017} surrounded by well-mixed convective regions above and below. While the mean density overall decreases with height, a very shallow density inversion appears on either side of the interface, as a result of the slight difference between the mean interface thicknesses of the temperature and composition profiles (see more on the interface thickness in \S\ref{sec:interface-thickness}). This density inversion ultimately drives the convection in the layers.

Figure \ref{fig:interface-structure}({\it i}) shows the rms vertical velocity
of the flow (purple line) $w_{\rm rms}(z)$, computed from
\begin{equation}
    w_{\rm rms}(z) = \left(\overline{w^{2}}\right)^{1/2}.  \label{eq:wrmsdef}
\end{equation}
We see that $w_{\rm rms}$ almost drops to zero in the interface. Figure \ref{fig:interface-structure}({\it i}) also shows the diffusive (dashed), turbulent (dotted) and total (dot-dashed) horizontally and time-averaged  density fluxes, defined as
\begin{eqnarray}\label{eq:flux_defs_forPlots}
f^{\rm turb}_\rho(z) = -  f^{\rm turb}_T(z) + f^{\rm turb}_C(z), \mbox{ and } f^{\rm diff}_\rho(z) = -  f^{\rm diff}_T(z) + f^{\rm diff}_C(z),  \\
f^{\rm tot}_\rho(z) = f^{\rm turb}_\rho(z) + f^{\rm diff}_\rho(z), \label{eq:flux_defs_forPlots_rho}
\end{eqnarray}
with
\begin{eqnarray}
f^{\rm turb}_T(z) = \overline{wT},   \label{eq:FTturb_timeHavg} \quad 
f^{\rm turb}_C(z) = \overline{wC}  \label{eq:FCturb_timeHavg}\\ 
f^{\rm diff}_T(z) = - \frac{d\bar T}{dz}, \label{eq:FTdiff_timeHavg} \\ 
f^{\rm diff}_C(z) = -\tau \frac{d\bar C}{dz} .\label{eq:FCdiff_timeHavg}
\end{eqnarray}

Consistent with the system having an overall stably stratified density field, and being in a statistically stationary state, we see that the total density flux is negative and more-or-less independent of height (small remaining fluctuations are inevitable because the time average takes place over a finite rather than infinite interval). The convective flux is large and negative in the convection zone, but becomes slightly positive in the ‘wings' of the interface before dropping to almost zero in the interface itself. The diffusive flux, by contrast, is negligible in the convective zone, but large and negative in the interface. A key property of the D-type interfaces (see also \S\ref{sec:interfacetype}) is that the diffusive flux through the interface is very close  to the convective flux through the layers -- in other words, the interface is thin enough to transport diffusively all of the flux it receives from the neighboring convection zone.  

We note that the sign of the turbulent density flux is an important diagnostic for the energy budget in the simulation, and establishes whether the interface acts, on average, as a kinetic energy sink or source. To see this, we construct the total kinetic energy equation by dotting the momentum equation with $\bu$, and integrating it over the domain and in time. We obtain
\begin{equation} \label{eq:totalKEeq}
 0 = -   \Pr \langle  w \rho \rangle_t - \Pr \langle | \nabla \bu|^2 \rangle_t .
\end{equation}

The second term on the right-hand-side represents the kinetic energy dissipation due to viscosity, which is always an energy sink, while the first term represents the conversion of potential to kinetic energy, which overall acts as an energy source in this system as $\langle w \rho \rangle_t < 0$. However, given that 
\begin{equation}\label{eq:KE_toVAflux}
    \langle w  \rho \rangle_t = \frac{1}{L_z} \int f^{\rm turb}_\rho(z) dz , 
\end{equation}
we see in figure \ref{fig:interface-structure}({\it i}) that while the convective zones are indeed kinetic energy sources, the interface (especially its wings) acts as a kinetic energy sink, with $f^{\rm turb}_\rho(z) > 0$. This was expected, as  the interface is strongly stably stratified. 

Figures \ref{fig:interface-structure}({\it d-f}) show the same quantities for a simulation with $\Pr = \tau = 0.1$, $R_\rho = 2.2$ and $L_z = 100$. We see that the interface is overall much thinner, as well as more turbulent. In particular, it is no longer flat, but oscillates up and down instead. We henceforth refer to these interfaces as O-type ‘oscillatory' interfaces, and show them with half-filled, half-open  symbols in all of the data figures.  The temperature and composition profiles no longer have a linearly stratified core, but are instead well-approximated by hyperbolic tangent profiles. As in the D-type interface, however, the difference in the thicknesses of the mean temperature and composition profiles continues to drive a weak density inversion near the top and  bottom of the interface. We see in figure \ref{fig:interface-structure}({\it f}) that the rms vertical velocity within the interface is generally larger, consistent with the observed oscillations. We also see that $f^{\rm turb}(z) > 0$ within the interface, as in the D-type case, showing that it is still an energy sink for the system. 

Figures \ref{fig:interface-structure}({\it a,b,c}) show the interface structure for $\Pr=  \tau = 0.1$, $R_\rho = 1.4$ and $L_z = 100$. In this weakly stratified  case, the interface is even more turbulent, and is regularly pierced by strong up-flows and down-flows. Consequently, its mean thickness is quite large (see more on this below). The rms vertical velocity within the interface is correspondingly also quite large, and, by contrast with the other two cases, we now see that $f^{\rm turb}_\rho(z)<0$ everywhere in the domain. In other words, the interface is too weak to act as an energy absorber for the system, and is intrinsically unstable. Turbulent interfaces for which $f^{\rm turb}(z) < 0$ for all $z$ are marked with filled symbols in all of the data figures, and are henceforth referred to as U-type ‘unstable' interfaces. We believe that while they are forced to exist in these triply periodic simulations, they would rapidly disappear via mergers in a staircase with more realistic boundary conditions \citep{Tulekeyev2024}.

\begin{figure}
 \centering
    \includegraphics[width=\textwidth]{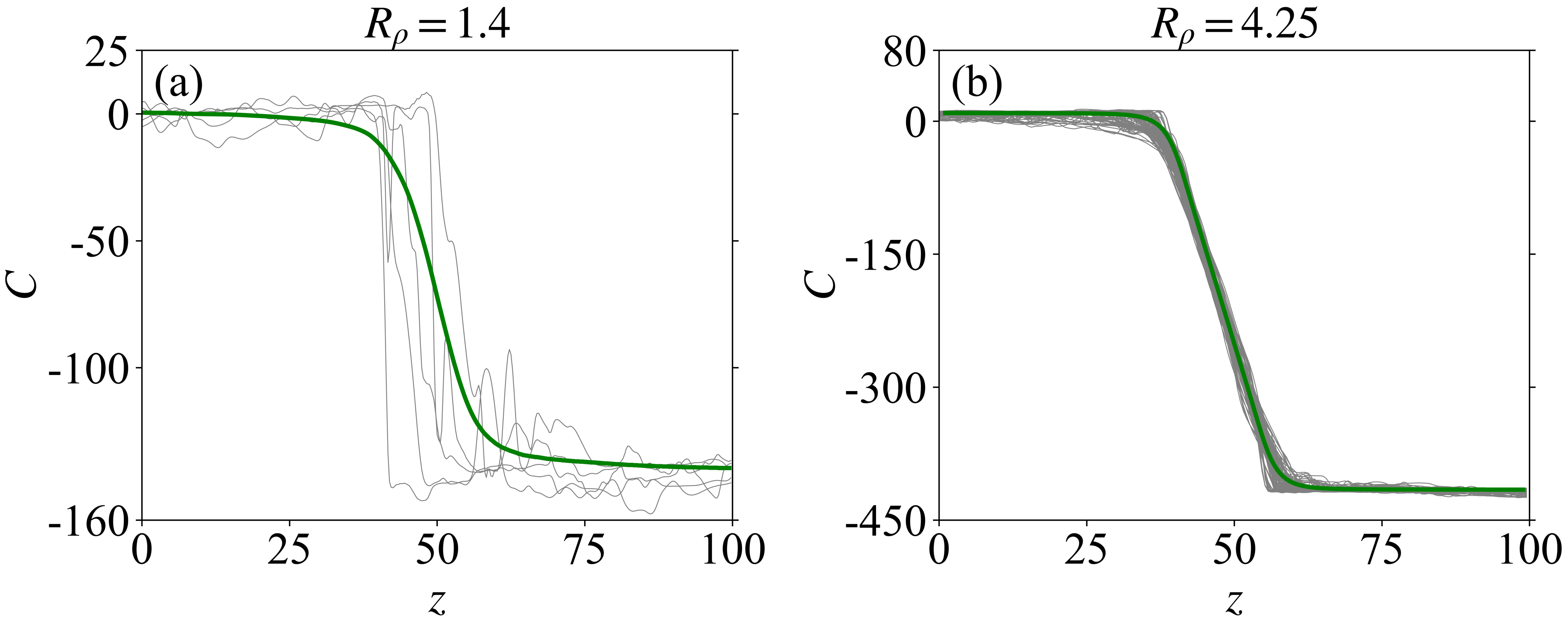}
    \caption{Comparison of the composition profiles for ({\it a}) a U-type interface $(\Pr = \tau = 0.1, L_z = 100, R_\rho = 1.4)$ and ({\it b}) a D-type interface $(\Pr = \tau = 0.1, L_z = 100, R_\rho = 4.25)$. In both panels, the thick green curve shows the mean vertical profile $\bar C(z)$. The thinner grey curves show local instantaneous vertical profiles $C(x,y,z,t)$ at randomly selected values of $x,y,t$ in the statistically stationary state.}
    \label{fig:interface-profiles}   
\end{figure}

Comparing the U-type interface in figure \ref{fig:interface-structure}({\it a-c}) with the D-type interface in figure \ref{fig:interface-structure}({\it g-i}), we see that the horizontally-averaged profiles $\bar T(z)$ and $\bar C(z)$ in both simulations have fairly thick ‘mean' interfaces, but for fundamentally different reasons. This is clarified in figure \ref{fig:interface-profiles}, which compares the mean profile $\bar C(z)$ (thick green lines) with selected local instantaneous profiles $C(x,y,z,t)$ (thin grey lines) in both cases. For the D-type interface ($R_\rho = 4.25$) shown in figure \ref{fig:interface-profiles}({\it b}) we see that the mean and local instantaneous profiles are quite similar, and both take the form of piecewise linear functions with a wide interface. For the U-type interface ($R_\rho = 1.4$) in figure \ref{fig:interface-profiles}({\it a}), by contrast, we see that the interface inferred from the mean profile vastly overestimates the local, instantaneous interface thickness. The mean interface only appears to be thick in this case because the local interface position varies significantly in space and time (see more on this topic in figure \ref{fig:dzmid}).  

\subsection{Interface properties}
\label{sec:interface-thickness}

We now look at the interface properties more quantitatively. We begin with properties related to the interface in the mean profiles in \S\ref{subsec:meanthick}-\ref{subsec:Ridata}, then study the properties of the local instantaneous interfaces in \S\ref{subsubsec:localdTdC}.

\subsubsection{Mean interface thickness}
\label{subsec:meanthick}

We measure the mean interface thicknesses $\delta_T$ and $\delta_C$ associated with the mean temperature and composition profiles $\bar T(z)$ and $\bar C(z)$ as follows. For U-type and O-type interfaces, we fit hyperbolic tangent profiles to $\bar T(z)$ and $\bar C(z)$, as 
\begin{equation}
    \bar T(z) \simeq \bar T_i + \frac{\Delta T}{2} \tanh\left(\frac{\bar z_i - z}{\delta_T}\right), \mbox{  and  } \bar C(z) \simeq \bar C_i + \frac{\Delta C}{2} \tanh\left(\frac{\bar z_i - z}{\delta_C}\right).
    \label{eq:tanhmodel}
\end{equation}
The quantity $\bar z_i$ is the mean interface position, which is assumed to be the same for the two profiles. 
Because the interface can be in different positions from one simulation to another, we need to fit the constants $\bar T_i$, $\bar C_i$ and $\bar z_i$ as well as $\delta_T$ and $\delta_C$. The interfacial jumps $\Delta T$ and $\Delta C$, by contrast, are known and given in equation \eqref{eq:iterface-jumps}.  

\begin{figure}
    \centering
    \includegraphics[width=0.6\textwidth]{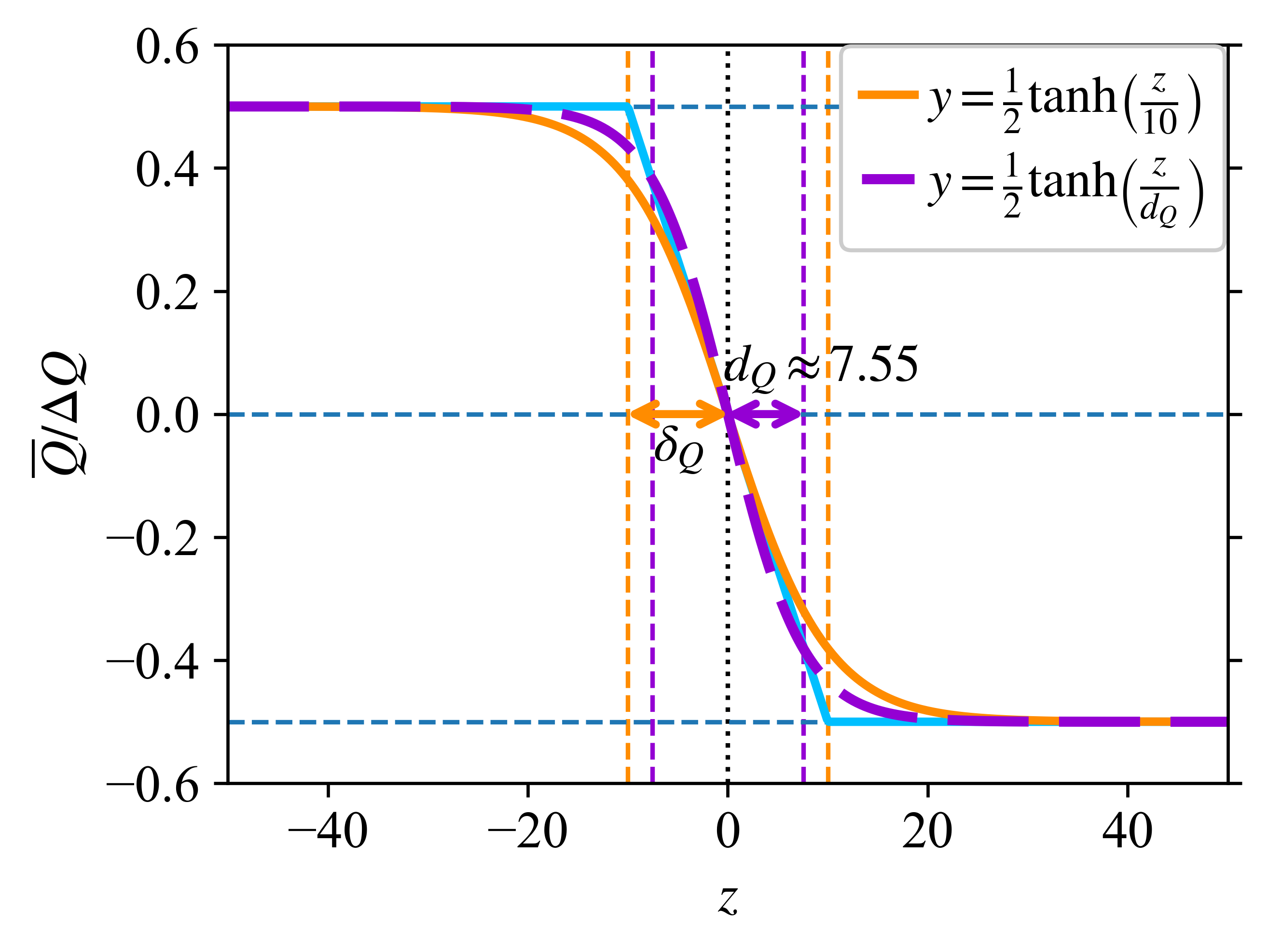}
    \caption{Illustration of the interface thickness measurement protocol. The quantity $Q$ represents either  $T$ or $C$. The solid orange line, $\bar Q (z)/ \Delta Q  = 0.5\tanh(z/10)$,  represents a model O-type interface with half-thickness $\delta_Q = 10$. The light blue solid line is a piece-wise linear function representing a typical D-type interface with the same half-thickness  $\delta_Q = 10$. They have the same $Q$ gradient at the interface midpoint.  The purple line shows the best-fit $\bar Q (z)/ \Delta Q  = 0.5\tanh(z/d_Q)$  function to the D-type interface profile, which underestimates the actual thickness by finding $d_Q = 7.55$.}
    \label{fig:interfacemodelprofile}
\end{figure}

With these definitions, $\delta_T$ and $\delta_C$ correspond to half the actual interface thickness in either temperature or composition, see figure \ref{fig:interfacemodelprofile}. The temperature and composition gradients at the midpoint of the interface are then given by 
\begin{equation}
    \left. \frac{d\bar T}{dz}\right|_{z= \bar z_i} \equiv \frac{d\bar T_i}{dz} = -\frac{\Delta T}{2 \delta_T} \mbox{ and } \left. \frac{d\bar C}{dz}\right|_{z= \bar z_i} \equiv \frac{d\bar C_i}{dz} = -\frac{\Delta C}{2 \delta_C}.
        \label{eq:tanhmodelgradient}
\end{equation}

For D-type interfaces, attempting to fit a hyperbolic tangent profile to a piecewise linear one has a tendency to underestimate the interface thickness. This is illustrated in figure \ref{fig:interfacemodelprofile}, which shows the normalized mean profile $\bar Q(z)/\Delta Q$ of a quantity $Q$ (that could be $T$ or $C$). The solid orange line  illustrates a hyperbolic tangent profile with half-thickness $\delta_Q = 10$, and the solid light blue line illustrates a piecewise linear profile with the same half-thickness. Attempting to fit a hyperbolic tangent to the piecewise linear profile results in the  purple dashed curve, which has a half-thickness of $d_Q = 7.55$, that is, about $25\%$ smaller than that of the actual profile. The gradient of $\bar Q$ at the midpoint of the interface is correspondingly over-estimated by the same amount. 

For this reason, we must use a different method to define and extract $\delta_T$ and $\delta_C$ in D-type interfaces. We first fit linear functions to the core of the interface only, to extract $d\bar T/dz$ and $d\bar C/dz$ within, and call these $d\bar T_i/dz$ and $d\bar C_i/dz$. We then define  
\begin{equation}
    \delta_T = \frac{\Delta  T}{2 |d\bar T_i/dz|}, \mbox{  and  } \delta_C = \frac{\Delta C}{2 |d\bar C_i/dz|}.
\label{eq:deltadefs}
\end{equation}
The factor of 2 in this definition ensures consistency with \eqref{eq:tanhmodelgradient}: $\delta_T$ and $\delta_C$ actually represent half the interface thickness. 

\begin{figure}
    \centering
    \includegraphics[width=\textwidth]{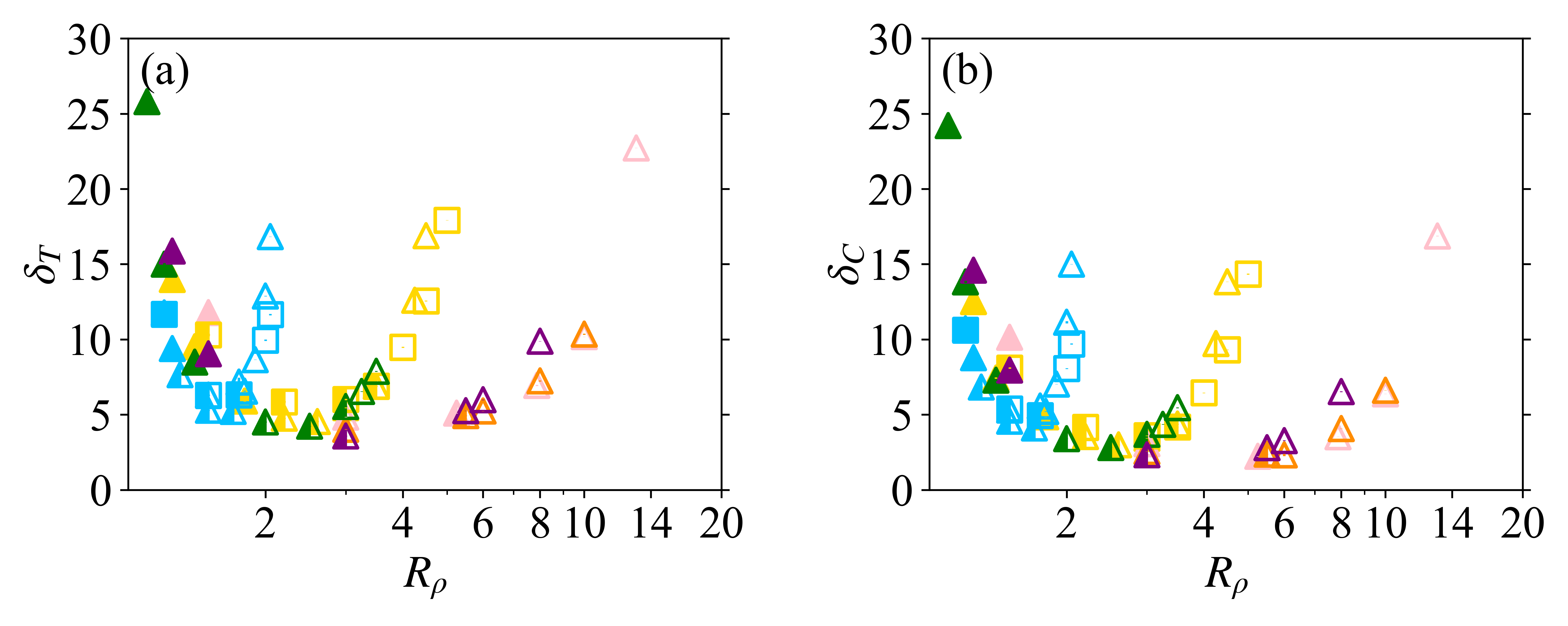}
    \caption{({\it a}) Interface thickness  $\delta_{T}$ of the mean temperature profile $\bar T(z)$ as a function of $R_\rho$. ({\it b}) Interface thickness $\delta_{C}$ of the mean composition profile $\bar C(z)$, as a function of $R_{\rho}$. See table \ref{tab:simulations} for the symbol legend. Errorbars on the fitted values of $\delta_T$ and $\delta_C$ are smaller than the symbol size.}
    \label{fig:Thicknessdata}
\end{figure}

Figure \ref{fig:Thicknessdata} shows $\delta_T$ and $\delta_C$ extracted in this manner, for all available simulations. We see that at fixed $\Pr$, $\tau$, and $L_z$, $\delta_T$ and $\delta_C$ 
have minima at intermediate values of $R_\rho$ for O-type interfaces, and increase both as $R_\rho \rightarrow 1$ for U-type interfaces (filled symbols), and when  $R_\rho$ increases for D-type interface (open symbols). We also find that both $\delta_T$  and $\delta_C$ are relatively independent of the total domain height $L_z$ (i.e. compare triangles and squares). Finally, $\delta_T$ and $\delta_C$ seem to depend more on $\tau$ than on $\Pr$.  

The high $R_\rho$ limit is easy to understand, at least qualitatively. Recall from figure \ref{fig:TCfluxes} that both temperature and composition fluxes through the domain decrease with increasing $R_\rho$ in that region of parameter space. Because the interface is diffusive, decreasing the flux through the interface simply requires decreasing the magnitude of the interfacial temperature and composition gradients $d\bar T_i/dz$ and $d\bar C_i/dz$. As the total jumps $\Delta  T$ and $\Delta C$ across the interface are fixed by the input parameters (see equation \ref{eq:iterface-jumps}), decreasing the magnitude of the gradients simply requires increasing the interface width. 

To understand the low $R_\rho$ limit, recall that oscillations of the interface position (see figure \ref{fig:interface-profiles}({\it a})) 
lead to an apparent increase in the mean interface thickness for U-type interfaces. The amplitude of these oscillations is larger at lower $R_\rho$ because the interface is more weakly stratified, hence $\delta_T$ and $\delta_C$ both increase as $R_\rho \rightarrow 1$. This effect is further investigated in \S\ref{subsubsec:localdTdC}. 

\subsubsection{Diffusive vs. turbulent interfaces}
\label{subsec:diffturbinterfaces}
 
Using the measured mean interface thicknesses $\delta_T$ and $\delta_C$, or equivalently, the measured interfacial temperature and composition gradients $d\bar T_i/dz$ and $d \bar C_i/dz$,  we define the diffusive flux across the interface using 
\begin{equation}\label{eq:diffFlux_atInterface}
    F^{\rm diff}_{i,T} = - \frac{d\bar T_i}{dz} =  \frac{\Delta T}{2 \delta_T} \mbox{ and }  F^{\rm diff}_{i,C} =  - \tau \frac{d\bar C_i}{dz} = \tau \frac{\Delta C}{2 \delta_C}.
\end{equation}
This definition is valid throughout the core of the D-type interfaces, and at the midpoint of the interface in the $\bar T$ and $\bar C$ profiles for O-type and U-type interfaces.

\begin{figure}
        \centering
    \includegraphics[width=\textwidth]{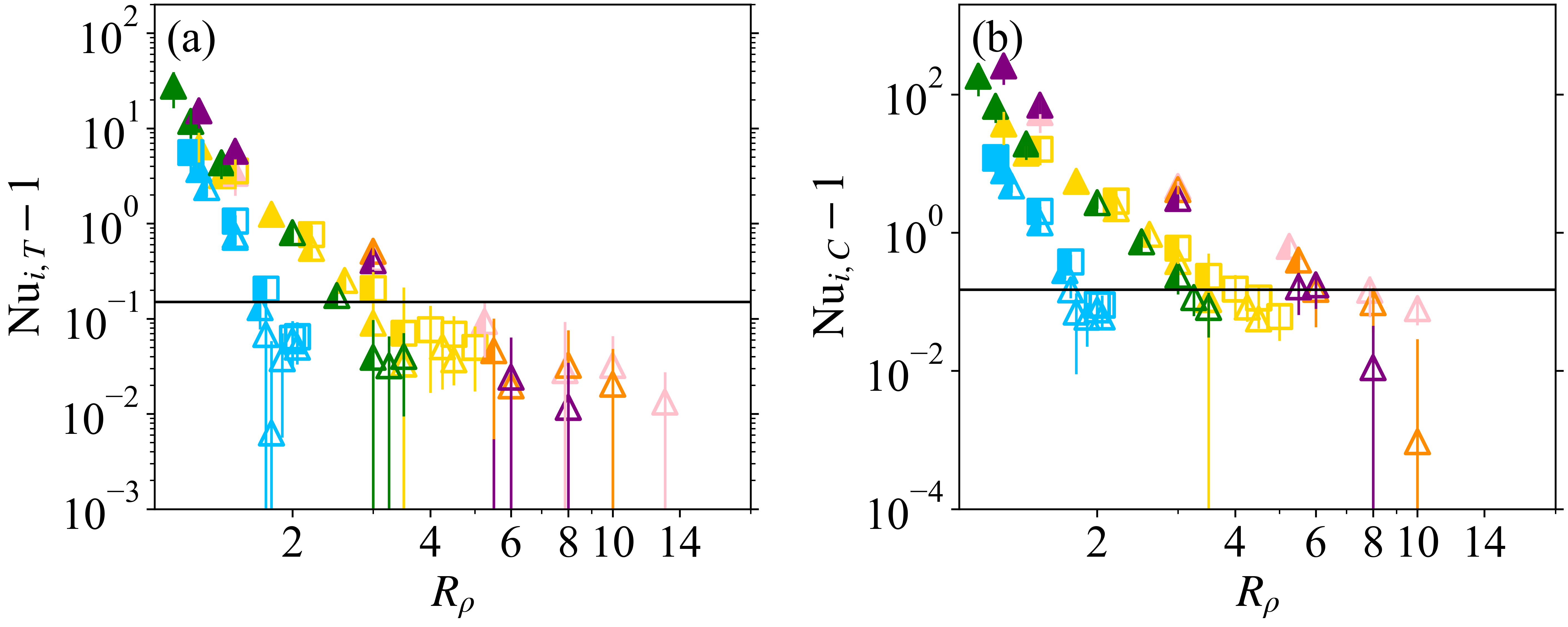}
    \caption{Interfacial Nusselt numbers ({\it a}) $\Nu_{i,T}-1$ and   ({\it b}) $\Nu_{i,C}-1$ as functions of $R_{\rho}$. We  define D-type interfaces to have $\Nu_{i,C}-1< 0.15$ (horizontal black line). See table \ref{tab:simulations} for the symbol legend.}
    \label{fig:interfacenusselt}
\end{figure}

Figure \ref{fig:interfacenusselt} shows the interface Nusselt numbers for temperature and composition, each defined as the ratio of the total flux through the domain (which is also the total flux through the interface in a statistically stationary state), divided by the diffusive flux through the interface: 
\begin{equation}\label{eq:Nusselt_defs}
    \Nu_{i,T} = \frac{F^{\rm tot}_T}{F^{\rm diff}_{i,T}} \mbox{  and  } \Nu_{i,C} = \frac{F^{\rm tot}_C}{F^{\rm diff}_{i,C}}.
\end{equation}
As expected, $\Nu_{i,T}$ and $\Nu_{i,C}$ both decrease as $R_\rho$ increases. Interestingly, we also find that both Nusselt numbers appear to be independent of $L_z$, and are only weakly dependent on $\Pr$. 

For large $R_\rho$, we see that both Nusselt numbers tend to 1, a limit which indicates a fully diffusive interface. 
We thus formally define an interface to be diffusive (D-type) when 
\begin{equation}\label{eq:diffInterface_defNuC}
    \Nu_{i,C} \le 1.15, 
\end{equation}
or, in other words, when the turbulent composition flux is less than 15\%  of the diffusive composition flux. The 15\% threshold was selected somewhat  arbitrarily, and we note that several O-type interfaces lie very close to it, showing that they are ‘almost' fully diffusive. 

\subsubsection{Mean interface stratification}
\label{subsec:Ridata}

\begin{figure}
    \centering
    \includegraphics[width=\textwidth]{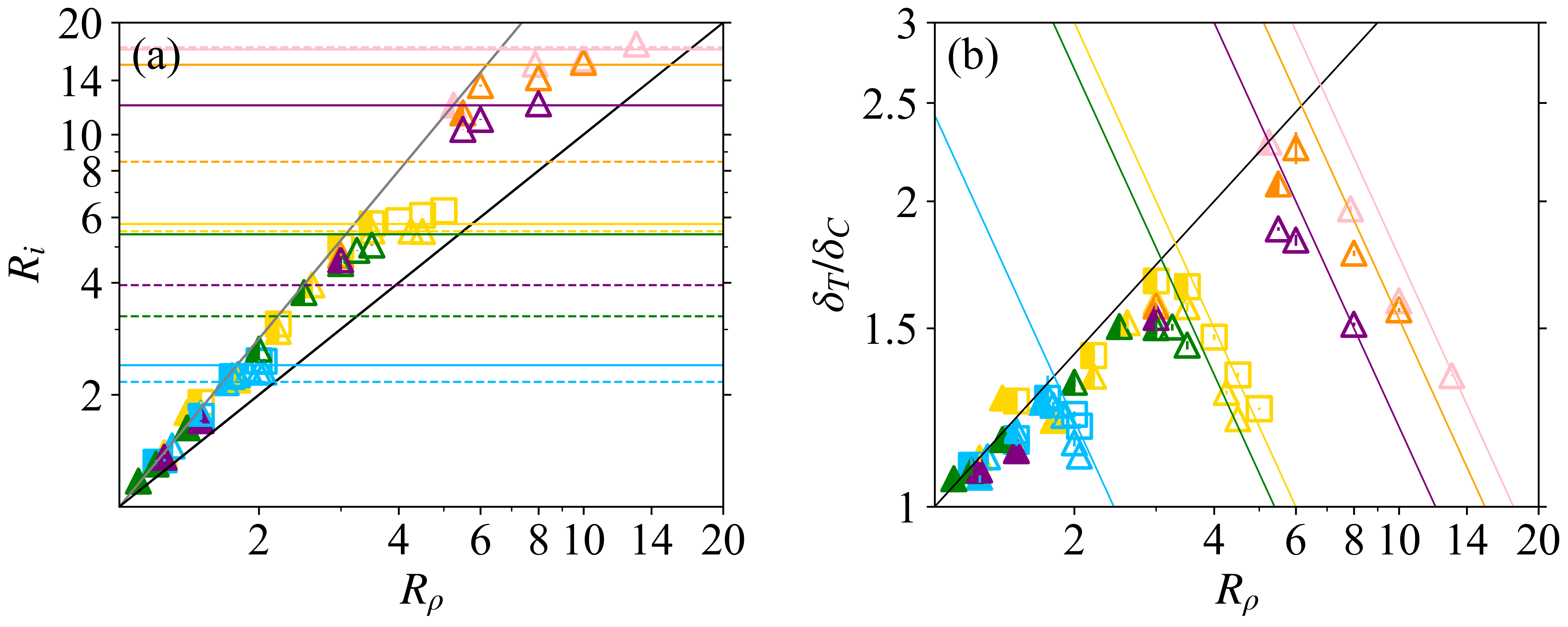}
    \caption{({\it a}) Interfacial density ratio  $R_{i}$ as a function of $R_{\rho}$. The grey line shows the power law  $R_i = R_\rho^{3/2}$ discovered by \citet{Molletal2017}, and the black line shows $R_i = R_\rho$, the upper limit for the  existence of an interface. Dashed coloured lines show $R_i = R_c$, while solid coloured lines show $R_i = \gamma_p/\tau$ (for $L_z = 100$ simulations only, with $ \gamma_p$ given in table \ref{tab:gammac}). ({\it b}) Mean interface thickness ratio $\delta_T / \delta_C$ as a function of   $R_{\rho}$. The black solid line shows $\delta_T / \delta_C = R_{\rho}^{1/2}$, and the coloured lines represent $\delta_T / \delta_C =  \gamma_p/\tau R_{\rho}$, where $\gamma_p$ is given in table \ref{tab:gammac}. All U- and O-type interfaces follow the black line, and all D-type interfaces collapse on the  $\delta_T / \delta_C = \gamma_p/\tau R_{\rho}$ line of the corresponding colours. In both panels, the vertical error bars on $\delta_T / \delta_C$ are often much smaller than the marker size. For both panels see table \ref{tab:simulations} for the colour and symbol legend. }
    \label{fig:Ri}
\end{figure}

\citet{Molletal2017} argued that while the total jumps in temperature and composition across the interface are fixed (see equation \ref{eq:iterface-jumps}) in this triply-periodic model setup, the effective stratification of the interface might be better characterized by the interfacial gradient density ratio, defined as 
\begin{equation}\label{eq:interfaceR}
    R_i = \frac{d\bar C_i/dz}{d\bar T_i/dz} = R_\rho \frac{\delta_T}{\delta_C},
\end{equation}
where the second expression is derived using equation (\ref{eq:deltadefs}).
The quantity $R_i$ is shown in figure \ref{fig:Ri}({\it a}) as a function of $R_\rho$ for all available simulations.  
We confirm the findings of  \citet{Molletal2017}, who had shown that $R_i$ increases as $R_\rho^{3/2}$ for all turbulent interfaces (filled and semi-filled symbols), then reaches a plateau when the interfaces become diffusive (open symbols). However, \citet{Molletal2017} only studied cases where $\Pr = \tau$, and incorrectly concluded from the data available at the time that the diffusive plateau satisfies $R_{i} \simeq R_c = (\Pr + 1)/(\Pr + \tau)$. This value is shown as dashed lines of the same colour as the corresponding symbols for each set of parameters. We see that their conclusion does not hold for cases with $\Pr > \tau$ (see in particular the dark green, dark purple, and orange symbols and dashed lines), suggesting the 
good match with the data for $\Pr = \tau$ could have been a strange coincidence. 

We propose a different interpretation for this ‘diffusive plateau'. First, recall that the interfacial flux ratio for D-type interfaces is both equal to the total flux ratio $\gamma$ in the whole domain (since the fluxes through the layers are the same as the fluxes through the interface), and equal to the ratio of the diffusive fluxes in the interface (because the interface is diffusive), so that 
\begin{equation}\label{eq:gamma_tauRint}
    \gamma = \tau R_i, \mbox{ for all D-type interfaces.} 
\end{equation}

We also saw in figure \ref{fig:TCfluxes}({\it c}) that $\gamma$ is approximately constant and equal to  $ \gamma_p(\Pr,\tau,L_z)$ at large $R_\rho$. We can measure $ \gamma_p$ from the data shown in figure \ref{fig:TCfluxes}({\it c}) by averaging $\gamma$ (for given $\Pr$, $\tau$ and $L_z$) over the intermediate range of $R_\rho$ for which it is approximately constant. Table \ref{tab:gammac} presents the results. We note in particular that $\gamma_p$ increases with $L_z$ (consistent with our findings from figure \ref{fig:TCfluxes}({\it c})). As the interface thickness is mostly independent of $L_z$, we conclude that $\gamma_p$ must be a property of the convective region itself (whose size increases with $L_z$). In other words, convectively-driven turbulence controls $\gamma_p$ within the convective layers, and the interface must adjust itself to accommodate the  flux ratio that it receives from layer below. 

\begin{table}
    \centering
    \begin{tabular}{c|c|c|c|c|c}
    $\Pr$ & $\tau$ & $L_z$ & Range & $ \gamma_p$ & $R_{L}$\\
     \hline
     
      0.3  & 0.3 & 100 & $[1.3,2.1]$ & $0.72 \pm 0.02$ & 2.4 \\
       0.3  & 0.3 & 200 & $[1.5,2.1]$ & $0.765 \pm 0.01$ & 2.55 \\
       0.3  & 0.1 & 100 & $[1.8,5]$ & $0.54 \pm 0.015$ & 5.4 \\
       0.3  & 0.03 & 100 & $[2,9]$ & $0.38 \pm 0.05$ & 12.0  \\
     \hline
       0.1  & 0.1 & 100 & $[2,5]$ & $0.59 \pm 0.02 $ &  5.9 \\
       0.1  & 0.1 & 200 & $[1.5,6] $& $0.65 \pm 0.02$ &  6.5 \\
       0.1  & 0.03 & 100 & [1.5,11] & $0.465 \pm 0.015$ & 15.5 \\
     \hline
       0.03  & 0.03 & 100 &  $[2,11]$ & $0.52 \pm 0.03$ & 17.67\\
     \hline
    \end{tabular}
    \caption{For each parameter set $(\Pr,\tau,L_z)$, this table presents the values of $\gamma_p$ extracted from the data shown in figure \ref{fig:TCfluxes}({\it c}),  the range of $R_\rho$ over which the average was taken, and the maximum density ratio for  layer formation $R_{L} =  \gamma_p/\tau$  inferred from this data (see equation \ref{eq:RLc}). Note that $\gamma_p$ increases slightly with $L_z$, so $R_{L}$ does too. }
    \label{tab:gammac}
\end{table}

This interpretation will be revisited in  more detail in \S\ref{sec:convlayers}, but in the meantime leads us to conclude that for D-type interfaces, $R_i$ should have the value 
\begin{equation}\label{eq:Rint_gammaTau}
R_i \simeq \frac{\gamma_p(\Pr,\tau,L_z)}{\tau}.
\end{equation}
Figure \ref{fig:Ri}({\it a}) shows the horizontal asymptotes corresponding to $R_i =\gamma_p / \tau$ in solid lines, using the values of $\gamma_p$ given in table \ref{tab:gammac}. To avoid crowding the figure, only the cases with $L_z = 100$ are shown. We see that these estimates provide a much better explanation for the plateau in $R_i$, which is not surprising since $\gamma_p$ was measured from the simulations. We note, however, that $R_i$ was computed from $\delta_T$ and $\delta_C$, obtained by fitting the mean $\bar T$ and $\bar C$ profiles, while $ \gamma_p$ was obtained from the volume- and time-averaged flux ratio. The fact that $R_i \simeq \gamma_p / \tau$ for D-type interfaces, therefore, confirms that the staircase is indeed in a statistically stationary state, with convective fluxes in the layers matching the diffusive fluxes through the interface. 

We acknowledge that this explanation remains largely unsatisfactory in as much as it shifts the question to why should $\gamma$ be roughly constant ($\gamma \simeq \gamma_p$) at large $R_\rho$. Furthermore, it does not yet provide a theory to explain the dependence of $ \gamma_p$ (and therefore of $R_i$) on $\Pr$, $\tau$ or $L_z$. This will be discussed in \S\ref{sec:convlayers} and \S\ref{sec:discussion}.

An interesting consequence of \eqref{eq:Rint_gammaTau}, however, is that the ratio of the mean interface thicknesses should satisfy
\begin{equation}
\frac{\delta_T}{\delta_C} =  \frac{ \gamma_p}{R_\rho \tau}
\label{eq:interfaceratio}
\end{equation}
for D-type interfaces. Figure \ref{fig:Ri}({\it b}) shows $\delta_T/\delta_C$ as a function of $R_\rho$. We see that this ratio initially increases with $R_\rho$ roughly as $R_\rho^{1/2}$ (consistent with $R_i \propto R_\rho^{3/2}$ for U-type and O-type interfaces), reaches a maximum, then decreases again as $ \gamma_p/\tau R_\rho$ (see solid lines) when the interfaces become diffusive, as predicted. For sufficiently large $R_\rho$, we see that $\delta_T/\delta_C\rightarrow 1$. When this happens, $R_i = R_\rho$ (shown in the black line in figure \ref{fig:Ri}({\it a})), implying that the interface is no longer more strongly stratified than the layers and its thickness grows to occupy the whole domain (see figure \ref{fig:Thicknessdata}), while the layers shrink and ultimately disappear. Because $R_i \simeq \gamma_p/\tau$ for D-type interfaces, we conclude that layers must disappear when 
\begin{equation}
R_\rho \rightarrow   R_L = \frac{\gamma_p(\Pr,\tau,L_z)}{\tau},
\label{eq:RLc}
\end{equation}
which defines $ R_L$ as the critical density ratio above which layers cannot be sustained.  Using our measurements for $\gamma_p$, the predicted value of $R_L$ is also provided in table \ref{tab:gammac}.  Crucially, we see that $ R_L > R_c$ when $\Pr > \tau$ ($R_c$ is given in table \ref{tab:simulations}), which shows that layered solutions are indeed possible even the background state is linearly stable to ODDC. These layered states can only be obtained through finite amplitude perturbations \citep{Veronis1965}. 

 Finally, it is worth noting that this result is consistent with the predictions from the theory of  \citet{Linden_Shirtcliffe_1978} when applied to the $\Pr \ge 1$ limit: if we take as given that $\gamma_p = \tau^{1/2}$ in this case (as suggested by the data from \citet{TURNER1965} and \citet{Shirtcliffe_1973}) then \eqref{eq:RLc} naturally implies $ R_L = \tau^{-1/2}$, as they also find. The difference is in our interpretation of this result: in their model, $\gamma_p$ is set by the interfacial dynamics, while here we argue that it is set by the convective dynamics.

  \subsubsection{Local interface properties}\label{subsubsec:localdTdC}

We now shift our focus to the local properties of the interface. 
We measure the local instantaneous interface position $z_i(x,y,t)$ and thicknesses $d_T(x,y,t)$ and $d_C(x,y,t)$ by fitting hyperbolic tangent  functions with {\it SciPy}-package for {\it Python} (similar to those given in equation \ref{eq:tanhmodel}) to the local vertical temperature and composition profiles at each value of $x$, $y$, and $t$ available. Using the tanh profile for all cases underestimates $d_C$ and $d_T$ by about 25\% for D-type interfaces (see figure \ref{fig:interfacemodelprofile}), but we use this method here anyway because it can be automated more easily. If an interface is known to be of D-type, we then multiply $d_C$ and $d_T$ by 1/0.755 to correct (at least approximately) this measurement bias.
We also note that for U-type interfaces, some profiles have to be excluded because their interfaces wrap around convective parcels  of fluid (see figure \ref{fig:interface-structure}({\it a})), causing the fitting procedure to fail.

\begin{figure}
    \centering
    \includegraphics[width=0.8\textwidth]{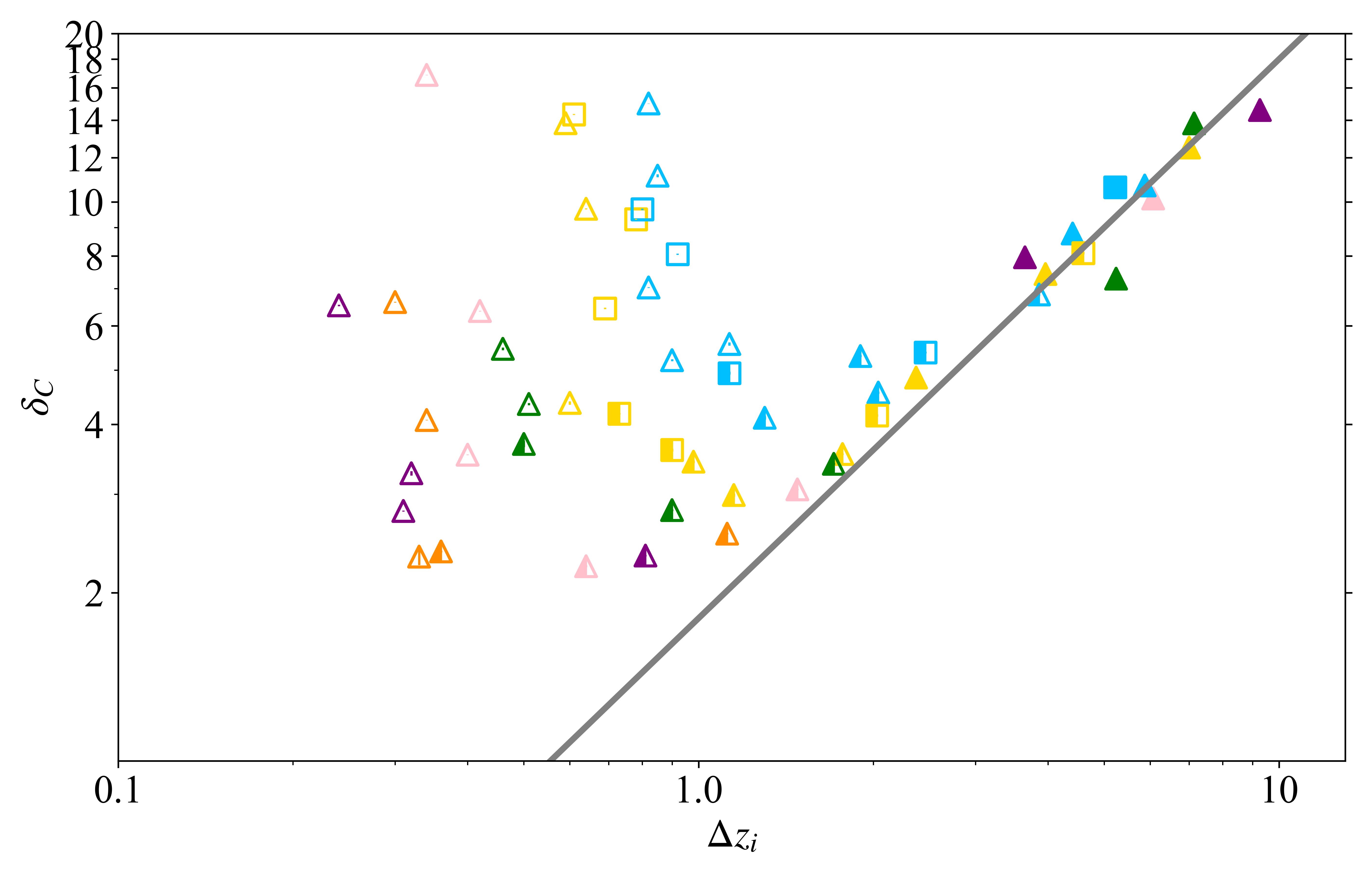}
    \caption{Mean interface thickness $\delta_{C}$ as a function of rms variability $\Delta z_i$ of the instantaneous local interface center position $z_{i}$. The  grey solid line represents $\delta_C = 1.8 \Delta z_i$. The vertical error bars on $\delta_{C}$ are present but often are smaller than the marker size. }
    \label{fig:dzmid}
\end{figure}

From the measurements of $z_i(x,y,t)$, we can compute the rms variability of the interface position $z_i$ around the mean $\bar z_i$, as
\begin{equation}
\Delta z_i = \left[\overline{(z_i(x,y,t)  - \bar z_i)^2}\right]^{1/2}.
\end{equation}
Figure \ref{fig:dzmid} shows the mean compositional interface thickness $\delta_C$ against $\Delta z_i$ for all available simulations. We see that for D-type interfaces and a few O-type interfaces, $\Delta z_i\lesssim 1$, and $\delta_C$ is independent of  $\Delta z_i$. This is not surprising: the interfaces in this limit are very stable, and do not oscillate much. Their mean thickness is independent of the amplitude of the oscillations, and significantly larger as well. By contrast, for U-type interfaces and some of the O-type interfaces, $\Delta z_i\gtrsim1$, and $\delta_C \simeq 1.8 \Delta z_i$. The fact that these two quantities are proportional confirms our intuition that in this limit, the apparently large width of the mean composition interface is primarily due to the oscillations in space and time of a much thinner local interface. These oscillations have larger amplitudes when the interface is weaker, i.e., when 
$R_\rho$ decreases. This confirms that the mean interface thickness is not a good estimate of the actual interface thickness at low $R_\rho$.

We are also interested in the probability distribution functions (pdfs) of $d_T(x,y,t)$ and $d_C(x,y,t)$, called $P(d_{C})$ and $P(d_{T})$, respectively, as well as their joint pdf $P(d_{T}|d_{C})$. We  define the mean values of $d_{C}(x,y,t)$ and $d_{T}(x,y,t)$ in the statistically stationary state as:
\begin{eqnarray}
&&\overline{d}_{C} =\frac{1}{(t_{f}-t_{s})A} \int_{t_{s}}^{t_{f}} \int_{A} d_{C}(x,y,t)dA dt, \\
    &&\overline{d}_{T} =\frac{1}{(t_{f}-t_{s})A} \int_{t_{s}}^{t_{f}} \int_{A} d_{T}(x,y,t)dA dt.
\end{eqnarray}
These quantities are expected to be more or less equal to the interface thicknesses measured from the mean profiles ($\delta_C$ and $\delta_T$, respectively) for steady interfaces, but thinner for interfaces that regularly undergo large vertical excursions.

\begin{figure}
        \centering
    \includegraphics[width=\textwidth]{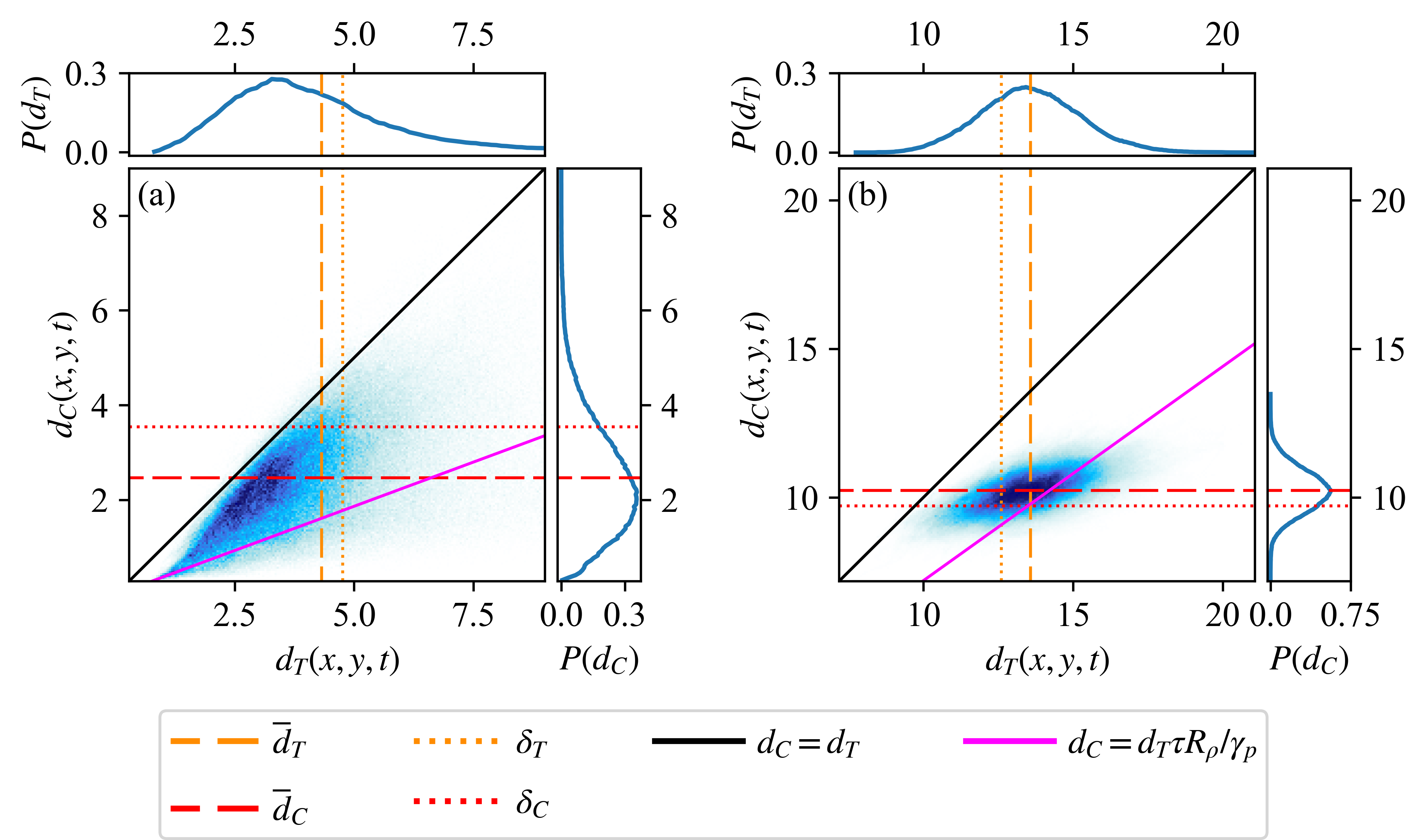}
    \caption{Joint pdfs $P(d_T|d_C)$, with the corresponding marginal pdfs $P(d_T)$ and $P(d_C)$ shown adjacent to the corresponding axis, for two simulations. ({\it a}) O-type interface from a simulation at $\Pr = \tau = 0.1$, $R_{\rho}=2.2$. ({\it b}) D-type interface from a simulation at $\Pr = \tau = 0.1$, $R_{\rho}=4.25$. Note that raw measurements of $d_C$ and $d_T$ for this simulation have been rescaled by 10/7.55 as explained in the main text. The dashed lines are $\overline{d}_{T}$ (orange) and $\overline{d}_{C}$ (red). The dotted lines are the interface thicknesses measured from the mean profiles,  $\delta_{T}$ (orange) and $\delta_{C}$ (red). The solid black line corresponds to $d_{C} = d_{T}$, and the solid pink line corresponds to $d_{C} = d_{T}\tau R_{\rho}/\gamma_{p}$. Note that $\delta_T \simeq {\bar d}_T$ in both cases (within errorbars), and similarly $\delta_C \simeq {\bar d}_C$ for the D-type interface. By contrast $\delta_C$ is significantly greater than ${\bar d}_C$ for the O-type interface.}
\label{fig:dCdT_joinPDF}
\end{figure}

Figure \ref{fig:dCdT_joinPDF} presents  $P(d_{T}|d_{C})$ for two of the sample simulations  shown in figure \ref{fig:interface-structure}: the O-type interface at $\Pr = \tau = 0.1$, $R_{\rho} = 2.2$ on panel ({\it a}) and the D-type interface at $R_{\rho} = 4.25$ on panel ({\it b}). The cumulative pdfs $P(d_T)$ and $P(d_C)$ are also shown in each panel.
The red dashed lines show $\overline{d}_{C}$ (horizontal lines), and the orange dashed lines show $\overline{d}_{T}$ (vertical lines). These can be compared with $\delta _T$ (orange dotted line) and $\delta_C$ (red dotted line).  The solid black line indicates where $d_{C} = d_{T}$. Lastly, the pink solid line corresponds to $d_{C} = d_{T}R_{\rho}\tau/ \gamma_p$ (where $\gamma_p$ is given in table \ref{tab:gammac}). Interfaces which satisfy this equation can diffusively transport all of the flux presented to them by the convective layers. 

We observe that the joint pdfs of $d_{C}$ and $d_{T}$ occupy a broad region bounded by the solid black and pink lines. In the case of $R_{\rho}=2.2$ (panel ({\it a})), the joint pdf is more concentrated near the black line, which might indicate that O-type interfaces are regularly subject to small-scale turbulent mixing events that locally equalize the thicknesses of the respective interfaces in the composition and temperature profiles. In contrast, we see in the case of $R_{\rho}=4.25$ (panel ({\it b})) the joint pdf is closer to the diffusive limit $d_{C} = d_{T}R_{\rho}\tau/ \gamma_p$ shown in the pink line. This was expected as the interface in this simulation is of D-type. We also see that the pdf occupies a more compact region of the plot, consistent with the fact that its thickness is much less variable in time and space. 

We also see that O-type interfaces tend to be significantly thinner than D-type interfaces in both $T$ and $C$ fields, which is consistent with our results from \S\ref{subsec:meanthick}.  On the panel  {\it (a)}, we see that $\overline{d}_{T}$ is close to the value of $\delta_{T}$, while $\overline{d}_{C}$ is considerably smaller than $\delta_{C}$.
More specifically, we find that 
\begin{equation}
    \frac{\left| \delta_{T} - \overline{d}_{T}\right|}{\frac{1}{2}( \delta_{T} + \overline{d}_{T})}\approx 9.7\%, \quad \frac{\left| \delta_{C} - \overline{d}_{C}\right|}{\frac{1}{2}( \delta_{C} + \overline{d}_{C})}\approx 35.8\%.
\end{equation}
This shows that the wave-like structure of O-type interface is mostly present in the $C$ field but not in the $T$ field. In hindsight this is expected, as temperature diffuses much faster than composition, so oscillations in $T$ rapidly average out. By contrast, on panel {\it (b)} we note that $\overline{d}_{T} \simeq \delta_{T}$ and $\overline{d}_{C}\simeq \delta_{C}$:
\begin{equation}
    \frac{\left| \delta_{T} - \overline{d}_{T}\right|}{\frac{1}{2}( \delta_{T} + \overline{d}_{T})}\approx 7.5\%, \quad \frac{\left| \delta_{C} - \overline{d}_{C}\right|}{\frac{1}{2}( \delta_{C} + \overline{d}_{C})}\approx 5.1\%,
\end{equation}
which is consistent with D-type interfaces being more  stable.

In figure \ref{fig:dCdT_joinPDF}, we additionally plot the marginal pdfs $P(d_{C})$ and $P(d_{T})$ for both simulations. 
 In the case of O-type interfaces (panel {\it (a)}), $P(d_{C})$ and $P(d_{T})$ have elongated ‘tails' that shift the mean values $\overline{d}_{C}$ and $\overline{d}_{T}$ to the right of the peaks of $P(d_{C})$ and $P(d_{T})$. These tails are caused by small-scale turbulent events that locally increase the local interface thickness in both $T$ and $C$. Such tails are absent in $P(d_{C})$ and $P(d_{T})$ on panel {\it (b)}, whose shapes resemble a normal distribution instead. Since D-type interface are more stable to local turbulent events, the mean $\overline{d}_{C}$ and $\overline{d}_{T}$ of D-type local interfaces on the panel {\it (b)} are close to the peaks of $P(d_{C})$ and $P(d_{T})$.

\subsection{Properties of the convective layers}
\label{sec:convlayers}

\subsubsection{Density anomaly at the edge of the interface}
\label{subsec:drhodata}

\begin{figure}
        \centering
    \includegraphics[width=\textwidth]{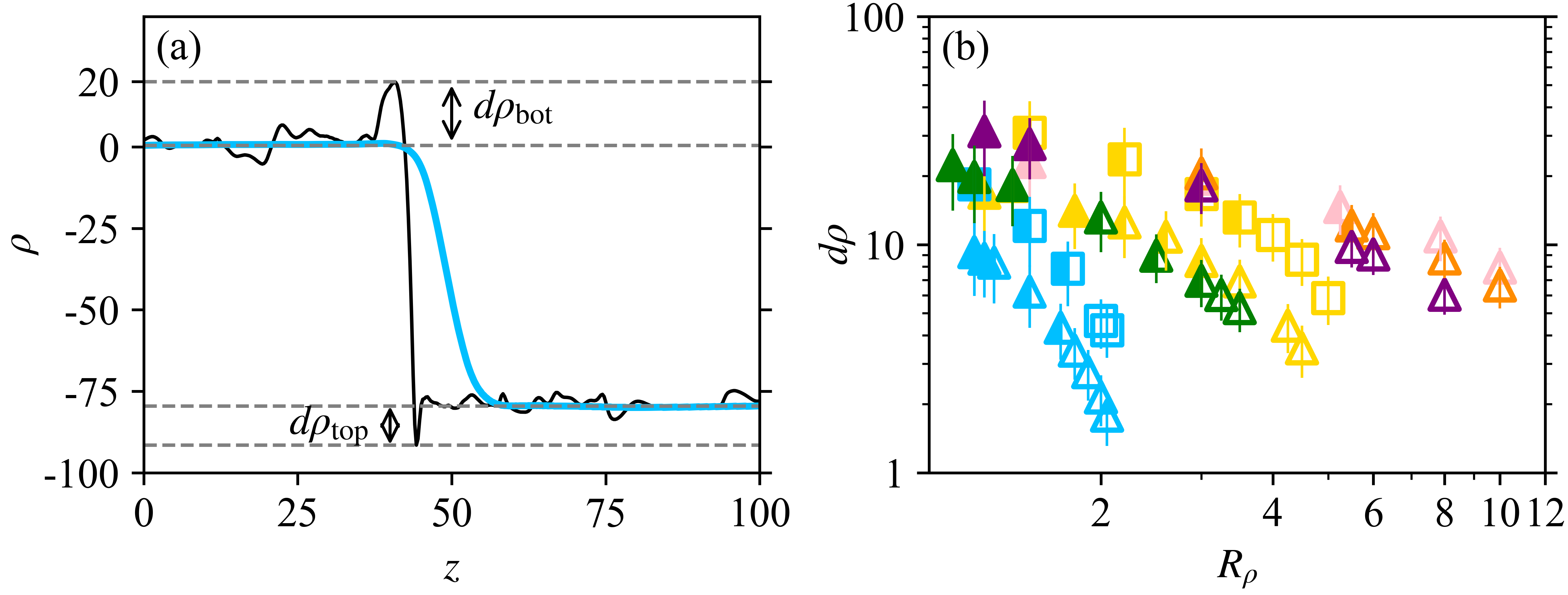}
    \caption{({\it a}) Mean density profile $\overline{\rho}(z)$ (blue line) and a sample density profile $\rho (x,y,z)$ (black line) from a typical simulation at $\Pr = \tau = 0.1, R_{\rho}= 1.8$, and $L_z = 100$, demonstrating the presence of strong local density inversions of  amplitude $d\rho_{\rm top}$ and $d\rho_{\rm bot}$ even when the mean profile does not have any. ({\it b})  Characteristic density anomalies $d\rho$ as a function of $R_{\rho}$, for all simulations, see  table \ref{tab:simulations} for symbol legend. Errorbars correspond to the rms variability around the mean.}
    \label{fig:drhoextract}
\end{figure}

In LDC, the overall background density stratification is stabilizing (i.e. $d\rho^*_b/dz^* < 0$), by contrast with regular convection where it is destabilizing. Convection within the layers is driven by small density anomalies generated at the edges of the interface by thermal diffusion. More specifically, the warmer layer below the interface heats the fluid above it, lowering its density, and ultimately causing it to become buoyant and rise. The process is anti-symmetric across the interface: the colder layer above the interface cools 
the fluid below, increases its density, and causes it to sink. A similar diffusion process also takes place for the composition field (with a stabilizing effect on the density), but it is slower and therefore only has moderate impact on the dynamics especially when $\tau \ll 1$.   
In figure \ref{fig:interface-structure} \citep[see also][]{Molletal2017}, we saw how this process sometimes drives small inversions in the mean density profile just above and below each interface. However,  the mean profiles can be misleading. In the simulation shown in figure \ref{fig:drhoextract}({\it a}), we see that the mean density profile (thick light blue line) has no such inversions. Instead, the convection is driven by the strong density inversions visible in the local density profiles only (e.g. thin black line).  

In what follows, we estimate the characteristic magnitude $d\rho$ of these local density anomalies using the procedure illustrated in figure \ref{fig:drhoextract}({\it a}). We first measure from the data the mean density of the two convective zones above and below the interface $\bar \rho_{\rm top}$ and $\bar \rho_{\rm bot}$. Then, for each available density profile $\rho(x,y,z,t)$, we define the maximum local density anomaly in the bottom convective layer to be $d\rho_{\rm bot}(x,y,t) = \max_z \rho - \bar \rho_{\rm bot}$, and similarly in the top layer $d\rho_{\rm top}(x,y,t) = \bar \rho_{\rm top} - \min_z \rho$. Finally the characteristic value of $d\rho$ is obtained by averaging the local density anomalies thus obtained in $x,y$ and $t$ over the statistically stationary state, over both top and bottom layers. 

The results are shown in figure \ref{fig:drhoextract}({\it b}). We see that $d\rho$ decreases with increasing $R_\rho$, all other parameters being fixed, and increases with $ L_z$. As with most other quantities extracted so far, $d\rho$ depends more strongly on $\tau$ than on  $\Pr$. In hindsight this is not surprising here, as this density anomaly can only form because of the differential diffusion of temperature and composition \citep[see, e.g.][]{Linden_Shirtcliffe_1978}. We therefore expect and find that $d\rho$ is larger for smaller values of $\tau$ at constant $\Pr,R_{\rho}$ and $L_{z}$.

\subsubsection{From density anomaly to rising plumes.}
\label{subsec:snaps}
 
\begin{figure}
    \centering
    \includegraphics[width=0.9\linewidth]{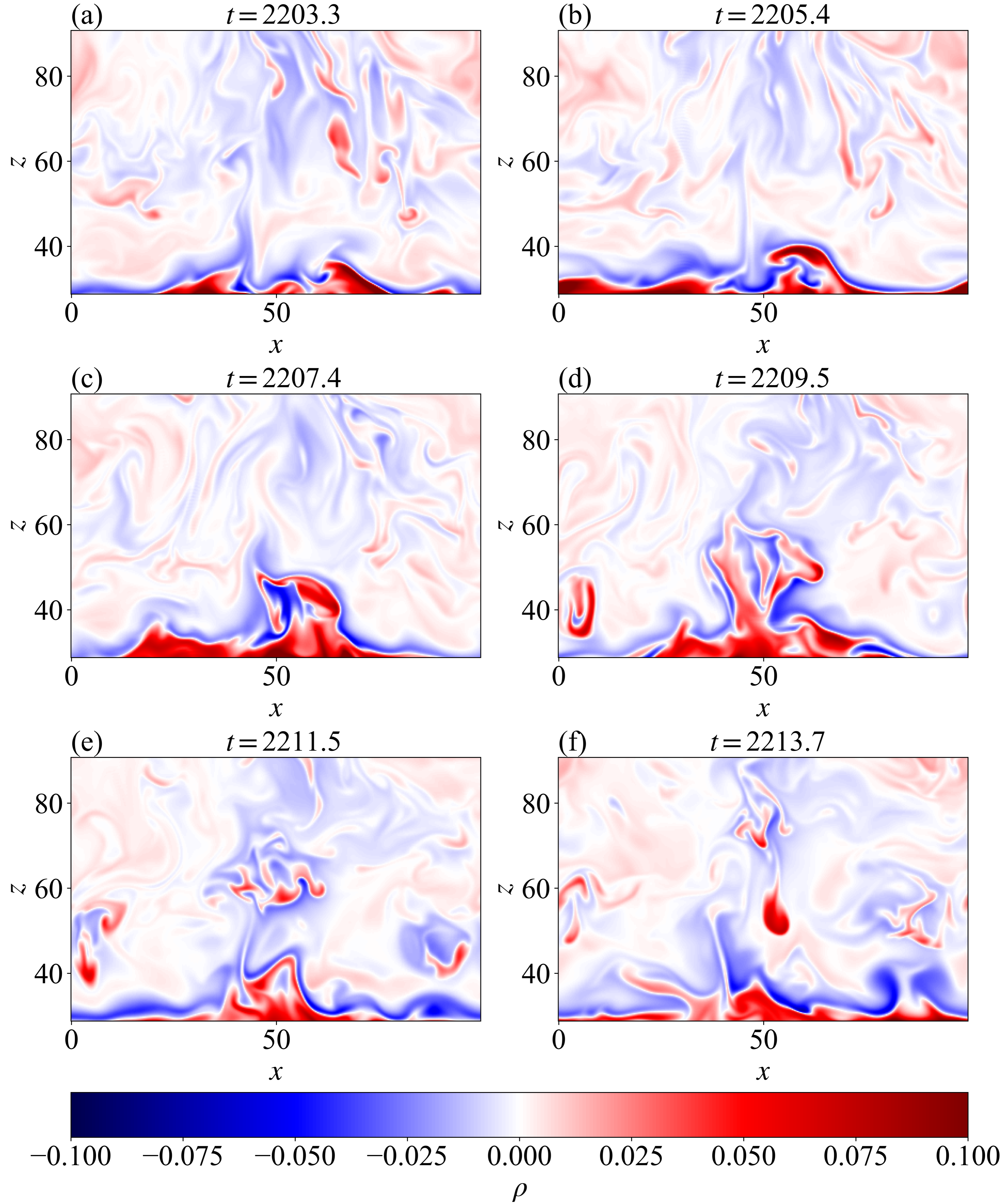}
    \caption{Time series of  density snapshots taken in the $y=0$ plane from a sample simulation at $\Pr = 0.3, \tau = 0.1, R_{\rho} = 3$ and $L_z = 100$, illustrating the dynamics of a fluid parcel, located around $x = 50$, rising from the interface located on average at $\bar z_{i} = 25$. The mean density of the top layer in this simulation is $\bar \rho_{\rm top}  = 0$, so the blue colour represents fluid that is more buoyant than its surrounding, while red represent fluid that is denser than its surrounding.}
    \label{fig:blobs}
\end{figure}

The density anomalies discussed in the previous section form density parcels that are more/less buoyant than the surrounding fluid above/below the interface, driving them  away from the interface into the bulk of the convection zone. To illustrate the evolution of one of the buoyant parcels above the interface, figure \ref{fig:blobs} shows a time series of density snapshots from a sample simulation at $\Pr =0.3, \tau = 0.1$, $R_{\rho} = 3$ and $L_z = 100$. The snapshots are cropped between $z = 30$ and $z = 90$ in the convective zone above the interface (which is located at $\bar z_{i} \approx 25$)  for better visualization of the density field.

On the early snapshots at $t = 2203.3$, $t = 2205.4$ and  $t = 2207.4$, we observe the formation of a lighter density fluid parcel (blue colour) around $x = 50$ that starts rising  and entraining higher density fluid (red colour) in its vicinity through viscous effects. The higher density fluid has both higher $T$ and  higher $C$ than the surrounding. It continues to rise while surrounded by a low density sheath of fluid,  which it maintains by heating (through thermal diffusion) from its warm core.  

On the subsequent snapshots at $t = 2209.5$ and $t = 2211.5$, we see that the parcel separates from the interface, rises further up but is broken apart into smaller parcels by the ambient turbulence, while still surrounded by a sheath of low density fluid. This low density sheath eventually disappears when the temperature of the smaller parcels equalizes with the surrounding fluid. This can be seen on the last snapshot at $t = 2213.7$: the dense core of the parcel has lost its low density sheath and starts to sink back towards the interface. 

Overall, this complex process of diffusively-driven rise, combined with turbulent entrainment and breakup, followed by partial fall-back of composition-rich fluid, sets the ratio of composition and temperature fluxes $\gamma$ through the convective layer. We see that while most of the heat content of the rising parcel ends up mixed into the convective zone, not all of the compositional content does,  as some of it returns to the lower layer. As a result, the flux ratio $\gamma$ in the convective zone depends much less on the initial properties of the fluid parcels, than on diffusion and on their interaction with the surrounding fluid. The ambient turbulence, which causes the parcels to break up into smaller parcels, accelerates the temperature diffusion, and therefore the likelihood that high $C$ fluid falls back down. At the same time, smaller parcels are more easily entrained upwards in flow updrafts, and diffuse their high $C$ content into the convection zone more rapidly.  The flux ratio  thus fundamentally depends on the diffusion parameters $\Pr$ and $\tau$, and on the turbulent motions in the convective layer.  We believe that this is the primary reason why $\gamma$ is more-or-less independent of $R_\rho$, and instead depends more obviously on $\Pr$, $\tau$ and $L_z$. Furthermore, this interpretation provides a qualitative rationale for this dependence. For instance, 
it is more difficult to irreversibly mix compositionally rich material into the bulk of the convection zone at low $\tau$ than at high $\tau$, which explains why $\gamma$ decreases with decreasing $\tau$ when all other parameters are being held fixed. At fixed $R_\rho$, $\Pr$ and $\tau$, convective velocities are higher for taller layers (see \S\ref{sec:convective_w}), so the compositionally rich dense parcels can be maintained aloft in the bulk of the convection zone for longer, providing more chances for irreversible mixing, thus increasing the flux ratio $\gamma$. Finally, the dependence on $\Pr$ (with all other parameters being fixed) is more subtle. On the one hand, a higher $\Pr$ corresponds to higher convective velocities (see \S\ref{sec:convective_w}), which would increase $\gamma$ by the same reasoning. However, a higher $\Pr$ also prevents the breakdown of the fluid parcels into smaller ones, and the more massive parcels of compositionally rich fluid are harder to maintain aloft and are more likely to sink back to the lower layer. It appears the second effect dominates over the first.

\subsubsection{Vertical velocities of  convective parcels.}\label{sec:convective_w}

In this section, we now study quantitatively the properties of the convective layers, starting with the characteristic (rms) vertical velocity of the fluid parcels.
In figure \ref{fig:interface-structure}, we already saw that the rms vertical velocity of the parcels, $w_{\rm rms}$, is not constant within the convective layer. In figures \ref{fig:interface-structure}({\it h,g})  for example, we see that the dynamical structure of the convective layers consists of ‘transition' zones, attached to the interfaces ($25\lesssim z \lesssim 75$), and ‘bulk' zones ($ 75\lesssim z$ and $z\lesssim 25$), where $w_{\rm rms}$ becomes uniform.  

\begin{figure}
        \centering
    \includegraphics[width=0.9\textwidth]{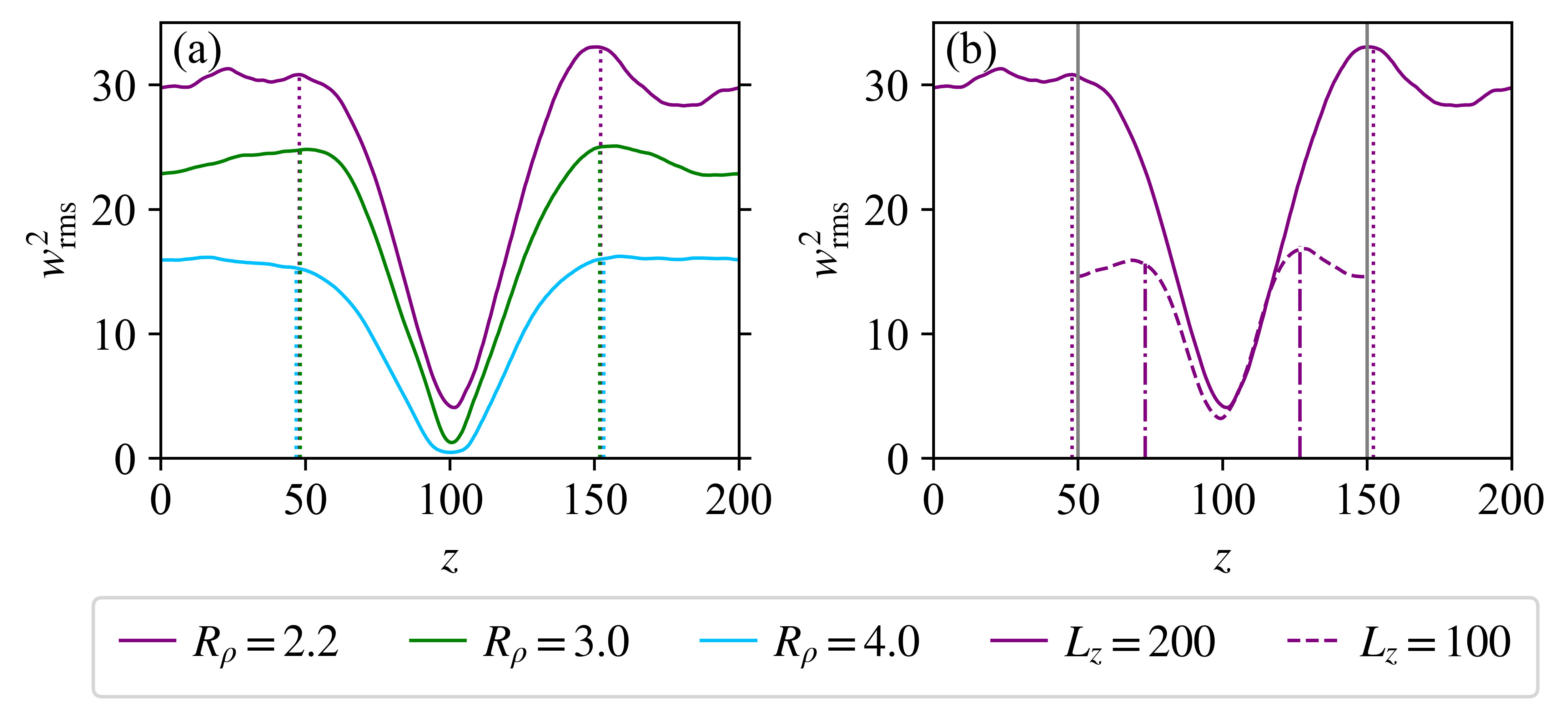}
    \caption{({\it a}) Profiles of $w_{\rm rms}^{2}(z)$ at fixed $\Pr =\tau = 0.1$, $L_{z} = 200$ and varying $R_{\rho}$. ({\it b}) Profiles of $w_{\rm rms}^{2}(z)$ at fixed $\Pr=\tau = 0.1$, $R_{\rho} = 2.2$, and varying $L_{z}$. For ease of comparison, all of the profiles are shifted to have the midpoint of the interface at $z = 100$. The vertical grey lines mark the bounds of $L_{z}=100$ simulation at $z = 50$ and $z = 150$. On both panels, coloured vertical lines mark the end of the transition region (where $w_{\rm rms}^2$ increases with distance from the interface) and the start of the bulk of convection zones (where $w_{\rm rms}^2$ is roughly constant), see \S\ref{sec:convective_w}. 
    }
    \label{fig:wrms}
\end{figure}

To see how these results depend on input parameters, figure \ref{fig:wrms} presents vertical profiles of $w_{\rm rms}^{2}$, computed using equation (\ref{eq:wrmsdef}), in various simulations. 
In figure \ref{fig:wrms}({\it a}), we compare $w_{\rm rms}^{2}$ profiles for $\Pr = \tau = 0.1$ and $L_{z}=200$ for increasing $R_{\rho}$.
All curves have a similar structure: $w_{\rm rms}^{2}$ is at a minimum in the core of the interface at $\bar z_i \simeq 100$, then linearly increases with distance away from $\bar z_i$, approaching a plateau in the ‘middle' of the convective layers.  
Not surprisingly, we find that $w_{\rm rms}^{2}$ is greater in the more weakly stratified simulation at $R_{\rho} =2.2$ (purple line), than in the more strongly stratified simulations at $R_{\rho} =3$ (green) and $R_{\rho} =4$ (blue).

Figure \ref{fig:wrms}({\it b}) compares two simulations at low  $R_{\rho} = 2.2$ and different domain heights: $L_{z} = 100$ (dashed purple) and $L_{z} = 200$ (solid purple). We see that $w^2_{\rm rms}(z)$ is almost identical near the interface in the two simulations, and increases with distance from the interface at approximately the same rate. However, $w^2_{\rm rms}$ reaches a plateau much earlier in the smaller domain. 
Consequently, we find that $w^2_{\rm rms}$ in the bulk of convective layers is greater in taller domains. In practice, we find that the ‘bulk' of the convective zones occupies about a half of the total convection zone size in our simulations, suggesting that the transition zone ends because plumes falling from the bottom of the upper interface  meet plumes rising from the top of the lower interface.

In what follows, we wish to measure salient quantities in the bulk of the convective layer, which requires us to define its approximate location. Having measured the compositional interface thickness $\delta_{C}$ and position $\bar z_i$ as explained in \S \ref{sec:interface-thickness}, we define the broader convection zone (including both the transition regions near the interface, and the bulk) as any $z$ outside of the interval $[\bar z_i - \delta_C,\bar z_i + \delta_C]$. The bulk of the convective zone itself is then defined as the middle-half of the broader convection zone (taking periodicity into account if needed), and denoted as $z \in \Omega_{\rm cz}$. In figure \ref{fig:wrms}, where the interface is in the middle of the domain, the bulk region thus identified is the region outside of the vertical lines shown. The transition regions are each $0.25 (L_z - 2\delta_C)$ thick.

Using this, we can define $w_{\rm rms}^{\rm cz}$ to be the rms vertical velocity in the bulk of convective zone only as:
\begin{equation}
    w_{\rm rms}^{\rm cz} =  \left[\frac{2}{L_{z} - 2\delta_{C}}  \int_{\Omega_{\rm cz}} w^2_{\rm rms}(z)dz\right]^{1/2}.
        \label{eq:wrmsL}
\end{equation}

In a similar manner, we define the rms density fluctuation in the bulk of the convective zone, $\rho_{\rm rms}^{\rm cz}$:
\begin{equation}
    \rho_{\rm rms}^{\rm cz} = \left[ \frac{2}{L_{z} - 2\delta_{C}}  \int_{\Omega_{\rm cz}}\overline{\left[\rho (x,y,z,t) - \overline{\rho}(z) \right]}^{2} dz \right]^{1/2},
    \label{eq:rhormsL}
\end{equation}
where $\overline{\rho}(z,t)$ is the mean density at a given height $z$ and time $t$, which is almost constant in the convective zone. Finally we also define the convective density flux $F_{\rho}^{\rm cz}$ in the bulk of the convection zone as:
\begin{equation}\label{eq:frho_layer}
    F_{\rho}^{\rm cz} =  \frac{2}{L_{z} - 2\delta_{C}} \int_{\Omega_{\rm cz}} \overline{\left[-wT + wC\right]}dz.
\end{equation}

\begin{figure}
        \centering
    \includegraphics[width=\textwidth]{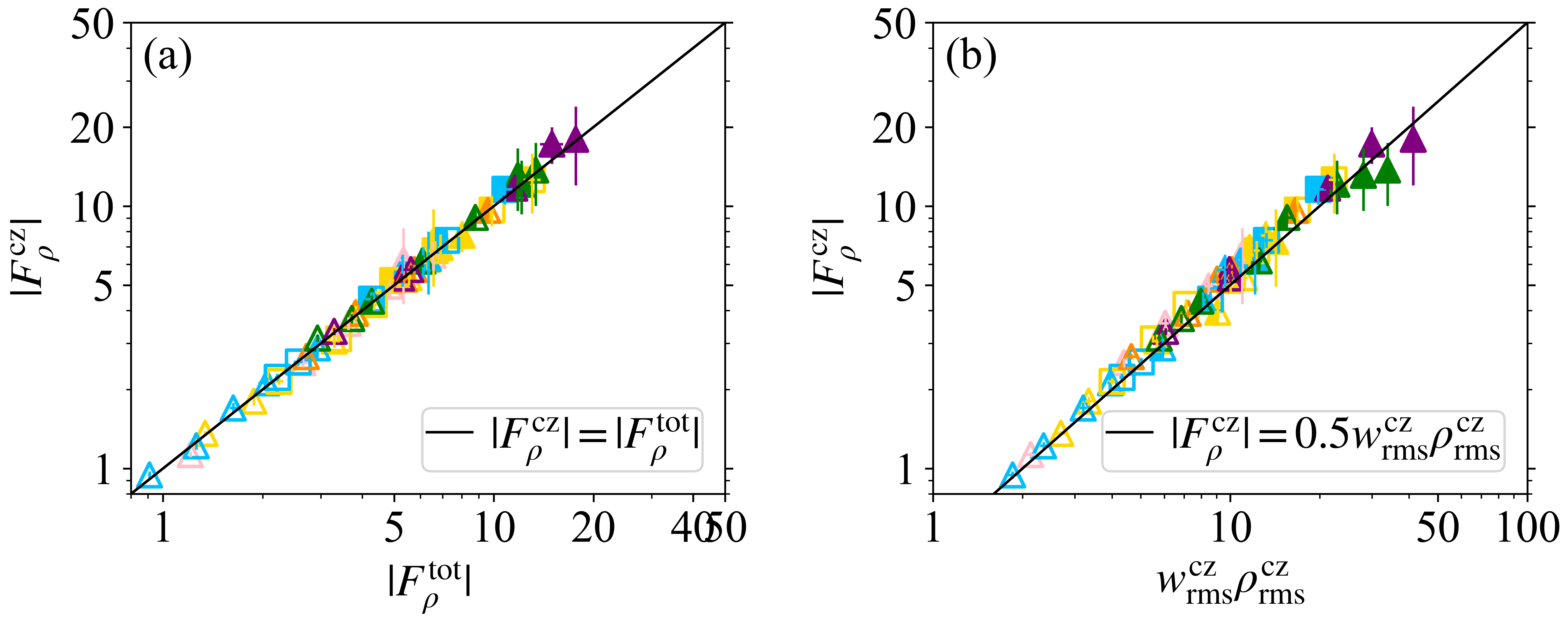}
    \caption{({\it a}) Density flux in the bulk of the convective layer, $|F_{\rho}^{\rm cz}|$, against $|F_{\rho}^{\rm tot}|$ for all simulations with corresponding error bars. The black line is $|F_{\rho}^{\rm cz}| = |F_{\rho}^{\rm tot}|$. The horizontal error bars on $|F_{\rho}^{\rm tot}|$ are often smaller than the marker size. ({\it b}) $|F^{\rm cz}_{\rho}|$ against the product $w_{\rm rms}^{\rm cz} \rho_{\rm rms}^{\rm cz}$, measured in the bulk of the convective zone with corresponding error bars. The solid black line represents $|F_{\rho}^{\rm cz}| = 0.5 w_{\rm rms}^{\rm cz} \rho_{\rm rms}^{\rm cz}$. The horizontal error bars  $w_{\rm rms}^{\rm cz} \rho_{\rm rms}^{\rm cz}$ are often smaller than the marker size. See table \ref{tab:simulations} for symbol legend. This  shows that separately measured $w_{\rm rms}^{\rm cz}$ and $\rho_{\rm rms}^{\rm cz}$ can be used to obtain a reasonable estimate of the density flux through a uniform staircase.}
    \label{fig:rhorms}
\end{figure}

In figure \ref{fig:rhorms}({\it a}), we first compare the convective density flux $F^{\rm cz}_{\rho}$ on the vertical axis against the total density flux, $F_{\rho}^{\rm tot}$  (see equation \ref{eq:flux_defs_forPlots_rho}), on the horizontal axis, measured from all our simulations. Note that both are negative, so we show their absolute value instead. The black solid line represents $|F_{\rho}^{\rm tot} |= |F_{\rho}^{\rm cz}|$, and we see that all markers follow this line closely, which confirms that the convective layers are indeed fully convective for all simulations; this was naturally expected for all U-type and O-type interfaces, but we see that it is also true of the D-type interfaces. In other words, even when the system is overall very strongly stratified, the convection is vigorous enough within the layers to carry all of the flux advectively.

Figure \ref{fig:rhorms}({\it b}) presents $|F_{\rho}^{\rm cz}|$ (vertical axis) against the product of $\rho_{\rm rms}^{\rm cz}$ and $w_{\rm rms}^{\rm cz}$ (horizontal axis) measured in the bulk of the convective zone. The black  line corresponds to $|F_{\rho}^{\rm cz}| = 0.5 w_{\rm rms}^{\rm cz} \rho_{\rm rms}^{\rm cz}$, and we see the data collapses on this line remarkably well for all cases, suggesting that $F_{\rho}^{\rm cz}$ can indeed be predicted if $\rho_{\rm rms}^{\rm cz}$ and $w_{\rm rms}^{\rm cz}$ are known. Thus, if we were somehow able to model $\rho_{\rm rms}^{\rm cz}$ and $w_{\rm rms}^{\rm cz}$ as functions of input parameters, we could use equation \eqref{eq:frho_layer} to also predict $F_{\rho}^{\rm cz}$.

A simple possible model for $\rho_{\rm rms}^{\rm cz}$ and $w_{\rm rms}^{\rm cz}$ can be created assuming that parcels with density anomaly $d\rho$ travel ballistically from the interface until they reach the bulk of the convective layer. If that were the case, we would expect that:
\begin{equation}
    \rho_{\rm rms}^{\rm cz} \propto d\rho,
\end{equation}
and 
\begin{equation}\label{eq:wRMSL_Ballisticscaling}
    w_{\rm rms}^{\rm cz} \propto  \sqrt{d\rho \Pr L_{z}},
\end{equation}
in non-dimensional units, assuming that the interface thickness $\delta_{C} \ll L_{z}$.

\begin{figure}
        \centering
    \includegraphics[width=\textwidth]{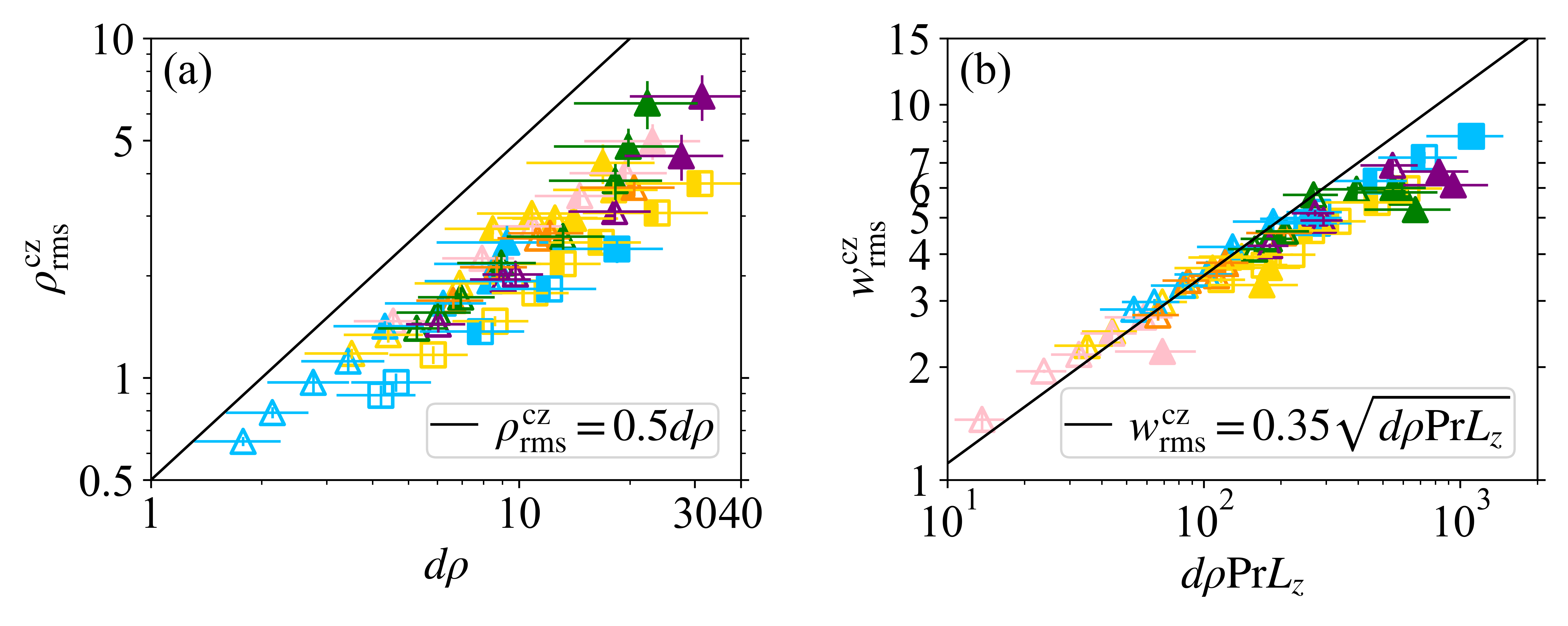}
    \caption{({\it a}) $\rho_{\rm rms}^{\rm cz}$ as a function of  $d\rho$ with corresponding errorbars, extracted from the DNS as explained in \S\ref{subsec:drhodata} and equation \eqref{eq:rhormsL}. The black line shows the $\rho_{\rm rms}^{\rm cz} = 0.5 d\rho$ line for comparison, illustrating that $\rho_{\rm rms}^{\rm cz}$ is not proportional to $d\rho$ in general. ({\it b}) $w_{\rm rms}^{\rm cz}$ as a function of  $d\rho \Pr L_{z}$.  We see that the proposed  ballistic argument (see  equation \ref{eq:wRMSL_Ballisticscaling}) is appropriate  for low $w_{\rm rms}^{\rm cz}$, with $w_{\rm rms}^{\rm cz} \simeq 0.35 \sqrt{d\rho \Pr L_z}$ (black line), but fails for high values of $w_{\rm rms}^{\rm cz}$. In both panels, see table \ref{tab:simulations} for the symbol legend. In both panels, the vertical error bars are often smaller than the marker size.}
    \label{fig:wrms-layer}
\end{figure}

Unfortunately, this model does not work as well as we had hoped for. In figure \ref{fig:wrms-layer}({\it a}), we show the measured $\rho_{\rm rms}^{\rm cz}$ against the measured density anomaly $d\rho$ for all simulations. We see that
while  
$\rho_{\rm rms}^{\rm cz}$ does seem to be a power law function of $d\rho$ for a given set of parameters $(\Pr, \tau, L_z)$, the exponent appears to be smaller than one, and the prefactor clearly depends (weakly) on $\tau$, $\Pr$, and strongly on layer height $L_{z}$. 
Similarly, we show the measured value of $w_{\rm rms}^{\rm cz}$ as a function of $\Pr d\rho L_{z}$ in figure \ref{fig:wrms-layer}({\it b}). The black line corresponds to $w_{\rm rms}^{\rm cz} = 0.35 \sqrt{d\rho \Pr L_z}$, which would match the predicted scaling law from equation \eqref{eq:wRMSL_Ballisticscaling}. Here, we see that many of the data points do collapse on this line at low $w_{\rm rms}^{\rm cz}$, suggesting that the $\sqrt{\Pr d\rho L_{z}}$ scaling captures the dependence of $w_{\rm rms}^{\rm cz}$ on $\Pr$  directly and the dependence on $R_{\rho}$ and $\tau$ through $d\rho$ implicitly. However, the model prediction overestimates $w_{\rm rms}^{\rm cz}$ for large values of $d\rho$, and we also see, as in figure \ref{fig:wrms-layer}({\it a}), that the predicted $L_{z}$ dependence is not quite right: square and triangle markers, which correspond to different values of $L_{z}$, do not collapse on the same curve. 

We believe that these discrepancies can all be attributed to the turbulent disruption of the fluid parcels as they rise from the interface towards the bulk of the convection zone and entrain (and mix with) the ambient fluid. This disruption is larger in taller layers (because the fluid parcel has to travel further), and is also larger for more turbulent flows (where larger convective velocities disrupt the ballistic rise and more rapidly destroy the parcels). Thus, creating a satisfactory ad-hoc model for $w_{\rm rms}^{\rm cz}$ and $\rho_{\rm rms}^{\rm cz}$ will require a better theory of plume transport in a turbulent double-diffusive environment. This is beyond the scope of the present work, and, as we now show, may not actually be necessary for the purpose of modeling the density flux.

\subsection{Relationship between the density anomaly and the density flux}
\label{subsec:densityfluxlaw}

\begin{figure}
    \centering
    \includegraphics[width=0.95\linewidth]{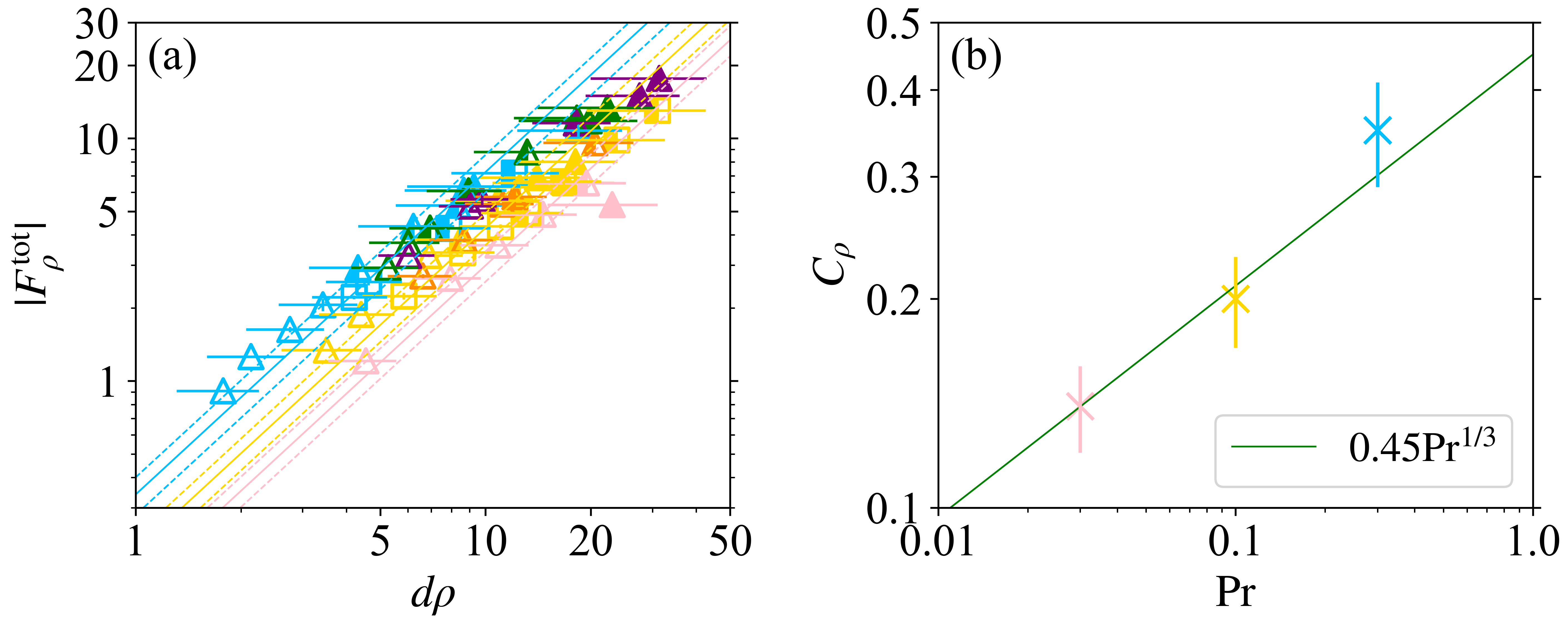}
    \caption{({\it a})  $|F_{\rho}^{\rm tot}|$ as a function of $d\rho$ for all available simulations. Note how all simulations with similar $\Pr$ more or less collapse on the same lines (i.e. $\Pr = 0.3$ for blue, green and purple symbols, $\Pr = 0.1$ for orange and yellow symbols, and $\Pr = 0.03$ for pink symbols). Coloured solid lines represent the empirical scaling law in equation \eqref{eq:Frhofunction}, with $C_\rho$ given in \eqref{eq:Crho},  for three different values of $\Pr$. The dashed lines of the same colour show the impact of the estimated errors in $C_{\rho}$. The vertical error bars on $|F_{\rho}^{\rm tot}|$ are smaller than the marker size. ({\it b}) Fitted values of $C_{\rho}$ as a function of  $\Pr$ with corresponding errorbars. The green line shows $C_\rho = 0.45 \Pr^{1/3}$ for comparison. } 
    \label{fig:rhoflux}
\end{figure}

While the naive model presented above fails to capture the exact dependence of $w_{\rm rms}^{\rm cz}$ and $\rho_{\rm rms}^{\rm cz}$ (and therefore $F_{\rho}^{\rm cz}$ or equivalently $F_{\rho}^{\rm tot}$) on the input parameters, we serendipitously found that $F_{\rho}^{\rm tot}$ itself is closely related to the characteristic density anomaly $d\rho$.
This relationship is shown in figure \ref{fig:rhoflux} (see also figure \ref{fig:compareTCfluxmodel} showing $F^{\rm tot}_T$ and $F^{\rm tot}_C$ as functions of $d\rho$). The errorbars on $F^{\rm tot}_\rho$ and $d\rho$ represent their rms temporal variability around the mean.

We find empirically that the data is consistent with the following scaling law: 
\begin{equation}
    F^{\rm tot}_\rho = F_\rho^{\rm cz} = - C_\rho(\Pr) d\rho^{4/3},
\label{eq:Frhofunction}
\end{equation}
for almost all cases, with the exception of some of the U-type interfaces.  This shows that the obvious dependence of the fluxes (or equivalently, the Nusselt numbers) on $L_z$ and $\tau$ noted in figure \ref{fig:TCfluxes} can be almost fully accounted for by the dependence of $d\rho$ itself on these quantities. 
In particular, we find that the prefactor $C_\rho$ is almost independent of $L_z$ and $\tau$. To see this, compare squares and triangles, and note that all ‘cold' coloured symbols (blue, green and purple) for which $\Pr = 0.3$ fall on the same line, that both orange and yellow markers ($\Pr =0.1$) fall on the same line, and finally that the pink markers ($\Pr = 0.03$) fall on yet another line. We can fit the available data to find that 
\begin{eqnarray}
C_\rho(\Pr = 0.3) = 0.35 \pm 0.06, \nonumber \\
C_\rho(\Pr = 0.1) = 0.2 \pm 0.03 ,\nonumber  \\
C_\rho(\Pr = 0.03) = 0.14 \pm 0.02.
\label{eq:Crho}
\end{eqnarray}
The error bars on $C_\rho$ capture both the intrinsic error bars on the density flux, as well as a possible weak remaining 
dependence of $C_\rho$ on $\tau$ and $L_z$. It is interesting to note that, for the range of values of $\Pr$ considered here, $C_\rho(\Pr)$ can be modeled (approximately, but not perfectly) by the function
\begin{equation}
C_\rho \simeq 0.45 {\rm Pr}^{1/3},
\label{eq:Crhoscaling}
\end{equation}
see figure \ref{fig:rhoflux}({\it b}).
Whether this 
formula continues to hold beyond the range of parameters explored here will need to be confirmed with additional data at lower and higher $\Pr$. In the meantime, the simple scaling law found in \eqref{eq:Frhofunction} motivates the modeling approach adopted in the next section.

\section{A model for turbulent fluxes in uniform layered convection at low $\Pr$}
\label{sec:model}

In the previous section, we presented extensive data on the properties of interfaces and layers in uniform diffusive-convective staircases at low Prandtl number. In this section, we gradually construct a model for the density, temperature and composition fluxes through such staircases.  We focus on the moderately and strongly stratified cases (intermediate and high $R_\rho$), containing O-type or D-type interfaces. Indeed, as shown by \citet{Woodal13} and \citet{Tulekeyev2024}, the strongly turbulent U-type interfaces at low $R_\rho$ do not live very long, and  undergo instead rapid mergers. On longer time scales, the evolution of a staircase would only be controlled by fluxes through O-type and D-type interfaces.

\begin{figure}
    \centering
    \includegraphics[width=0.9\linewidth]{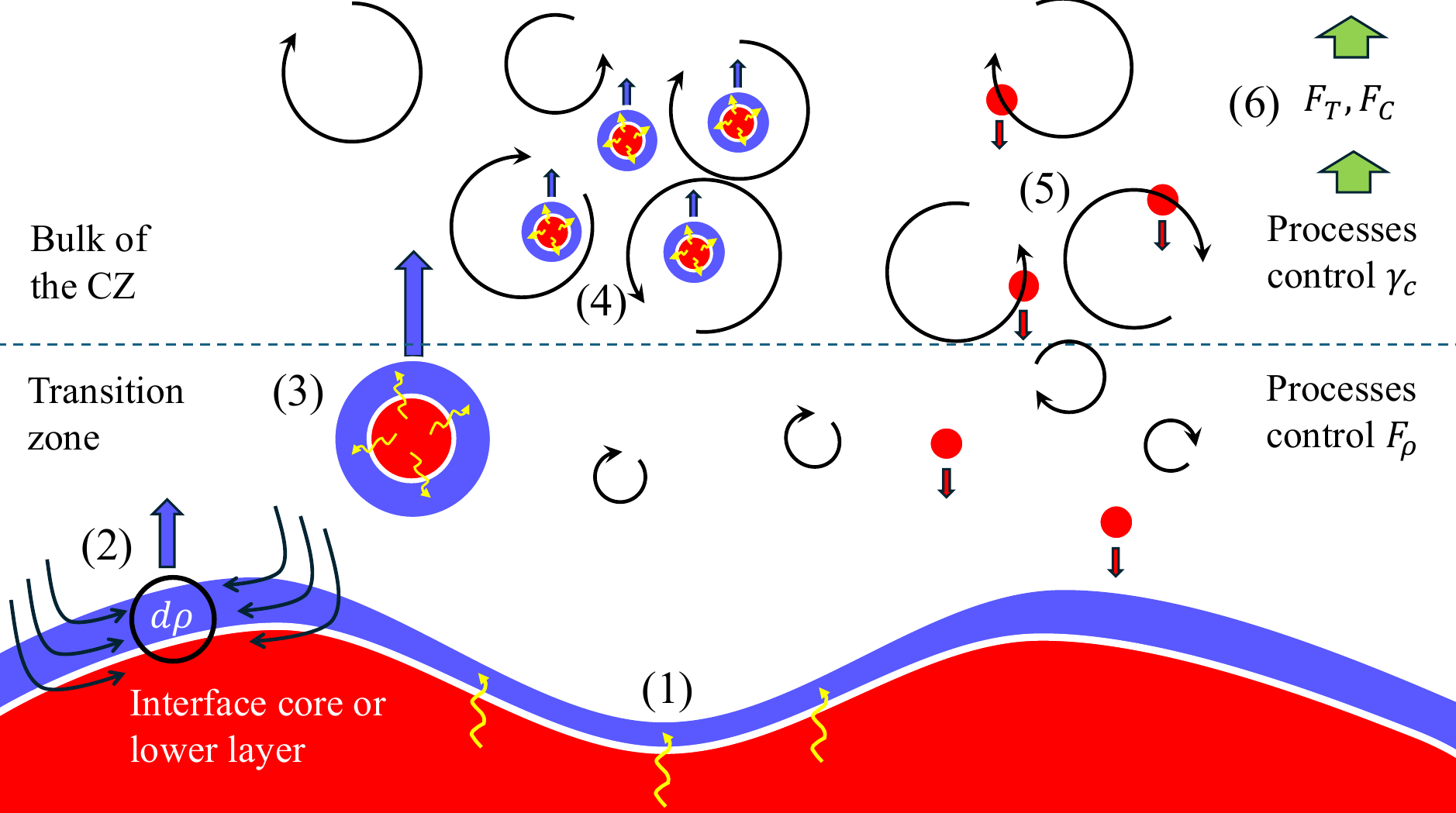}
\caption{Conceptual illustration of the proposed model, see text for detail. Blue colours represent lower density fluid, and red colours represent higher density fluid.}
    \label{fig:illustration}
\end{figure}

 Our proposed conceptual view of the system is illustrated in figure \ref{fig:illustration}, which is inspired by the sequence of events observed in figure \ref{fig:blobs}. The region in red near the bottom of the figure is a high-density region, which corresponds to the lower layer (in the case of an O-type interface) or the top of the interface core (in the case of a D-type interface). Here, it shown to be wavy to depict an O-type interface, but would be mostly horizontal for a D-type interface. The fluid in this layer has higher $T$ and $C$ than in the convective layer above (shown in white). In step (1), negative density anomalies (in blue colour) develop just above the interface due to heat diffusion from below (yellow squiggly arrows). In step (2), a buoyant parcel rises from this density anomaly, and viscously entrains some of the higher density fluid below. The fraction of higher-density fluid entrained is presumably a function of the Prandtl number. In step (3), the parcel rises more or less ballistically in the relatively quiescent lower region of the convective layer (the transition region). This is supported by the data from \S\ref{sec:convective_w}. Heat diffusion from its core continuously maintains a sheath of lower density fluid around it, which entrains the parcel upwards, as seen in figure \ref{fig:blobs}({\it c}). This process sets the density flux through the staircase. In step (4) the parcel encounters more vigorous turbulence in the bulk of the convective layer and is divided into smaller parcels. Each of them continues to be surrounded by a low-density sheath due to heat diffusion, again as seen in figure \ref{fig:blobs}({\it e}). In step (5), the temperature of the small parcels has equalized with that of the convective layer, so their low density sheath has disappeared. They  now behave as dense ‘particles' in a turbulent fluid. Some of them fall back to the lower layer, and some of them are maintained aloft long enough to be mixed with the convective layer, see figure \ref{fig:blobs}({\it f}). This process sets the total temperature and composition fluxes through the convective zone, and therefore their flux ratio $\gamma_p$. In step (6), by virtue of having an infinite uniform staircase, the same fluxes must also be carried diffusively through the interface above (not shown), thereby constraining the gradients of temperature and composition in its core. Finally, we note that the whole process is antisymmetric with respect to  interface, namely, denser anomalies similarly drive convection from the bottom of the interface (not shown). 

Based on this idea, we begin in \S\ref{subsec:drhomodel}  by creating an ad-hoc model for the density anomaly $d\rho$ as a function of input parameters, and compare this model to the data. In parallel,  \S\ref{subsec:fluxmodel} provides a tentative rationale for the scaling law discovered in \S\ref{subsec:densityfluxlaw}, namely  $F^{\rm tot}_\rho = -C_\rho(\Pr) d\rho^{4/3}$. From there, knowledge of the flux ratio $\gamma$ is required to infer the temperature and composition fluxes $F^{\rm tot}_T$ and $F^{\rm tot}_C$ separately. In \S\ref{subsec:FTFC}, we use the empirically found value of $\gamma_p$ (which we interpret to be the convective flux ratio, as explained above) and combine all our results
 to provide a model for the fluxes of density, temperature and composition across a uniform staircase that only depends on the input parameters $R_\rho$, $\Pr$, $\tau$ and $L_z$, and compare the predictions with the data.

\subsection{Interface model, and prediction of $d\rho$}
\label{subsec:drhomodel}
 
\begin{figure}
    \centering
    \includegraphics[width=0.8\linewidth]{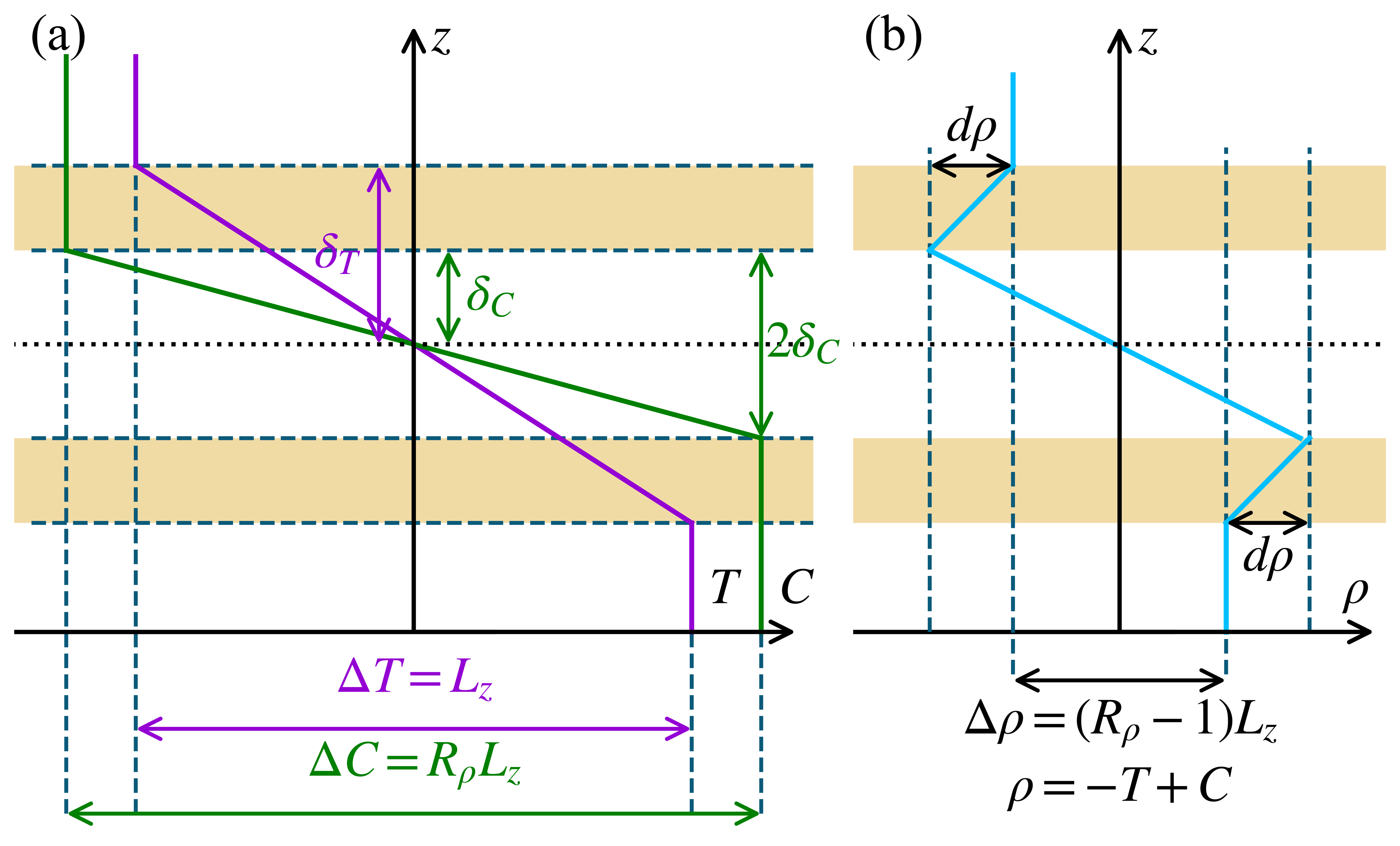}
    \caption{({\it a}) Illustration of the piecewise linear model for a D-type interface located at $z_{i} = 0$. The profile of $T(z)$ is shown in purple, and has an interface of full thickness $2 \delta_T$. Similarly, $C(z)$ is shown in green.  $T(z)$ and  $C(z)$ are constant below and above the interface with a total temperature jump of  $\Delta T = L_{z}$ and composition jump of $\Delta C = R_{\rho}L_{z}$. ({\it b}) Illustration of the resulting density profile, $\rho(z)$, in the same model with a total density jump $\Delta \rho = (R_{\rho}-1)L_{z}$. Density anomalies, where the gradient of $\rho$ is inverted, form at $\delta_T >z >\delta_C$ above the center of the interface and  $ - \delta_C>z > - \delta_T$ below the center of the interface. }
    \label{fig:diffinterfacemodel}
\end{figure}

To model the typical density anomaly $d\rho$ that is created at the edge of the interface by heat diffusion, we begin by focusing on the diffusive (D-type) interfaces. We saw evidence in \S \ref{sec:interfacetype} that the local and mean profiles of temperature and composition for D-type interfaces are roughly the same, and are well-approximated by piecewise linear functions (see figure \ref{fig:interface-profiles}({\it b})). In this section only, we call them for simplicity $T(z)$ and $C(z)$. They are illustrated in figure \ref{fig:diffinterfacemodel}({\it a}), assuming that the interface is located at $z_i = 0$ (this is without loss of generality since the computational domain itself is periodic and the interface can be placed anywhere). Mathematically,  $T(z)$ is modeled as  
\begin{eqnarray}
&& T(z) = \frac{\Delta T}{2} \mbox{  if  }  z < - \delta_T, \nonumber \\
&& T(z) = - \frac{\Delta T}{2\delta_T} z \mbox{  if  }    - \delta_T < z <  \delta_T, \nonumber \\ 
&& T(z) = - \frac{\Delta T}{2} \mbox{  if  }   z>  \delta_T,
\label{eq:linTmodel}
\end{eqnarray}
with an equivalent expression for the composition profile obtained by replacing $\Delta T$ with $\Delta C$, and $\delta_T$ with $\delta_C$. 
Using these expressions, we can construct a model for the  density profile $\rho(z) = -T(z)+C(z)$:
\begin{eqnarray}
&& \rho(z) = -\frac{\Delta T}{2} + \frac{\Delta C}{2} \mbox{  if  }  z < - \delta_T, \nonumber \\
&& \rho(z) = - \frac{\Delta T}{2\delta_T}z  + \frac{\Delta C}{2}  \mbox{  if  }    - \delta_T < z <  -\delta_C, \nonumber \\ 
&& \rho(z) = - \frac{\Delta T}{2\delta_T}z + \frac{\Delta C}{2\delta_C}z  \mbox{  if  }    - \delta_C < z <  \delta_C, \nonumber \\ 
&& \rho(z) = \frac{\Delta T}{2\delta_T}z  - \frac{\Delta C}{2}  \mbox{  if  }    \delta_C < z <  \delta_T, \nonumber \\ 
&& \rho(z) =  \frac{\Delta T}{2}  - \frac{\Delta C}{2} \mbox{  if  }   z>  \delta_T.
\label{eq:linRhomodel}
\end{eqnarray}
It is shown in figure \ref{fig:diffinterfacemodel}({\it b}). We see that having  $\delta_T > \delta_C$ combined with $\Delta T < \Delta C$ leads to the appearance of density minima and maxima just above and below the interface, respectively. From equation \eqref{eq:linRhomodel} and from figure \ref{fig:diffinterfacemodel}({\it b}), it is easy to show that the size of the density inversion $d\rho$ in this piecewise linear model is exactly
\begin{eqnarray}
    d\rho_D = \left( 1 - \frac{\delta_C}{\delta_T}\right) \frac{\Delta T}{2} \label{eq:drholinmodel1}
    \\ 
    =  \left( 1 - \frac{R_\rho\tau}{\gamma}\right) \frac{ L_z}{2}, \label{eq:drholinmodel2}  
\end{eqnarray}
where the second equality is obtained using (\ref{eq:interfaceratio}), and the fact that $\Delta T = L_z$ (see equation \ref{eq:iterface-jumps}). The first expression is useful if $\delta_C$ and $\delta_T$ are already known, while the second expression only requires knowledge of the input parameters $R_\rho$, $\tau$ and $L_z$, as well as the value of  $\gamma$ (which we argue can be approximated by $ \gamma_p$, reported in table \ref{tab:gammac}). 

\begin{figure}
    \centering
    \includegraphics[width=\linewidth]{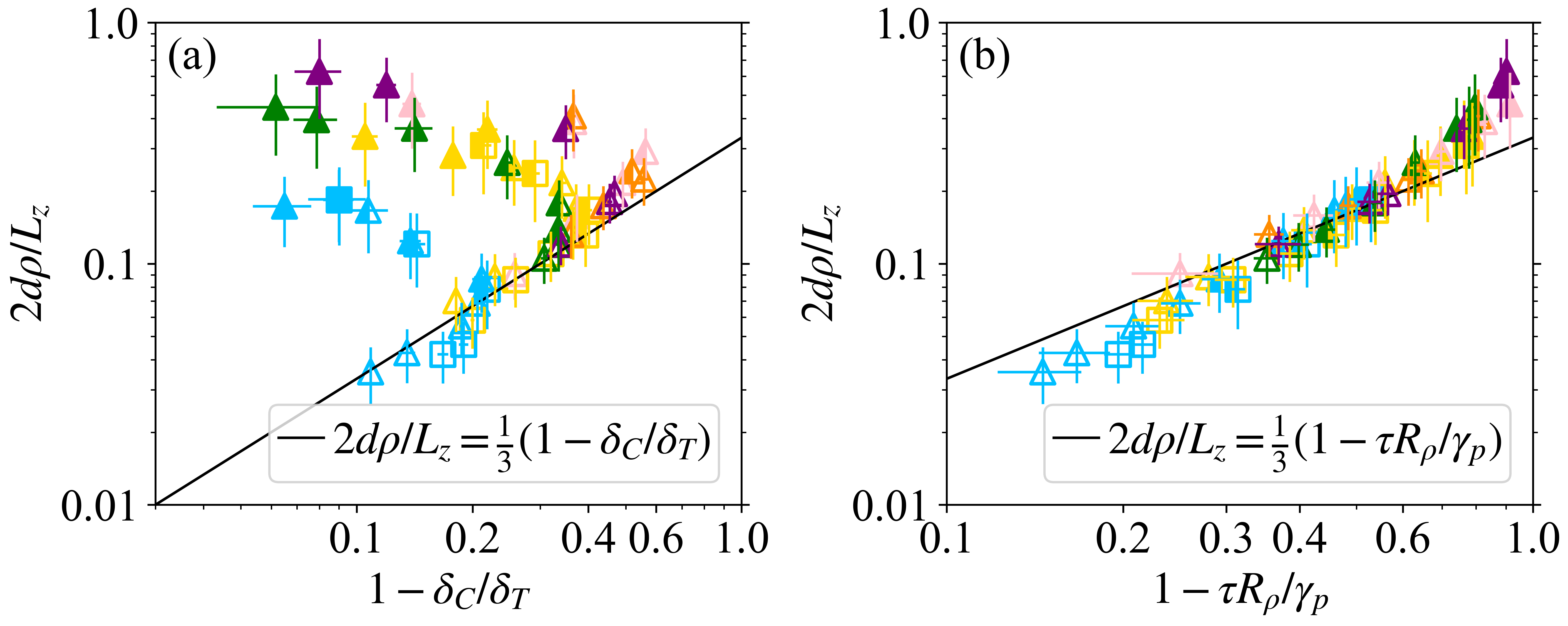}
    \caption{Scaled density anomaly $2 d\rho/L_{z}$ against two possible models. ({\it a}) Against $1 - \delta_{C}/\delta_{T}$. ({\it b})  Against $1-\tau R_{\rho}/ \gamma_p$, where $\gamma_p$ is given in table \ref{tab:gammac}. The latter provides a good fit to the data for a wider range of parameters. See table \ref{tab:simulations} for legend.}
    \label{fig:testlindrho}
\end{figure}

Figure \ref{fig:testlindrho} compares both predictions with the data. Figure \ref{fig:testlindrho}({\it a}) shows $2 d\rho/L_z$ against  $1-\delta_C/\delta_T$ to test equation (\ref{eq:drholinmodel1}), where $d\rho$, $\delta_T$ and $\delta_C$ are measured from the simulations as described in \S\ref{subsec:meanthick} and \S\ref{subsec:drhodata}. We see that equation (\ref{eq:drholinmodel1}) is a good model for  $d\rho$ for D-type interfaces (open symbols), and a few of the O-type interfaces, provided this expression is rescaled by a factor of approximately $1/3$. The fact that $d\rho$ is a little less than, but proportional to the predicted value can be explained by noting that the rising parcel entrains some of the fluid from the interface core (or lower layer) (see figure \ref{fig:illustration}), which partially reduces the density anomaly. This effect could in principle depend on the Prandtl number, but we are unable to establish this with certainty given the available data.

 We also see that the same expression fails to model $d\rho$ for many O-type and all of the U-type interfaces. In hindsight this is not surprising, as 
the mean thicknesses ($\delta_T$ and $\delta_C$) measured in O-type and U-type interfaces are more representative of the interface motion than of its structural properties (see figure \ref{fig:dzmid}). 

Figure \ref{fig:testlindrho}({\it b}) tests the second model for $d\rho$, namely equation (\ref{eq:drholinmodel2}), using $\gamma = \gamma_p$ (given in table \ref{tab:gammac}) for simplicity -- this is a good approximation for most O-type and D-type interfaces (see figure \ref{fig:TCfluxes}({\it c})). As expected, it does just as well as equation (\ref{eq:drholinmodel1}) for the D-type interfaces (after applying the same scaling factor of $1/3$), but now fits the data much better for all O-type interfaces and even for some of the U-type interfaces, which was unexpected. 

We can rationalize this surprising result a posteriori. An O-type interface
has a local vertical structure that is essentially very similar to the one shown in figure \ref{fig:interfacemodelprofile}({\it a}), but the midpoint of the interface and the respective thicknesses of the local temperature and composition profiles now vary with position and time. We still expect equation (\ref{eq:drholinmodel1}) to apply locally and instantaneously, however, as long as we replace $\delta_T$ with $d_T(x,y,t)$, and $\delta_C$ with $d_C(x,y,t)$. 
Furthermore, transport through the core of the interface remains close to being 
fully diffusive in O-type interfaces (see figure \ref{fig:interfacenusselt}), and as usual the interface structure must adjust to accommodate the flux ratio $\gamma$ coming from the convective zone below. To do this, the interface must locally satisfy
\begin{equation}
    \gamma = \tau \frac{\partial C/\partial z(x,y,z_i,t)}{\partial T/\partial   z(x,y,z_i,t)} = \tau R_\rho \frac{d_T(x,y,t)}{d_C(x,y,t)}. 
    \label{eq:gammaloc}
\end{equation} 

We see that this immediately leads to the same prediction (\ref{eq:drholinmodel2}) for $d\rho$ when expressed in terms of $\gamma$. We argue this is the main reason why \eqref{eq:drholinmodel2} is a better model for $d\rho$ than \eqref{eq:drholinmodel1} for most O-type interfaces, although the truth is a little more complicated, as explained in Appendix \ref{appA}. 

\begin{figure}
    \centering
    \includegraphics[width=0.8\linewidth]{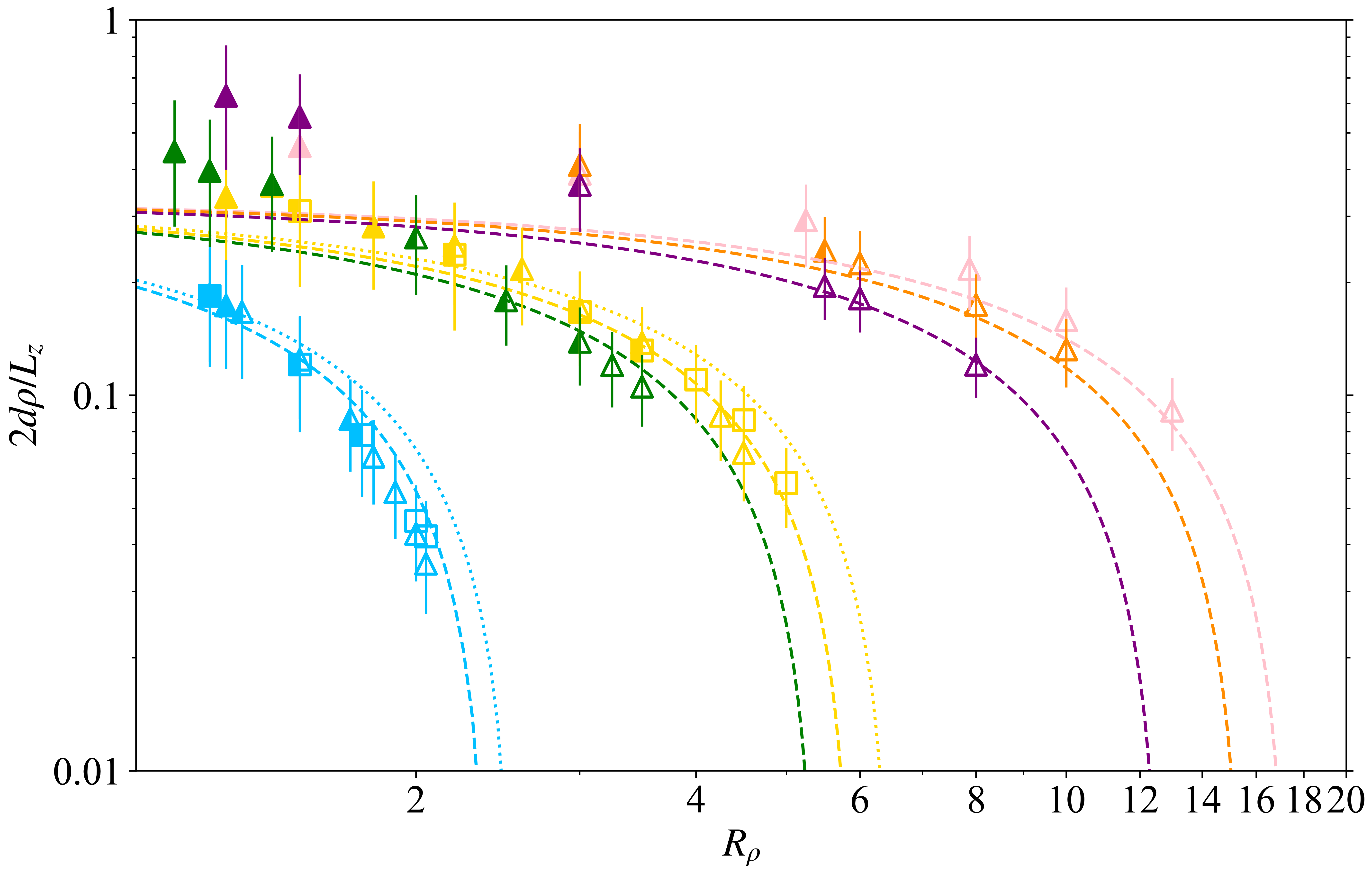}
    \caption{Scaled density anomaly $2 d\rho/L_{z}$ as a function of $R_{\rho}$ from all available simulations, where the symbol legend is given in table \ref{tab:simulations}. The dashed lines show the prediction obtained using $d\rho = \alpha_D d\rho_D$ from equation (\ref{eq:drholinmodel2}), with $\gamma = \gamma_p$ from table 2, $L_{z} = 100$ and $\alpha_D = 1/3$. The dotted lines show the corresponding predictions for the $L_{z}=200$ cases (these differ because of the different values of  $\gamma_p$).}
    \label{fig:drhomodels}
\end{figure}

A more direct comparison of the model given in equation (\ref{eq:drholinmodel2}) and the data is shown in figure \ref{fig:drhomodels}, which compares $2d\rho/L_z$ (symbols) with $\alpha_D d\rho_D$ (dashed lines). Again, we use $\gamma = \gamma_p$ for simplicity, with  $\gamma_p$ given in table \ref{tab:gammac}, and $\alpha_D = 1/3$.
 We see that this simple model is capable of predicting $d\rho$ reasonably well for a wide range of values of the input parameters. In particular, it captures the dependence of $d\rho$ on $R_\rho$, $\Pr$,  $\tau$, and even $L_z$ for all D-type interfaces and most of the O-type interfaces. 
We see there may be a weak dependence of the prefactor $\alpha_D$ on $\Pr$, but ignore this for now as this dependence can be absorbed into the model for the density flux presented below. For some of the most weakly stratified O-type and U-type interfaces, equation (\ref{eq:drholinmodel2}) underestimates $d\rho$ slightly, especially at low $\tau$. 

\subsection{Density flux through the layers}
\label{subsec:fluxmodel}

Having a model for $d\rho$, we can now predict the density flux through the layers using the 
empirical law $F^{\rm tot}_{\rho} \propto d\rho^{4/3}$ discovered in \S\ref{subsec:densityfluxlaw}. 
This 4/3 scaling law is strongly reminiscent of the corresponding scaling law in Rayleigh-B\'enard convection, and 
suggests using a simple ‘marginal plume'  argument similar the one described by \citet{Doering2009} based on 
\citet{Howard1963}.

We first reproduce the original argument for pedagogical purposes, then apply it to LDC. We use dimensional quantities for additional clarity and ignore compositional effects for now. Consider a convective zone located above an impermeable horizontal wall held at temperature $\Delta T^*/2$.  In the statistically stationary turbulent state, 
the core of the convective zone has constant mean temperature, which can be set to zero without loss of generality. The  fluid near the boundary is warmed up by the wall in a thin boundary layer of thickness $\delta_T^{*}$. Now and then, small plumes (also of typical size $\sim \delta_T^*$) detach from the wall. The length scale $\delta_T^*$ 
is primarily set by the following force balance in the vertical momentum equation:
\begin{equation}
    \alpha^*_T g^* \frac{\Delta T^*}{2} \sim \nu^* \frac{ w^*}{\delta_T^{*2}}, 
    \label{eq:dimmombalance}
\end{equation}
which expresses that the plumes detach from the wall when their typical buoyancy is just strong enough to overcome viscous drag. 
The characteristic vertical velocity in this equation is set by arguing that in a  statistically stationary state, there must be a balance between the diffusive flux of heat from the wall into the boundary layer, and the convected flux of heat out of the boundary layer by the plumes, namely: 
\begin{equation}
    w^* \Delta T^* \sim \kappa_T^*\frac{\Delta T^* }{2\delta_T^*} \rightarrow  w^* \sim \frac{ \kappa_T^*}{2\delta_T^*}. \label{eq:advdiff}
\end{equation}
Combining \eqref{eq:dimmombalance} and \eqref{eq:advdiff}, we obtain the standard result:
\begin{equation} 
\alpha^*_T g \frac{\Delta T^*}{2}
\sim \frac{\nu \kappa_T}{2\delta_T^{*3}} \rightarrow 
    \delta_T^* \sim \left( \frac{\alpha_T g \Delta T^*}{\kappa_T\nu}\right)^{-1/3}.
\end{equation}

Once $\delta_T^*$ is known, we can compute the temperature flux as:
\begin{equation}\label{eq:Tflux_model}
    F^*_T \sim  \frac{\kappa^*_T \Delta T^*}{2\delta_T^*} \sim \left( \frac{\alpha_T g \Delta T^*}{\kappa_T\nu}\right)^{1/3} \frac{\kappa^*_T  \Delta T^*}{2} \propto \Delta T^{*4/3}.
\end{equation} 
We can also compute the Nusselt number in the usual fashion as the ratio of the total heat flux to the diffusive heat  flux. If, for instance, $\Delta T^*$ and the total height of the domain $L_z^*$ are fixed, we recover the well-known result
\begin{equation}\label{eq:Nusselt_model}
\Nu_T \sim \frac{- \kappa_T \frac{\Delta T^*}{2\delta_T^*}}{-\kappa_T \frac{\Delta T^*}{L_z^*} } \propto \frac{L_z^*}{2\delta_T^*} \sim \Ra^{1/3},
\end{equation}
where in this setup
\begin{equation}\label{eq:Rayleign_definition}
    \Ra =  \frac{\alpha_T g \Delta T^* L_z^{*3}}{\kappa_T\nu}.
\end{equation}

We note that this marginal plume argument is unable to capture the experimentally well-known  dependence of $\Nu_T$ on the Prandtl number \citep[see, for example][] {VERZICCO_CAMUSSI_1999}, which would in practice lead to the relationship:
\begin{equation}\label{eq:Nu_scaling}
    \Nu_T = f(\Pr) \Ra^{1/3}.
\end{equation}
This dependence on $\Pr$, which is fairly weak, presumably arises because the shape of the plumes, and their coherence as they rise out of the boundary layer both depend on the viscosity and the thermal diffusivity.

LDC is fundamentally  similar: convective plumes are driven by density anomalies $d\rho$, that are created by the diffusion of heat across the interface (see \S\ref{subsec:snaps}). 
 For D-type interfaces in particular, the analogy with Rayleigh-B\'enard convection is highly relevant, as the core of the diffusive interface is so strongly stratified that it can be viewed as an almost impermeable wall.
 We begin by defining the characteristic length scale of the density anomalies to be $\delta_\rho$. 
Balance in the vertical component of the momentum equation yields equation (\ref{eq:dimmombalance}) as before, written this time non-dimensionally, and in terms of $d\rho$ and $\delta_\rho$, as 
\begin{equation}
    d\rho \sim \frac{w}{\delta_\rho^2}.
\end{equation}
 
From here on, the argument differs slightly from the case of pure convection, because density does not diffuse in the same way temperature does. The result, however, is  ultimately quite similar. As the plume rises, heat continuously diffuses out of it, maintaining the sheath of low density fluid seen in figures \ref{fig:blobs} and \ref{fig:illustration}. This requires a balance between advection and diffusion of temperature across the plume of size $\delta_{\rho}$, namely
\begin{equation}
    w \sim \frac{1}{\delta_\rho},
\end{equation}
when written non-dimensionally.
Combining the two we obtain
\begin{equation}
    d\rho \sim \frac{1}{\delta_\rho^3} \rightarrow \delta_\rho \sim d\rho^{-1/3},
\end{equation}
and use this to estimate the density flux through the staircase as
\begin{equation}
    |F_\rho| \sim 
    w d\rho \sim \frac{d\rho}{\delta_\rho} 
    \sim  d\rho^{4/3},
\end{equation}
where we reserve the notation $F_\rho$ specifically for the model prediction.
Noting that the prefactor is expected to depend on $\Pr$ (as it does for Rayleigh-B\'enard convection) finally explains the observed scaling law $F^{\rm tot}_\rho = - C_\rho(\Pr) d\rho^{4/3}$ discovered in figure \ref{fig:rhoflux}, see equation (\ref{eq:Frhofunction}), and provides a plausible explanation for why $C_\rho$ seems to be independent of $R_\rho$, $\tau$ or $ L_z$. 
In \S \ref{subsec:densityfluxlaw}, we tentatively found that $C_\rho \sim \Pr^{1/3}$ in the intermediate range of $\Pr$ used in the DNS, but we note that this may not apply at large $\Pr$ or very low $\Pr$.

\subsection{Temperature and composition fluxes through the layers}
\label{subsec:FTFC}

In order to derive a model for the temperature and composition fluxes $F_T$ and $F_C$ from $F_\rho$, we need to know the flux ratio $\gamma$ since 
$F_\rho = - F_T + F_C = -F_T (1-\gamma)$, and $F_C = \gamma F_T$. (Note that as in the previous section, we use the notation $F_T$ and $F_C$ to refer to the model, reserving $F^{\rm tot}_T$ and $F^{\rm tot}_C$ for the data). 

We saw in figure \ref{fig:TCfluxes}({\it c}) that the dependence of $\gamma$ on input parameters is complex. In \S\ref{subsec:snaps}, we provided a rationale for that complexity by noting that $\gamma$ in the convective layers is set by the interplay between turbulence and double diffusion. 
Because of this inherent complexity, we do not try to develop an ad-hoc model for $\gamma$ in this paper, leaving this task for future work. Instead, we leverage the fact that $\gamma$ is roughly independent of $R_\rho$ for O-type and D-type interfaces, and 
assume that $\gamma \simeq \gamma_p(\Pr,\tau,L_z)$ for all $R_\rho$, where  $\gamma_p$ is empirically obtained from the data, and given in table \ref{tab:gammac}. 
We then have
\begin{eqnarray}
F_T = - \frac{1}{1- \gamma_p} F_\rho =  \frac{C_\rho(\Pr)}{1-\gamma_p}  d\rho^{4/3}, \label{eq:FTmodel} \\
F_C = -\frac{\gamma_p}{1-{\gamma_p}} F_\rho = \frac{\gamma_p C_\rho(\Pr)}{1-\gamma_p}  d\rho^{4/3}. \label{eq:FCmodel} 
\end{eqnarray}
Given that $d\rho \propto L_z \propto \Delta T$, these laws recover the well-known result $F_{T,C} \propto \Delta T^{4/3}$ of \citet{TURNER1965}, and additionally provide estimates for the prefactors.

\begin{figure}
    \centering
    \includegraphics[width=\linewidth]{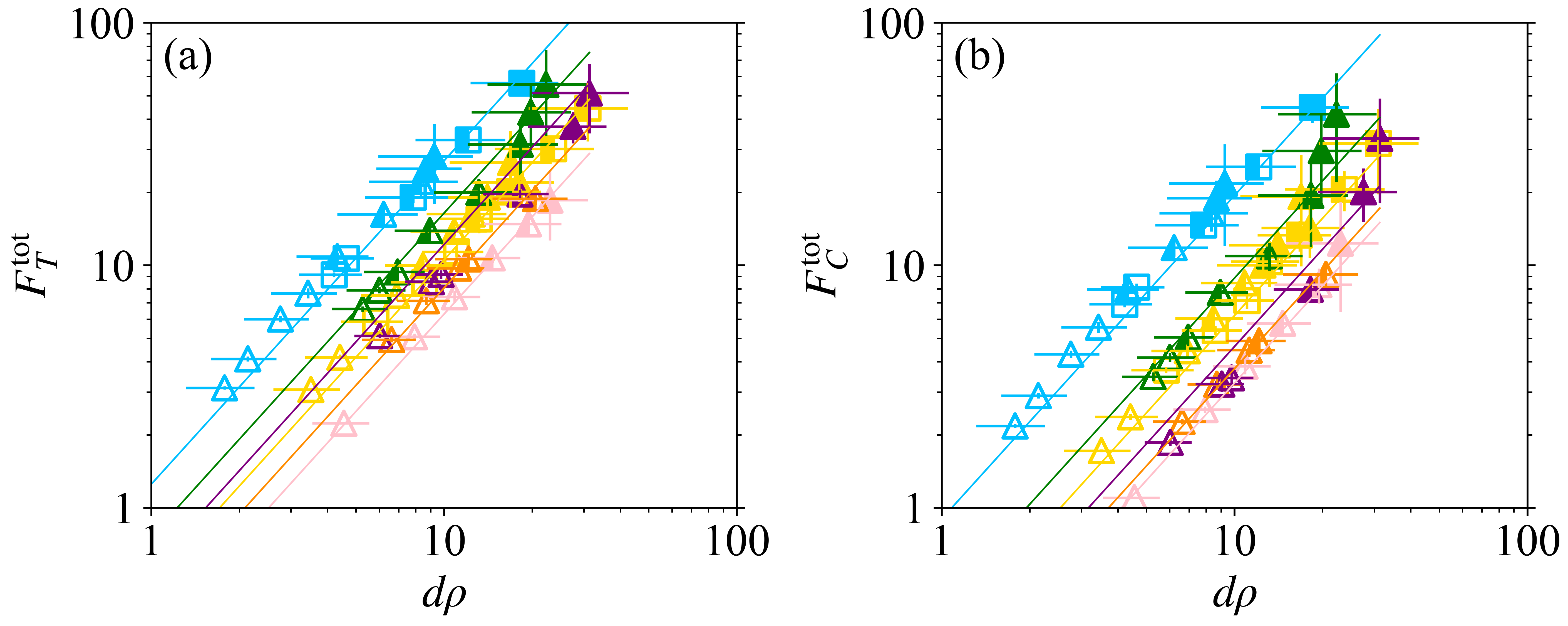}
    \caption{({\it a}) Measured temperature flux $F^{\rm tot}_{T}$ as a function of measured $d\rho$ with corresponding errorbars. ({\it b}) Measured  composition flux $F^{\rm tot}_{C}$ as a function of measured $d\rho$ with corresponding errorbars. In both panels,  solid lines of the corresponding colour show the models for $F_T$ and $F_C$ given in equations \eqref{eq:FTmodel} and \eqref{eq:FCmodel}, with $C_\rho(\Pr)$ given in \eqref{eq:Crho} and $ \gamma_p$ from table \ref{tab:gammac}.  }
    \label{fig:compareTCfluxmodel}
\end{figure}

Figure \ref{fig:compareTCfluxmodel} compares these predictions with the data, using $\gamma =  \gamma_p$ (with $ \gamma_p$ given in table \ref{tab:gammac})  for simplicity, and using $d\rho$ measured from the simulations. We see that, generally speaking, the model is a rather good fit to the data for both $F^{\rm tot}_T$ and $F^{\rm tot}_C$ (except for U-type interfaces, which we are not trying to model).

We can also use equations (\ref{eq:FTmodel}) and (\ref{eq:FCmodel}) to compute the Nusselt numbers for $T$ and $C$. We find:
\begin{eqnarray}
\Nu_T = \frac{F_T}{(\Delta T/L_z)} =  \frac{C_\rho}{1-\gamma_p} d\rho^{4/3} = \frac{C_\rho}{1-\gamma_p}  a_\rho^{4/3} \Ra^{1/3}, \label{eq:NuT1}  \\
\Nu_C = \frac{F_C}{(\tau \Delta C/L_z)} =  \frac{ \gamma_p C_\rho}{(1-\gamma_p)\tau R_\rho}  d\rho^{4/3} = \frac{\gamma_p C_\rho}{(1-\gamma_p)\tau R_\rho}  a_\rho^{4/3} \Ra^{1/3},  
\label{eq:NuC1}
\end{eqnarray}
where we have used our earlier finding that $d\rho \propto L_z$ (see equation \ref{eq:drholinmodel2}) to rewrite $d\rho$ as 
\begin{equation}
d\rho = a_\rho(R_\rho,\tau, \gamma_p)L_z, \mbox{ with }  a_\rho(R_\rho,\tau, \gamma_p) = \frac{\alpha_D}{2}\left(1 - \frac{R_\rho \tau}{\gamma_p}\right),
\label{eq:arho}
\end{equation}
 and the fact that in the chosen system of units the Rayleigh number is simply $\Ra = L_z^4$.
Because $C_\rho \propto  \Pr^{1/3}$ for this range of parameters (see equation \ref{eq:Crhoscaling}), one may naively conclude that $\Nu_T$ scales as  $(\Ra\Pr)^{1/3}$, and $\Nu_C$ scales as $(\Ra\Pr)^{1/3}/\tau R_\rho$. This would in fact recover the scaling laws advocated by \citet{Woodal13} from their DNS of LDC. 

However, it is important to recall that some of the other prefactors in \eqref{eq:NuT1} and \eqref{eq:NuC1}, including $a_\rho$, depend on $\gamma_p$, and that $\gamma_p$ is a function of $\tau, \Pr$ and  $L_z$ (see, e.g. figure \ref{fig:TCfluxes}({\it c})). As such, the dependence of the Nusselt numbers on $\Pr$, $\tau$, $R_\rho$ and even on $\Ra$ is more complex than what this naive argument suggests. 
\begin{figure}
    \centering
    \includegraphics[width=\linewidth]{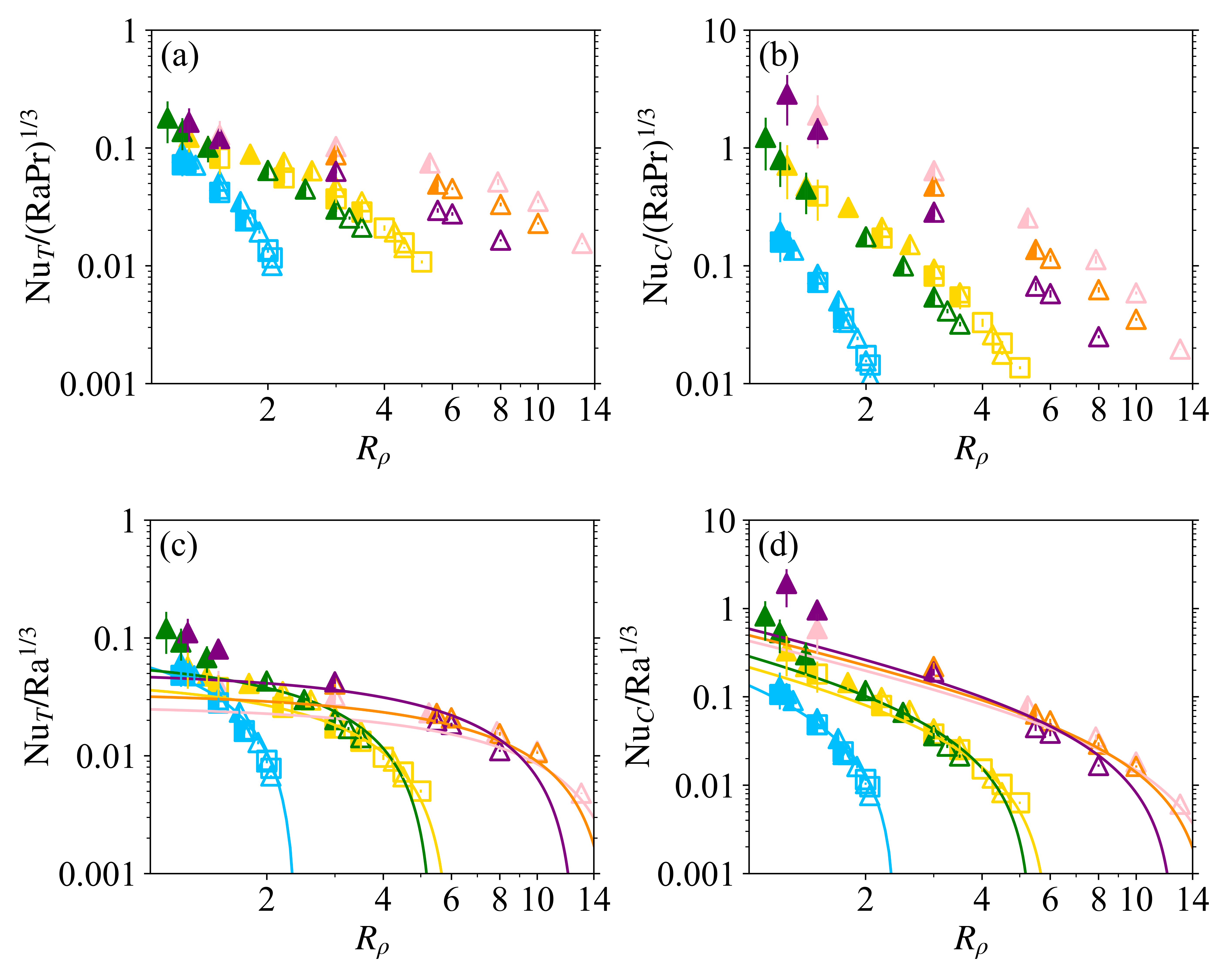}
    \caption{({\it a,b})  $\Nu_T/(\Ra\Pr)^{1/3}$ and $\Nu_C/(\Ra\Pr)^{1/3}$ as functions of $R_{\rho}$.  ({\it c,d}) $\Nu_T/\Ra^{1/3}$ and $\Nu_C/\Ra^{1/3}$ as functions of $R_{\rho}$. See table \ref{tab:simulations} for symbol legend. The solid lines in the lower panels show the model predictions for $\Nu_T$ and $\Nu_C$ given in equations \eqref{eq:NuT1} and \eqref{eq:NuC1}, respectively, with $\alpha_D = 1/3$, $\gamma_p$ given in table \ref{tab:gammac}, $C_\rho(\Pr)$ given in \eqref{eq:Crho}, and the predicted values of $d\rho$ provided by equation \eqref{eq:drholinmodel2}. In all panels, the vertical error bars are present but are often smaller than the marker size.}
    \label{fig:Nusselts}
\end{figure}
This can be verified in figure \ref{fig:Nusselts} which compares two rescalings of the Nusselt numbers with respect to $\Ra$ and $\Pr$: in the top row, we show $\Nu_T/(\Ra\Pr)^{1/3}$ and $\Nu_C/(\Ra\Pr)^{1/3}$  \citep[suggested by][]{Woodal13}, while in the second row we show $\Nu_T/\Ra^{1/3}$ and $\Nu_C/\Ra^{1/3}$. We see that $\Nu_{T,C} / \Ra^{1/3}$ seems to depend primarily on $\tau$ and $R_\rho$, with almost no remaining dependence on $\Pr$ or $L_z$.  

This apparent discrepancy between our conclusions and those of \citet{Woodal13} likely arises from the fact that they had only explored LDC at very low $R_\rho$, for which the interfaces are U-type, while most of our data has O-type and D-type interfaces. It thus seems that $\Nu_{T,C} \propto (\Ra\Pr)^{1/3}$ only applies for genuinely turbulent (U-type) interfaces, while $\Nu_{T,C} \propto \Ra^{1/3}$ is more appropriate for more stable O-type and D-type interfaces. Future DNS will be required to establish whether this remains true at much lower values of $\Pr$ and $\tau$ and larger layer heights.

Finally, we compare the model predictions for the Nusselt numbers $\Nu_T$ and $\Nu_C$ given in equations \eqref{eq:NuT1} and \eqref{eq:NuC1} (using the predicted $d\rho$ from equation \eqref{eq:drholinmodel2}, $C_\rho(\Pr)$ given in \eqref{eq:Crho}, and $ \gamma_p$ from table \ref{tab:gammac}) with the data. The predictions are shown as coloured solid lines in panels {\it (c)} and  {\it (d)} of figure \ref{fig:Nusselts}, using the same colour-coding as the corresponding markers (see table \ref{tab:simulations}). We can see that the predictions for $\Nu_{T}$ and $\Nu_{C}$ 
underestimate the fluxes for U-type interfaces.  This is not surprising, given that the $d\rho$ model assumes a diffusive interface core.  However, we also see that the model works extremely well for more stable O- and D-type interfaces, suggesting that it can be used to predict fluxes through an infinite uniform staircase at low $\Pr$ and $\tau$, and moderate-to-high density ratio $R_\rho$.

\section{Discussion and conclusion}\label{sec:discussion}

\subsection{Summary}
In this paper, we systematically studied the properties of statistically stationary infinite uniform diffusive-convective staircases at low Prandtl number using DNS with triply-periodic boundary conditions, for a wide range of parameters $\Pr$, $\tau$, $R_{\rho}$ and $L_{z}$ (see table \ref{tab:simulations}). We found that the interfaces between the layers can be categorized into three types depending on their properties. At the lowest $R_{\rho}$,  U-type interfaces are weakly stratified, strongly turbulent, and are thought to be unstable. For more moderate $R_{\rho}$, O-type interfaces are more strongly stratified, weakly turbulent and oscillate vertically.  The vertical temperature and composition profiles of O-type interfaces are well approximated by $\tanh$ functions with relatively small interface thicknesses. Finally the diffusive interfaces (D-type) exist at the highest $R_{\rho}$. D-type interfaces have a linearly-stratified purely diffusive core whose structure adapts to accommodate the fluxes coming from the convective zone below. Beyond a critical density ratio  $R_L$, convection in the layers becomes too weak, the interface thickness grows to fill the domain, and the layers disappear. Notably, we found that  $ R_L = \gamma/ \tau$, where $\gamma$ is the flux ratio coming from the convective layer below the interface. At low $\Pr$ and low $\tau$, $ R_L$  can be significantly larger than previously suggested by either \citet{Linden_Shirtcliffe_1978} (who argued that $ R_L = \tau^{-1/2}$) or \citet{Molletal2017} (who found that $ R_L = R_c$).

Using measured fluxes of density ($F_{\rho}^{\rm tot}$), temperature ($F_{T}^{\rm tot}$) and composition ($F_{C}^{\rm tot}$) for all of our simulations, we computed the flux ratio $\gamma$ and Nusselt numbers  $\Nu_{C}$ and $\Nu_{T}$. We found that $\gamma$ is roughly independent of $R_{\rho}$ for O- and D-type interfaces, and instead depends only on $\Pr, \tau$ and increases slowly with $L_{z}$. This strongly suggests that the value of $\gamma$ is set by the turbulent properties of the convection, rather than being a property of the interface. We also found that $\Nu_{C}$ and $\Nu_{T}$ are approximately proportional to $\Ra^{1/3}$ for O- and D-type interfaces, which differs from the $(\Ra\Pr)^{1/3}$ scaling proposed by \citet{Woodal13} for low $R_\rho$ (U-type) staircases. 
Convection in the layers is driven by small localized density anomalies that emerge from the differential diffusion of temperature and composition above and below the interfaces. We measured the characteristic amplitude $d\rho$ of these anomalies in all of our simulations and found that $F^{\rm tot}_{\rho}$, $F^{\rm tot}_{T}$ and $F^{\rm tot}_{C}$ all scale as $d\rho^{4/3}$. This scaling law can be explained by an argument of marginal stability of the boundary layer similar to that of \citet{Doering2009} \citep[see also][]{Howard1963} for Rayleigh-Bénard convection, up to a prefactor $C_\rho(\Pr)$  that can be fitted to the data.   

Assuming a locally diffusive interface structure (which could be flat for D-type interfaces or wavy for O-type interfaces), we  proposed a simple model for $d\rho$ that depends on the ratio of the local composition and temperature interface thicknesses (see equation \ref{eq:drholinmodel1}), which themselves are controlled by $\tau, R_{\rho}$ and $\gamma$. The model was shown to fit the data well as long as $\gamma$ is known. Given $d\rho$, we can then predict the fluxes through the uniform staircase, using equations \eqref{eq:FTmodel} and \eqref{eq:FCmodel}, and found a good agreement with the measured data as well, as long as $\gamma$ and $C_\rho$ are known.   

\subsection{Caveats}
While our model predictions are distinguishably good within the explored parameter range  (except for U-type interface), their applicability beyond this range remains uncertain. We have focused on a numerically accessible parameter regime where both $\Pr$ and $\tau$ range from $0.03$ to $0.3$, and it is unclear whether the model would hold for values of $\Pr$ and $\tau$ much smaller than $0.03$ or much larger than $0.3$ (i.e. as in the oceanographic case). In particular, the value of the flux ratio $\gamma$ (which is needed to predict $d\rho$ and disentangle the contributions of  $F^{\rm tot}_{T}$ and $F^{\rm tot}_{C}$ to the total density flux $F^{\rm tot}_{\rho}$) clearly has a non-trivial dependence  on $\Pr$, $\tau$, and $L_z$, but that dependence remains to be explained from first principles. 

Furthermore, again due to numerical constraints, our simulations have been limited to dimensionless domain heights $L_{z} \leq 200$. Since  the ODDC length scale $d^*$ (see \S\ref{sec:method}) ranges from a few centimeters  to a few hundreds of meters depending on the astrophysical application considered \citep{Garaud2020}, the simulated layer heights $L_z^* = L_z d^*$ are very small compared with the expected system size. Therefore, quantities that are observed to have a weak dependence on the layer height (such as $\gamma$) could be quite different in situ from the values empirically inferred from the simulation data. Notably, if $\gamma$ increases with $L_z$, but is bounded from above by 1 in LDC, we conclude that $\gamma$ will slowly approach 1 as $L_z$ increases. Similarly, we note that the Rayleigh number in our setup is $\Ra = L_{z}^{4}$, and was therefore limited to $\Ra \simeq O(10^{9})$, which is high but not asymptotically high. Given the ongoing debate in Rayleigh–B\'enard convection theory regarding a possible transition from $\Nu \propto \Ra^{1/3}$  to an `ultimate’ regime in which $\Nu \propto \Ra^{1/2}$ for $\Ra \ge 10^{15}$ \citep{Heetal2012,ShishkinaLohse2024}, the $\Ra^{1/3}$ scaling observed in our simulations may not persist for larger $L_{z}$. This potential transition is relevant, since $\Ra = 10^{16}$ corresponds to $L^*_z = 10^4d^*$, which is still significantly smaller than the planet’s or star’s radius.  Another limitation that we must also bear in mind when the layer height increases is that the Boussinesq approximation might cease to be valid. Lastly, it has been shown that magnetic fields and/or rotation impede convection and entrainment in LDC at low $\Pr$ \citep{MollGaraud2017,Sanghi_2022,Fuentes_2023}. These effects will need to be included in the future to provide a more realistic view of transport through diffusive-convective staircases in stars and giant planets.

\subsection{Future prospects}

Despite these caveats, our findings open several interesting prospects. In these simulations, we have forced the staircases to have uniform step heights and to be in a statistically stationary state by using triply-periodic boundary conditions. In reality, however, the steps do not always have equal heights, and undergo mergers over longer time scales in low $\Pr$ layered convection \citep{rosenblumal2011,Mirouh2012,Woodal13,Fuentes_2023,Tulekeyev2024}. Three possibilities arise for the ultimate outcome of the layer merger process: (1) the layers merge until an equilibrium height is reached and the staircase achieves a statistically stationary state; (2) the layers continue to merge but the merger time scale increases with layer height and ultimately becomes longer than the age of the system (in which case the staircase has the appearance of being in a statistically stationary state); and (3) the mergers proceed until the entire region is fully convective. \citet{radko2007} developed a theory to predict the merger time scale of layers in uniform density staircases, and showed that this time scale can be estimated knowing the dependence of the density flux $F_\rho^{\rm tot}$ on the layer height $L_z$ and on the density jump across the interface $\Delta \rho$. The flux laws derived here provide precisely the information needed to apply Radko's merger theory. In Tulekeyev et al. 2026 (in prep), we thus use our findings to compute the merger time scale for layered convection in the interiors of planets such as the Solar System's gas giants, where the relevant ranges of $\Pr$ and $\tau$ are comparable to those considered in this study.  

 Our findings also have interesting prospects 
beyond astrophysical applications. By arguing that the flux ratio $\gamma$ is first and foremost controlled by the convective dynamics within each layer (supported by our findings that $\gamma$ depends weakly on the layer height), which then sets the structure of the interface above, we are proposing a new paradigm for transport in LDC. It would therefore be interesting to see if $\gamma$ also depends on layer height for LDC in lakes and in the ocean, or in DNS at $\Pr \ge O(1)$ \citep[as predicted by][]{GoughToomre1982}. To our knowledge, this has not been investigated yet as most laboratory and numerical experiments have been run in roughly similar domain heights, and in-situ measurements do not always provide information on $\gamma$. We also believe that the proposed boundary layer argument relating the characteristic density anomaly $d\rho$ to the density flux $F^{\rm tot}_\rho$, as well as the proposed model for the density anomaly itself, should both remain valid at higher $\Pr$. Using a high precision microstructure profiler (or using DNS) it ought to be possible to measure $d\rho$, as well as the rms vertical velocity $w_{\rm rms}^{\rm cz}$ and rms density fluctuation $\rho_{\rm rms}^{\rm cz}$ (whose product is closely related to $F^{\rm tot}_\rho$, see \S\ref{sec:convective_w}), to check whether equations (\ref{eq:Frhofunction}) and (\ref{eq:drholinmodel2}) continue to apply. If that is the case, then the temperature and salt fluxes through Arctic thermohaline staircases should be adequately predicted by equations \eqref{eq:FTmodel} and \eqref{eq:FCmodel}  (with the caveat that the flux ratio $\gamma$ and the prefactor $C_\rho(\Pr)$ need to be recalibrated).

\begin{appen}

\section{}\label{appA}

In this appendix, we take a closer look at why equation \eqref{eq:drholinmodel2} is a good model for $d\rho$ for both O-type and D-type interfaces. This model essentially assumes that the interface core is diffusive, and accommodates a flux ratio $\gamma$ (approximated by $\gamma_p$) coming from the layer below, so that the local interface thickness ratio is everywhere $d_{C}/d_{T} \simeq \tau R_{\rho}/ \gamma_{p}$. We saw in figure \ref{fig:dCdT_joinPDF} that, in practice, $d_C/d_T$ is not always close to $\tau R_\rho/\gamma_{p}$. More specifically, we found that it takes values in the range $\tau R_\rho/\gamma_{p}\lesssim d_C/d_T \lesssim 1$ for O-type interfaces. Consequently, the local density anomalies should vary in amplitude from $0$ (when $d_C = d_T$) to $d\rho_{\rm max} = (L_{z}/2)(1- \tau R_\rho/ \gamma_{p})$ (when $d_C/d_T = \tau R_\rho/ \gamma_{p}$).

Figure \ref{fig:RatioVSdrho_jPDFs} shows the joint probability density function $P(d\rho|d_{C}/d_{T})$, namely  
\begin{equation}
     P(d\rho|d_{C}/d_{T}) = \text{joint  p.d.f.  of $d\rho(x,y,t)$ and $d_{C}(x,y,t)/d_{T}(x,y,t)$}
\end{equation}
for the same simulations as in figure \ref{fig:dCdT_joinPDF}. In both panels, the horizontal solid red line represents the mean density anomaly $d\rho$ computed as explained in \S\ref{subsec:drhodata}, which is also the mean of the corresponding marginal pdf. The red dotted lines correspond to the error bars on $d\rho$ computed from the rms variability around that mean. As in figure \ref{fig:dCdT_joinPDF}, the pink line corresponds to $d_{C}/d_{T} = \tau R_\rho / \gamma_{p}$, and the black line corresponds to $d_{C}/d_{T} = 1$. Finally, the orange slanted line corresponds to the value of $d\rho$ predicted using equation (\ref{eq:drholinmodel1}) with $d_C/d_T$ replacing $\delta_C/\delta_T$ and including the prefactor $\alpha_{D} = 1/3$, namely $d\rho = \alpha_D (\Delta T/2)(1 - d_C/d_T)$. Our proposed model for the  density anomaly given in equation (\ref{eq:drholinmodel2}) is thus at the level of the intersection of the pink and orange lines, and is 
plotted as a horizontal dashed red line. 

\begin{figure}
    \centering
    \includegraphics[width=0.99\linewidth]{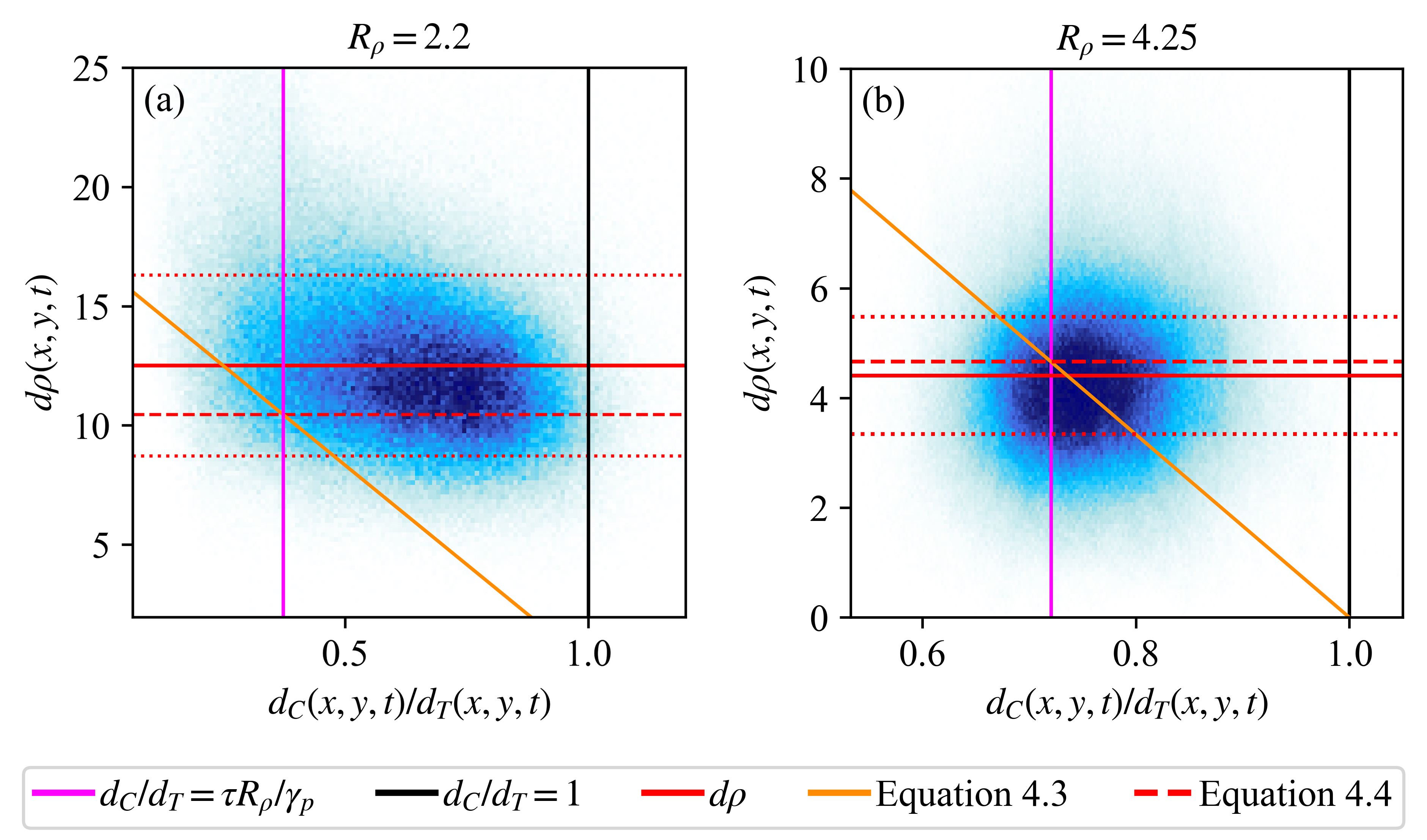}
    \caption{Colour maps of $P(d\rho|d_{C}/d_{T})$ from sample simulations at $\Pr = \tau = 0.1$, $L_{z}=100$ and ({\it a}) $R_{\rho} = 2.2$ (O-type interface) and ({\it b}) $R_{\rho} = 4.25$ (D-type interface). The horizontal solid red line shows the measured $d\rho$ with rms variability as dotted red lines. The vertical pink line
    corresponds the diffusive limit $d_{C}/d_{T} = \tau R_{\rho}/ \gamma_p$  and the black line shows $d_{C}/d_{T} =1$. The model in \eqref{eq:drholinmodel1} is shown as the orange line. The dashed red line corresponds to the value of $d\rho$ given in \eqref{eq:drholinmodel2}.} 
\label{fig:RatioVSdrho_jPDFs}
\end{figure}  

For the D-type interface in figure \ref{fig:RatioVSdrho_jPDFs}({\it b}), most measured values of $d_{C}/d_{T}$ lie  close to the intersection of the pink and orange lines as expected, and, correspondingly, the predicted value of $d\rho$ is close to the measured $d\rho$. For O-type interfaces in  figure \ref{fig:RatioVSdrho_jPDFs}({\it a}), by contrast, we see that $d_C/d_T$ is often substantially larger than the location of the pink line, but crucially, that the measured mean $d\rho$ remains close to the model prediction. We interpret this curious finding as follows: density anomalies with the predicted $d\rho$ primarily form when the interface is in its most laminar stage, i.e. when $d_C/d_T$ is smallest and $d\rho$ is largest, then begin to rise/fall. When the density anomaly detaches from the interface, it causes a little bit of mixing, which increases $d_C$ towards $d_T$ and explains the lateral shift in the centroid of the joint pdf. The effect is not as pronounced for D-type interfaces because their core is much thicker, so $d_C$ is much less affected by small mixing events.

\section*{Declaration of Interests}
The authors report no conflict of interest.

\end{appen}\clearpage

\bibliographystyle{jfm}
\bibliography{Arstanbib}

\end{document}